\pdfoutput=1
\documentclass[aps,pre,reprint,groupedaddress,nofootinbib]{revtex4-1}

\usepackage{graphicx}
\usepackage{float}
\usepackage{wrapfig}
\usepackage{subfigure}
\usepackage{color}
\usepackage{hyperref}
\usepackage{amsmath}
\usepackage{mathtools}
\usepackage{amssymb}
\usepackage{tabularx}
\usepackage{dcolumn} 

\usepackage{caption}

\usepackage{placeins}

\newcolumntype{d}[1]{D{.}{.}{#1}}

\begin{document}



\title{Self-Organization of Self-Clearing Beating Patterns\\
in an Array of Locally Interacting Ciliated Cells\\
Formulated as an Adaptive Boolean Network}


\author{Martin Schneiter}
\author{Jaroslav Ri\v{c}ka}
\author{Martin Frenz}

\email[]{martin.frenz@iap.unibe.ch}
\affiliation{Institute of Applied Physics, University of Bern, Sidlerstrasse 5, 
3012 Bern, Switzerland}


\date{\today}

\begin{abstract}
We present a two-dimensional array of locally interacting virtual
actuators, which exhibits self-organization towards 
self-cleaning spatio-temporal structures. 
Some of the elaborated results might be more general for dynamical 
systems based on local interactions. However, here, we would like to 
present our model under the aspect of mucociliary clearance. 
The observed spatio-temporal ciliary beat patterns leading to 
proper mucociliary transport on multiciliated epithelia 
are suspected to 
be the result of self-organizing processes on various levels. 
Our two-dimensional array of locally interacting system elements
can be seen as an oversimplified pluricellular epithelium model, 
which intends to make 
the self-organization of ciliary beating patterns as well as 
of the associated fluid transport across the airway epithelium plausible.
Ciliated cells are modeled in terms of locally interacting oscillating 
two-state actuators. 
The local interactions among these boolean actuators are triggered by seeded 
mucus lumps. 
In the course of a simulation the actuators' state and 
the associated 
mucus velocity field self-organize in tandem. 
We suggest to consider the dynamics on multiciliated epithelia in the context 
of adaptive (boolean) networks. 
Within the framework of adaptive boolean networks ciliated cells represent the 
nodes
and as the mucus establishes the local interactions among nodes, its 
distribution determines the topology of the network. 
Furthermore, we present the results and insights from comprehensive parameter 
studies.  
The results show evidence that 
so called deterministic update schemes, which are meant to represent 
intercellular signaling, lead to more
realistic and robust dynamics and may therefore be favored by nature. 
Finally, we suppose that unciliated cells introduce a modular network topology 
on ciliated epithelia 
causing the self-organization taking place simultaneously in each ``ciliated 
module''. 
This reasoning provides the first consistent explanation for the meaning of the 
observed patchy 
expression patterns of the mucus modulation wave fields. 
Modularity may therefore be seen as a modular construction plan of nature 
for ciliated epithelia whose 
number of cells range over several order of magnitudes.   
\end{abstract}


\maketitle


\section{Introduction}
\label{sec:intro}
Motivation for the present work is to contribute to the understanding of the 
fascinating  
and omnipresent phenomenon of mucociliary transport. \par 
The epithelium of our airways constitutes a self-cleaning surface protecting 
our 
lungs
from a variety of inhaled substances such as exhaust, dust, bacteria and other 
harmful substances of micro- and submicrometer size.  
These particles get entrapped by the mucus layer lining the inner surface of 
the 
tracheal 
and pulmonary airways, which is propelled by the coordinated oscillatory 
movement of millions of subjacent cilia.
Cilia are hair-like protrusions of the cell membrane, an overview of their
structure and function can be found in \cite{Linck2009, Satir2007}. \par
To gain insight into the mucociliary clearing mechanism, many experimental 
techniques 
on various length- and time-scales are conducted. Experimental studies 
typically 
focus 
either on a structural or functional component  
on a typical scale, like the remarkable insight into the detailed molecular 
structure
of the axoneme (see e.g. \cite{Sui2006,Burgess2003}), the orientation 
of the ciliary beating plane \cite{Satir2007}, 
the distribution of the different epithelial cell types (e.g. 
\cite{Oliveira2003, Plopper1983}) 
or the observation of mucociliary phenomena on the pluricellular level 
\cite{Ryser2007}.
Even though much work has been done on various scales, our understanding 
of the basic mesoscopic function of the system still appears rather limited.
Experimental methods face several challenges: until today, 
it is not possible to observe the details of the mucociliary dynamics in vivo.
Further, it is difficult to observe the mucociliary clearing mechanism under 
controlled conditions. And finally, 
it is highly complex and laborious to simultaneously measure structural and 
functional parameters at different scales. \par
Mathematical models are perfectly suited to conduct parameter studies in order 
to determine 
the effects of structural and functional parameters on mucociliary phenomena 
and 
therefore, 
serve as an alternative approach for the investigation of the intriguing 
mucociliary phenomena.  
Considering 
the implementation of mucociliary interactions 
the existing models can roughly be divided into two classes.\par
One class of models prescribes the motion of cilia, including their 
coordination, and concentrates on the
hydrodynamics and rheology of the system.
In these models the action of cilia is modeled as distributed oscillating 
momentum source, 
such as oscillating envelope \cite{Ross1974}, traction layer \cite{Lubkin2007}, 
active porous medium or 
oscillating array of cilia represented by a distribution of hydrodynamic 
singularities along the cilia's 
centerline \cite{Smith2007}.  
The aim of these studies is to 
elaborate the geometrical and rheological conditions under which the system 
achieves 
an efficient transport (e.g. \cite{Lee2011}).\par
More relevant to the context of the present study is a second class of models, 
aimed at the understanding of the emergence of the pattern of motion. 
In these models the details of the motion of cilia or their coordination 
are not prescribed explicitly.
The coordinated behavior rather emerges in the course of self-organization, 
during which the system components interact locally, what
drives the system from an initially uncoordinated state towards a globally 
coordinated state exhibiting cooperative behavior. \par
Self-organization may play a role on various time- and length-scales, at 
various 
levels for the generation 
of mucociliary transport. As discussed in \cite{Marshall2010} the interactions 
between 
cilia-generated fluid flow and planar cell polarity (PCP) signaling may lead 
to self-organization during the morphogenesis, which
establishes the alignment of the axonemes on the individual cells. 
Much of work has been done on the molecular level, concerning the  
molecular motors (dynein motor proteins), their organization in the axoneme as 
well as the hydrodynamics of the resulting model cilium 
(eg. \cite{Riedel-Kruse2007, Hilfinger2008a, Hilfinger2009}). 
The complex motion pattern of a single cilium is thought to result from 
the self-organization of many dynein motor proteins generating stresses on the 
elastic microtubules in the axoneme.  
The next higher level,
the cellular level, concerns the formation of metachronal waves by 
the self-organized synchronization of cilia covering a single 
cell \cite{Elgeti2013}. 
For a long time it remained an open question whether the synchronization 
is achieved by cellular signaling (membrane potentials and calcium waves), or 
if 
it emerges spontaneously, 
due to interactions between the individual cilia. Today, the computational 
models indicate that hydrodynamic
coupling is sufficient to induce synchronization \cite{Elgeti2013, Mitran2007, 
Gueron1997}. 
It is conceivable, however, that calcium signaling is needed 
for fine tuning the synchronization \cite{Salathe2007}.\par
Here we propose a first step to the next higher level, the
pluricellular level, on which the self-organization 
of ciliary activity among ciliated cells  
is thought to generate the global wave field and fluid transport 
on the airway epithelium.
In order to make the self-organized beat patterns as well as self-organized 
fluid transport across the airway epithelium plausible, we 
present an oversimplified model, 
which is intended to represent a virtual self-cleaning epithelium.  
Ciliated cells are modeled as actuators alternating between two possible states 
representing ciliary oscillations. 
Interactions between the two-state actuators are mediated by discrete mucus 
lumps and enmeshed dust particles.
Whenever it is possible, the mucus droplets get displaced by the action of 
actuators and they  
may block their motion in certain configurations, which is prescribed by local 
interaction rules. 
This highly simplified model based on locally interacting motors 
self-organizes towards a virtual self-cleaning epithelium: the initially 
randomly distributed phases 
erratically displace the mucus lumps at the beginning of the simulation. 
As time passes the motors  
self-organize, which is expressed by emergent global spatio-temporal 
structures, 
resembling the metachronal wavelets, which have been observed on the 
ciliated tracheal epithelium \cite{Ryser2007}. 
These metachronal wavelets efficiently  
transport the mucus lumps into a well defined direction.\par
This paper is organized as follows.  
In Sec.\ref{sec:Description} we introduce our abstract epithelium model 
based on symmetrically 
interacting two-state actuators. The local mucociliary interactions are first 
formulated in an intuitive way, before they are reformulated as
logical update functions of the local state and mucus distribution.  
The goal of Sec.\ref{sec:Description} is to formulate 
our epithelium model in terms of an adaptive boolean network.
Within the framework of adaptive boolean networks nodes are represented by 
actuators and  
the topology of the network (in effect the links between the nodes) is 
determined by the mucus distribution.
Furthermore, we introduce the involved model parameters encompassing 
in particular different morphologies, boundary conditions and update schemes. 
Thanks to the simplicity of our epithelium model, 
it is possible 
to investigate the impact of the model parameters on the system's 
behaviour by conducting comprehensive parameter studies. 
The corresponding simulation data
are presented in Sec.\ref{sec:results}.   
The simulation data have been screened for any suspicious coherences between
the model parameters and the network dynamics as well as between the 
parameters and their corresponding attracting states.
Furthermore, we aimed at finding the mechanisms driving the model efficiently 
towards properly self-cleaning states. 
Finally, in Sec.\ref{sec:discussion} we discuss our findings.
In particular, we hypothesize about the meaning of  
the discovered dynamical aspects (which seem to be universal 
for our specific model of locally interacting cells)
for the real biological system.
\section{Model Description}
\label{sec:Description}
\subsection{Symmetrically Interacting Two-State Actuators}
\begin{figure}[!phbt]
  \centering
  \includegraphics[trim= 0cm 0cm 0cm 
0cm,clip,width=0.9\linewidth]{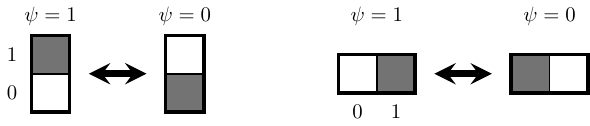}
  \caption{Ciliated cells are represented by two-state actuators. 
  An actuator's current state is expressed by the boolean 
state variable $\psi$. Each actuator provides two fields (0 and 1).}
  \label{fig:booleanstates}
\end{figure}

Our model is based on symmetrically interacting two-state actuators.
Each actuator represents a ciliated epithelial cell. 
By arranging many actuators in a parquet-like manner, 
the model represents a ciliated epithelium.   
Cilia corresponding to the same cell are assumed to move synchronously back and 
forth. 
This back and forth motion of cilia bundles is incorporated by the alternation 
between the actuators' two possible states, which is illustrated in 
Fig.\ref{fig:booleanstates}. 
The actual state of an actuator can thus be expressed by the boolean state 
variable $\psi \in \{0,1\}$.
\subsection{Morphology of Virtual Epithelia}
The morphology of the airway epithelium is thought to be the result 
of self-organizing processes during the morphogenesis \cite{Marshall2010}.
An important characteristic of the morphology is the distribution of the 
orientation of the ciliary beating plane, which 
can be determined by the orientation of the microtubules in the axoneme 
\cite{Satir2007}.
The impact of the (dis-)orientation of the ciliary beating plane i.e. of the 
axonemal orientation on mucociliary transport has been 
discussed e.g. in \cite{Mittleman1993, DeIongh1989}.
It has been concluded that a disorganized ciliary orientation 
may be a primary cause for mucociliary dysfunction and vice versa. 
Studies attempting to quantify the axonemal orientation of respiratory cilia 
usually determine 
the ciliary beating plane on a few cells derived from nasal brushing (e.g. 
\cite{Rautiainen1988}).
To the best of our knowledge, possible differences between the 
intra- and intercellular 
orientation of the ciliary beating plane is lacking so far in the literature. 
Therefore, we first of all assume that cilia belonging to a cell have a 
coinciding beating plane. 
As it is conceivable that the ciliary orientation on a cell might be more 
strict 
than between cells, 
the actuators can be oriented vertically or horizontally, as illustrated in 
Fig.\ref{fig:booleanstates}. 
Note that the two different possible orientations 
are not meant being perpendicular to each other, but to represent two 
distinct orientations of the ciliary beating plane differing by an arbitrary 
angle.

\begin{figure}[!phbt]
\begin{minipage}[t]{0.3\linewidth}
\centering unidir. square lattice (USL)
\includegraphics[trim= 0cm 0cm 0cm 0cm,clip,scale=0.75]{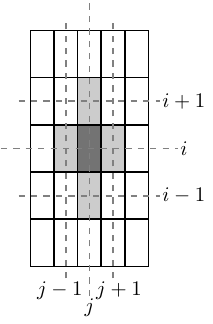}\\
\end{minipage}
\hfill
\begin{minipage}[t]{0.3\linewidth}
\centering unidir. hex. lattice (UHL) 
\includegraphics[trim= 0cm 0cm 0cm -0.2cm,scale=0.75]{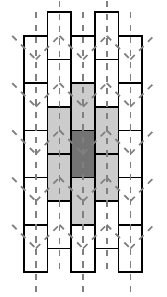}
\end{minipage}
\hfill
\begin{minipage}[t]{0.3\linewidth}
\centering bidir. hex. lattice (BHL)
\includegraphics[trim= -0.5cm 0cm 0cm 0cm,clip,scale=0.75,angle=-90]
{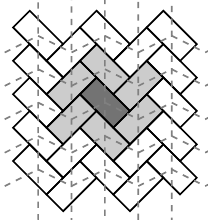}

\end{minipage}
\hfill
\caption{Three different cell alignments have been investigated. Bright gray 
colored cells label the neighborhood of a central cell colored 
in dark gray. Dashed grid 
lines indicate the mapping onto a two-dimensional array.}
\label{fig:arrangements}
\end{figure}

Here, we consider three different conceivable parquet-like cell 
alignments shown in Fig.\ref{fig:arrangements}. In the 
following we shall refer to these three cell alignments as (from the left to 
the 
right according to 
Fig.\ref{fig:arrangements}) unidirected square lattice (USL), unidirected 
hexagonal lattice (UHL) and 
bidirected hexagonal lattice (BHL).    

Another important characteristic of the morphology represents the 
population densities of ciliated and unciliated cells, whose 
role for the mucociliary dynamics has not been considered so far.    
According to electron microscopic studies \cite{Plopper1983, Oliveira2003} 
the area covered by ciliated cells roughly varies between one and two thirds of 
the total surface of the tracheal lining.    
Consequently, unciliated cells should be considered 
in pluricellular models. Therefore, we include the fraction of 
unciliated epithelial cells by the parameter 
$f \in [0,1]$, representing the proportion of randomly distributed empty sites 
in an array of actuators.\par
Finally, as indicated by the dashed grid lines in Fig.\ref{fig:arrangements} 
the virtual epithelium is represented by a two-dimensional 
array. For convenience, a site located at the $i$-th row and $j$-th column  
is denoted as $\psi_{ij}$. $\psi_{ij}$ either represents an actuator, then 
$\psi_{ij} \in \{0,1\}$, or 
indicates an empty site representing an unciliated cell, then $\psi_{ij}=NAN$.
Accordingly, the state of an array having $I$ rows and $J$ columns
can be denoted as: $\Psi = \{\psi_{ij}\}$, where $\psi_{ij} \in 
\{0,1,NAN\}$.

\subsection{Boundary Conditions}
\begin{figure}[!phbt]
\begin{minipage}[t]{0.38\linewidth}
\centering open (OP)
\includegraphics[trim= 0cm 0cm 0cm 
0cm,clip,width=0.75\textwidth]{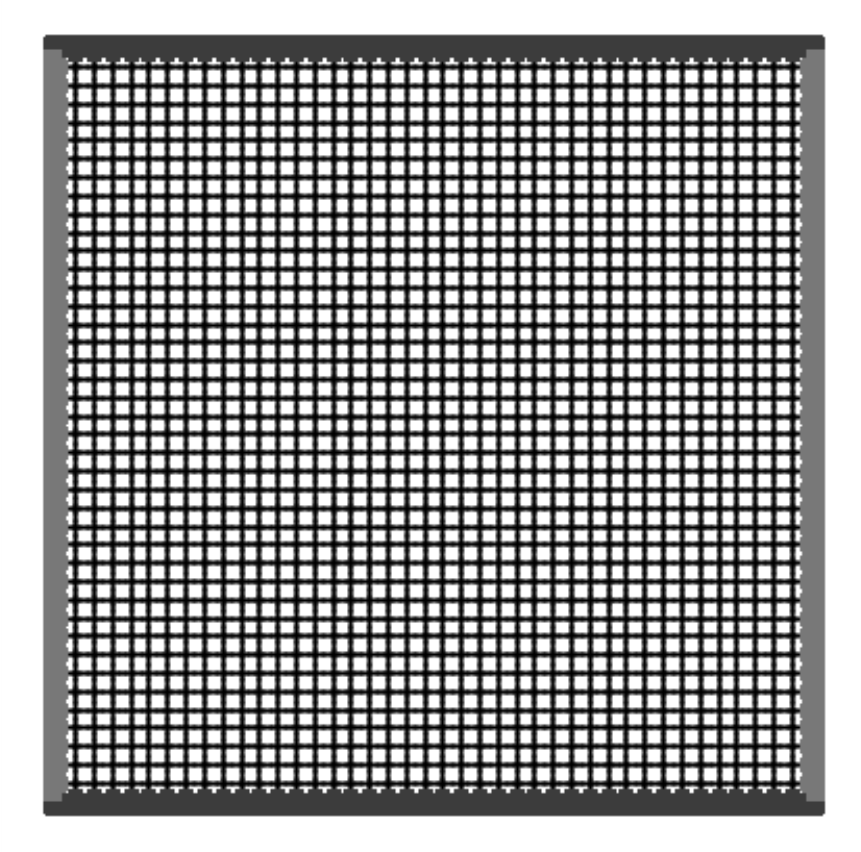}
\end{minipage}
\begin{minipage}[t]{0.38\linewidth}
\centering vertic. cylindric (VC)
\includegraphics[trim= 0cm 0cm 0cm 
0cm,clip,width=0.75\textwidth]{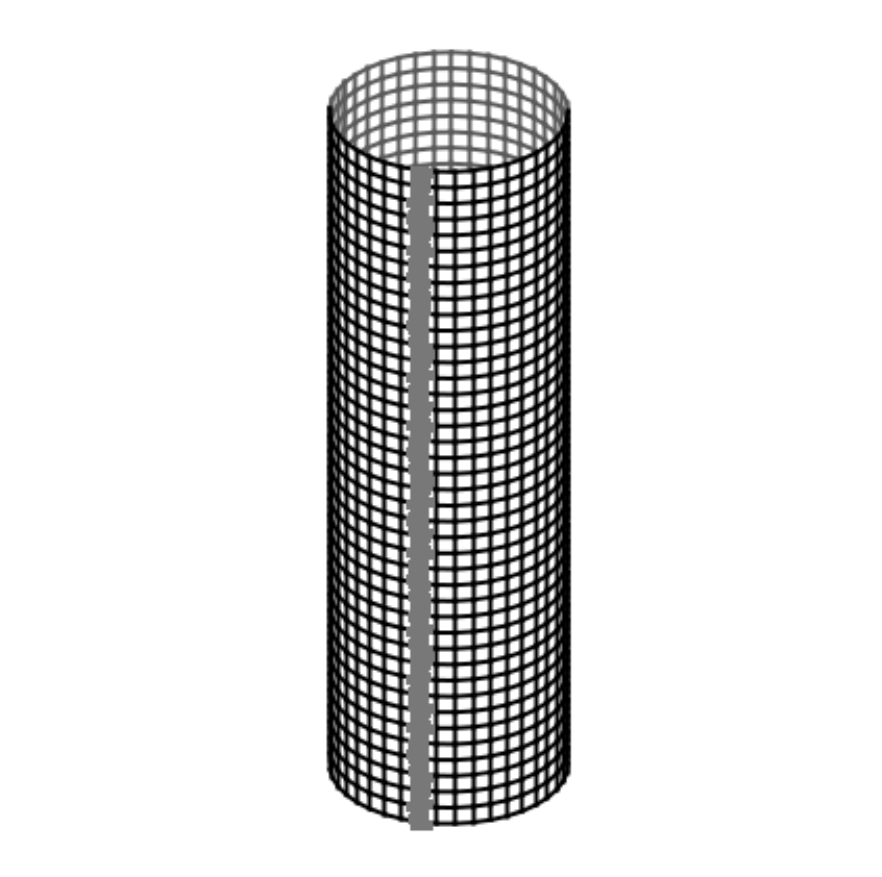}
\end{minipage}\\
\vspace*{0.7cm}
\begin{minipage}[t]{0.38\linewidth}
\centering horiz. cylindric (HC) 
\includegraphics[trim= 0cm 0cm 0cm 
0.5cm,clip,width=0.75\textwidth]{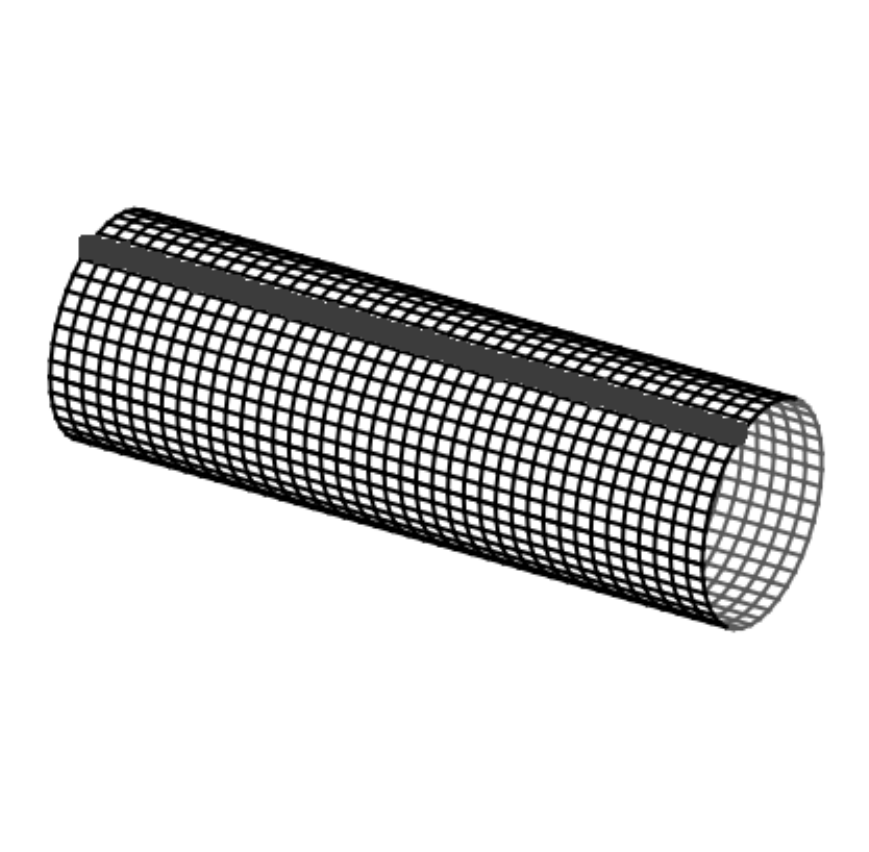}
\end{minipage}
\begin{minipage}[t]{0.38\linewidth}
\centering toric (TO)
\includegraphics[trim= 0cm 0cm 0cm 
0.5cm,clip,width=0.75\textwidth]{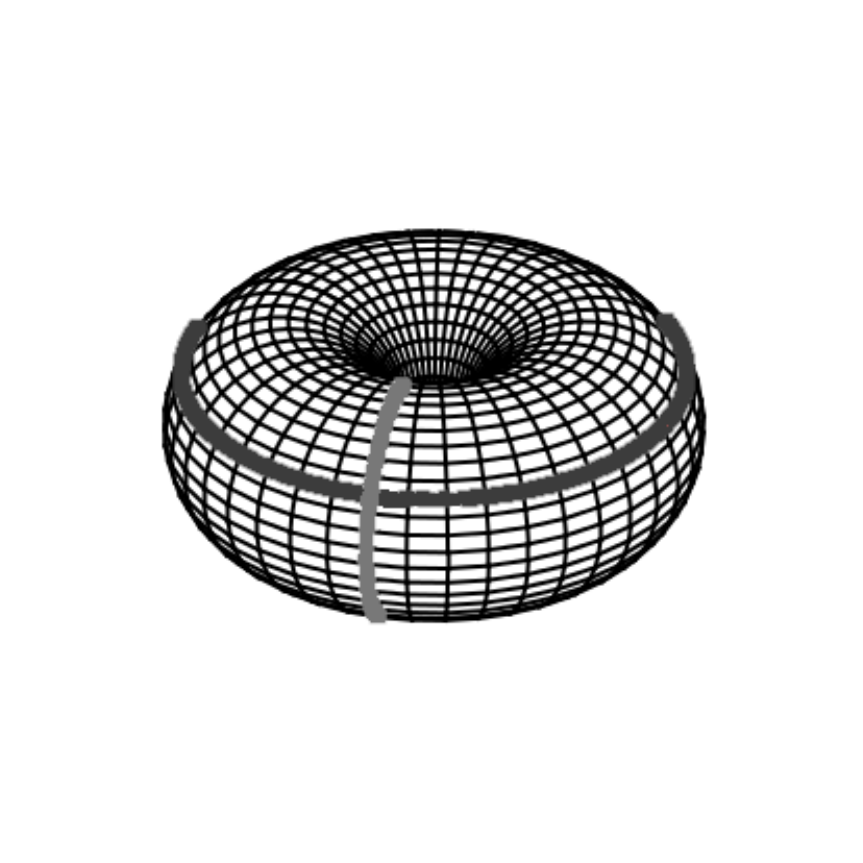}
\end{minipage}
\caption{Boundary conditions: open, vertical cylindrical, horizontal 
cylindrical 
and toric boundaries. The dark gray and bright gray lines indicate wrapped 
boundaries.}
\label{fig:boundcond}
\end{figure}

Experiments aiming at the characterization of spatio-temporal 
features of mucociliary phenomena on the tracheal ciliary epithelium are based 
on the approach of 
excising a rectangular piece of the cylindric trachea 
(e.g. \cite{Ryser2007, Lee2005, Yi2002, Yeates1993}), which changes the 
boundary conditions from cylindrical to open.  
The excision might 
influence the collective dynamical behavior of the system. 
Therefore, we consider four different boundary conditions, which are 
illustrated 
in Fig.\ref{fig:boundcond} and
shall be referenced in the following as: open boundaries (OP), 
vertical cylindric boundaries (VC), horizontal cylindric boundaries (HC) 
and toric boundaries (TO).
\subsection{Mucus}
Mucus is discontinuously incorporated by randomly seeding mucus droplets
of equal size. 
On the one 
hand, the discrete mucus model allows to simplify mucociliary interactions. On 
the other hand, it accounts for the fact that 
the mucus blanket is anyway made up of excreted mucus ``flakes'', ``plaques'' 
or 
''droplets`` \cite{VanAs1974}, which may coalesce into 
a continuous layer, if their density is 
sufficiently large \cite{Geiser1997}.\par
According to Fig.\ref{fig:booleanstates} each actuator provides an empty field
(0 or 1), 
which can be occupied by mucus droplets. 
On the other hand, empty sites (unciliated cells) provide two fields (0 and 1), 
which can be 
occupied by mucus droplets.  
Thus, the distribution of mucus droplets on the virtual epithelium is given by 
$M = \{m_{ij}\}$; if $\psi_{ij} = NAN$, then $m_{ij} = m_{0ij} + m_{1ij}$ .

\subsection{System Update}\label{sec:sysup}
In the course of a simulation the actuators are actuated sequentially. 
This means that only the state of the actuated actuator and its adjacent mucus 
configuration is 
updated while the states of adjacent actuators do not change.  
As soon as an actuator and its local mucus configuration has been updated, a 
subsequent actuator
is updated for which the changes of the prior step are taken into account.  
We shall label the sequence of update steps by the ``time'' superscript $t$.  
Furthermore, $t^{\prime} \doteq t / N$ labels the update of the 
whole network, consisting of $N$ actuators, in the following. 
In the context of discrete dynamical systems such sequential update schemes are 
called asynchronous.  
An excellent overview of the impact of different update schemes 
on the network dynamics of random boolean networks is provided in 
\cite{Gershenson2003, Gershenson2002}.   
In order to distinguish the different update schemes, we shall use 
a slightly modified form of the terms proposed by C. Gershenson 
\cite{Gershenson2002} 
(since we are not dealing with random boolean networks). 
We shall use the following update schemes 
(consider the corresponding illustrations in Fig.\ref{fig:updateschemes}): 
\begin{itemize}
\item	\textbf{DAU - Deterministic Asynchronous Update}\\
This update mode corresponds to a pre-defined sequence, in which the 
nodes/cells in the lattice are addressed. Keep in mind, however,
that determinism of addressing doesn't mean determinism of actuators motion. 
Whether an actuator moves or not depends on the mucociliary interactions, 
which are on a high degree stochastic, as we shall see in the proceeding 
sections.   	
\item	\textbf{RAU - Random Asynchronous Update}\\
At each time step a single node is chosen randomly and updated. The random 
choice of nodes has been
applied for sampling with and without replacement, as in 
\cite{Cornforth2005}. 
Random sequences of nodes generated without and with replacement 
are denoted as RAU1 and RAU2, respectively. 
\item 	\textbf{SRAU - Semi-Random Asynchronous Update}\\
The update scheme SRAU uses a wavelike activation: at each time step a strip of 
actuators gets addressed. Inside of each strip 
the actuators get addressed according to the 
RAU1 scheme. As the choice of the strip is predefined but inside the strips the 
sequence of actuators is chosen 
randomly, this addressing scheme has a semi-random character. 
This wave-like activation has been implemented for a wave traveling from the 
left to the right, 
as well as for a wave traveling from the top to the bottom, which we denote as 
SRAU1 and SRAU2, respectively.
\begin{figure*}[!phbt]
  \includegraphics[trim= 0cm 0cm 0cm 
0cm,clip,width=0.9\textwidth]{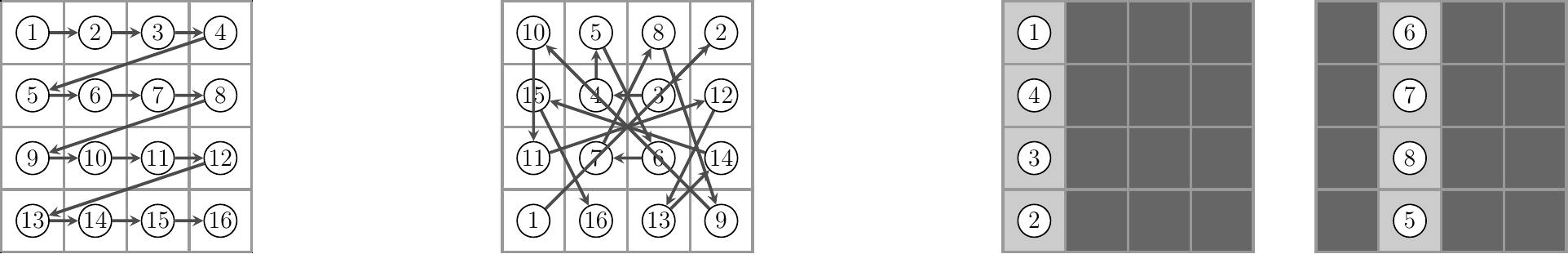}
  \caption{The figure illustrates different update schemes for a lattice 
consisting of 4$\times$4 cells. 
  The Numbers indicate the order of activation (for one certain time step). 
  From left to right: a pre-defined sequence (DAU), random cell selection 
without replacement (RAU1) 
  and a planar wave-like activation (SRAU1).}
  \label{fig:updateschemes}
\end{figure*}
\item \textbf{Mimicking Purcell's Two-Hinged Low-Reynolds-Number-Swimmer}\\ 
As non-reciprocal motions play an important 
role in a low-Reynolds-number environment 
the cell arrangement BHL has been used together with a prescribed 
addressing sequence of two neighboring actuators, V and H, 
forming the shape of an ``L''. 
The two actuators are addressed in the sequence VHVHVH, which would result 
in a four phase cyclic motion if the actuators were to move unhinderedly. 
This scheme, originally proposed by one of the authors 
in an  essay in a popular scientific journal \cite{ricka2010}, was motivated by 
the ``two-hinged low-Reynolds-number-swimmer'', which
has been introduced in \cite{Purcell1977} and is illustrated in 
Fig.\ref{fig:purcell}. 
The prescribed addressing sequence principally represents a special locally 
deterministic update scheme,
as it considers the actual state of two locally coupled cells to determine 
which actuator 
will be addressed. The selection of the ``L'' is based on the update schemes 
introduced above (RAU1, RAU2, SRAU1, SRAU2 and DAU). 
Consequently, we actually mixed a local update scheme (prescribing the 
four-phase sequence of an ``L'') with different 
global update schemes (selecting which ``L'' to address). 
Settings using the coupling of two cells forming an ``L'', 
will be accounted for a cell arrangement, referred to ``BHL+L''.
\begin{figure}[!phbt]
  \centering
  \includegraphics[trim= 0cm 0cm 0cm 0cm,clip,width=\linewidth]{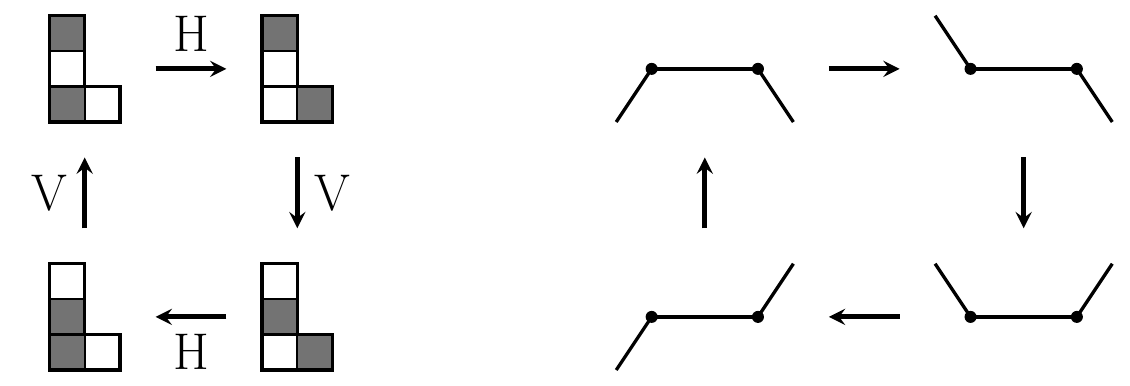}
  \caption{The prescribed cyclic four-phase motion of two adjacent
  actuators forming the shape of an ``L'' (left).
  The construction of this four-phase motion-sequence was motivated
  by Purcell's two-hinged swimmer (right).}
  \label{fig:purcell}
\end{figure}
\end{itemize}
The different update schemes have been applied in order to 
investigate the role of potential intercellular signaling mechanisms on the 
airway epithelium for the dynamical behavior of locally interacting
ciliated cells.  
\subsection{Local Mucociliary Interactions}\label{sec:localinteractions}
Hydrodynamical interactions between adjacent ciliated cells
are considered in a simplified fashion and implemented in terms of logical 
local 
decision rules 
induced by mucus droplets randomly seeded on the empty sites in the network. 
The system's evolution is achieved by 
attempting to move the individual actuators. 
As interactions between actuators occur 
only if mucus is located on the 
activated actuator, an activated actuator can switch its state unhinderedly as 
long as there is no mucus hindering its oscillation. 
In the presence of mucus, the actuator can 
shift and squeeze single droplets, or stacks of 
droplets, depending on the available energy.

\begin{figure}[!phbt]
  \centering
  \includegraphics[scale=0.8]{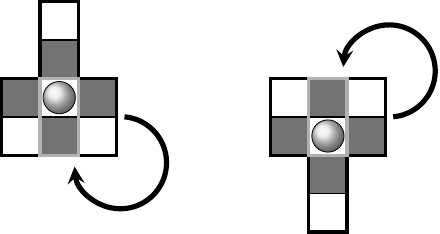}
  \caption{In certain state configurations the active actuator (indicated by 
the 
  bright surround) is locked and consequently, remains in its actual state: 
$\psi_{ij}^{t+1} \rightarrow \psi_{ij}^t$.}   	
  \label{fig:rule1}
\end{figure}

There are two possibilities that 
an activated actuator remains in its actual state. 
Either the active actuator has not enough energy to 
squeeze the mucus lumps on adjacent fields, 
or the actuator is situated in a locked configuration (see 
Fig.\ref{fig:rule1}). Note that there are always 
two possible locked configurations for each actuator, 
which we shall refer to in the following as $l_0$ and $l_1$. 

\begin{figure}[!phbt]
  \centering
  \includegraphics[scale=0.88]{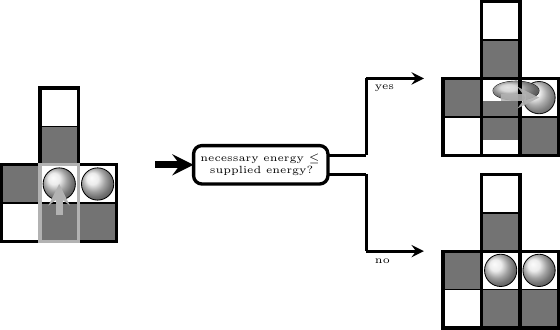}
  \caption{If the active actuator holds a mucus lump, the actuator is 
  either going to stagnate or it has enough energy available to  
  squeeze the mucus lump and switch its state.}	
  \label{fig:rule2}
\end{figure}

If the activated actuator is not situated 
in a locked configuration
it either flips its state by squeezing the concerned 
mucus on adjacent fields, or it stagnates and remains in its current state (see 
Fig.\ref{fig:rule2}).
The squeezing of mucus on adjacent fields is associated with a 
certain change of free energy.
Prior to its attempt to move, an actuator gets a certain 
amount of energy ascribed, which has been inverse sampled.  
If the actuator's ascribed energy exceeds the  
required moving energy, the actuator is able to move. 
This way,
the actuators' energy distribution and the situation-specific change in free 
energy  determines the probability 
for an actuator to flip its state by squeezing 
the mucus on adjacent actuators. 

\begin{figure}[!phbt]
  \centering
  \includegraphics[scale=0.88]{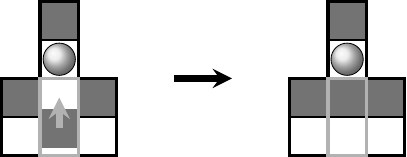}
  \caption{An activated mucus-free actuator 
  is always able to alternate its state.}
  \label{fig:nomucus}
\end{figure}

For computational purposes it is convenient to consider the mucus-free case 
($m_{ij}^t=0$) 
as a special case, in which an actuator is always able to 
switch its state (see Fig.\ref{fig:nomucus}).  

Intercellular coordination, or rather the emergence of global spatio-temporal 
patterns, 
are caused by stagnating actuators as well as locked configurations, as in 
these 
situations 
the activated actuator has to adjust its state according to the local state and 
mucus distribution.
\subsection{Boolean Network Representation}
In this study the mucociliary dynamics on the epithelium 
are represented in terms of an adaptive boolean network. 
Ciliated cells are imported in terms of boolean actuators 
and represent the nodes of the network. The links between nodes represent 
mucociliary interactions, which
we formulate in terms of boolean functions. \par
Formally, a boolean network consists of $N$ elements $\{\psi_1, \psi_2, ..., 
\psi_N\}$, 
each of which is a binary variable $\psi_i \in \{0,1\}$ representing a node in 
the network. 
In general \cite{Kadanoff2002}, the value of a node $\psi_i$ at time $t+1$ 
is given as a function $f_i$ of its $K_i$ controlling elements at 
time $t$:
\begin{equation}
	\psi_i^{t+1} = f_i(\psi_{{j_1}(i)}^t, \psi_{{j_2}(i)}^t ..., 
	\psi_{{K_i}(i)}^t)~.
\label{eq:general_bn}
\end{equation}
The boolean function $f_i$ as well as the number of controlling 
elements $K_i$ may be different for each node.
The dependence on node $i$ 
explicitly denotes
that the set of controlling nodes with indices $\{j_1,j_2,...,K_i\}$ 
generally varies from one node to the other.

In our case, $\psi_{ij}$ at time $t+1$ is 
given as a function $f_{ij}$ of  
the local state configuration 
$\{\psi_{pq}\}_{pq \in n_{ij}}$
and the local distribution of mucus droplets 
$\{m_{pq}\}_{pq \in n_{ij}}$ at time $t$:
\begin{equation}
  \psi_{ij}^{t+1} = f_{ij}\big( \{\psi_{pq}^{t}\}_{pq \in n_{ij}}, 
  \{ m_{pq}^t \}_{pq \in n_{ij}} \big) ~. 
  \label{eq:net}
\end{equation} 
The neighborhood $n_{ij}$
of the actuator $ij$ is illustrated in Fig.\ref{fig:arrangements}.
The function $f_{ij}$, i.e. representing the local 
mucociliary interactions, imports the actuators' available energy 
and will be specified in Sec.\ref{sec:energy}.
 
The oscillatory driving force of ciliated cells 
results in the permanent attempt of a node for state reversal 
and is represented by the logical XOR-function:  
\begin{equation}
  \psi_{ij}^{t+1} = \text{XOR}(\psi_{ij}^t,c_{ij}^t)~.
  \label{eq:xor}
\end{equation}
Here $c_{ij}^t \in 
\{0,1\}$ represents the local couplings among the nodes. 
If $c_{ij}^t=1$, that is if the actuator is capable of 
displacement, then the XOR-function inverts an actuator's actual state 
$\psi_{ij}^t$.
On the other hand, if the actuator is situated in a locked 
configuration (Fig.\ref{fig:rule1}) or stagnates due to an excessive 
mucus load (Fig.\ref{fig:rule2}), then $c_{ij}^t$ gets 
false/zero  and consequently, remains in its 
current state.

The actuator is capable of displacement if it's not blocked and if
the free energy $u$ available to the actuator is larger than the work $\Delta 
w$ needed to displace the mucus. Thus, we write the condition for 
a state alternation as: 

\begin{equation}
  c_{ij}^t=\text{NOR}(l_0,l_1) \cdot \Theta \big[ u - \Delta w \big] = 1~.
  \label{eq:arr} 
\end{equation}
$l_0$ and $l_1$ denote the two locked state configurations 
(Fig.\ref{fig:rule1}) and $\Theta(x)$ the Heaviside step function, such that 
$\Theta(0)=1$. 

In order to clarify the state alternation condition we shall reconstruct the 
intuitively formulated  local interactions, which have been illustrated in 
Fig.\ref{fig:rule1}-Fig.\ref{fig:nomucus}:

\begin{itemize}
\item Fig.\ref{fig:rule1}: In locked configuration case, either $l_0=1$ or 
$l_1=1$ and consequently, NOR$(l_0,l_1)=0$. 
As $c_{ij}^t \sim \text{NOR}(l_0,l_1)$, the activated actuator remains in its 
current state. 
\item Fig.\ref{fig:rule2}: In hindered motion case NOR$(l_0,l_1)=1$  and thus 
the condition for state 
alternation reads (according to Eq.\ref{eq:arr}): 
$u \geq \Delta w~$.   
\item Fig.\ref{fig:nomucus}: 
Obviously, mucus-free actuators are not locked and consequently: $l_0 = l_1 = 
0$ 
$\implies NOR(l_0,l_1) = 1$.
As there is no mucus to redistribute the required work vanishes ($\Delta w = 
0$) and thus, the state alternation occurs spontaneously.
\end{itemize}

It is important to realize that the distribution of mucus droplets 
determines the topology of the network, as the update of mucus-free nodes 
is not affected by their local environment.   
As soon as an actuator gets occupied by a mucus droplet, it gets functionally 
connected 
to its neighboring nodes, as its subsequent state
depends on the local state and mucus configuration. 
Consequently, in our network not only the network's state 
is exposed to dynamics, but also the network's topology, as the 
(re-)distribution of the 
droplets depends on the state transitions of the nodes and vice versa.
Networks exhibiting such a feedback loop between the network's state and its 
topology are 
called coevolutionary or adaptive networks \cite{Gross2008}.
\subsection{Actuator-Energy and Mucus Relaxation}\label{sec:energy}
\begin{figure}[!phbt]
  \includegraphics[trim= 0cm 0cm 0cm 
0cm,clip,width=0.99\linewidth]{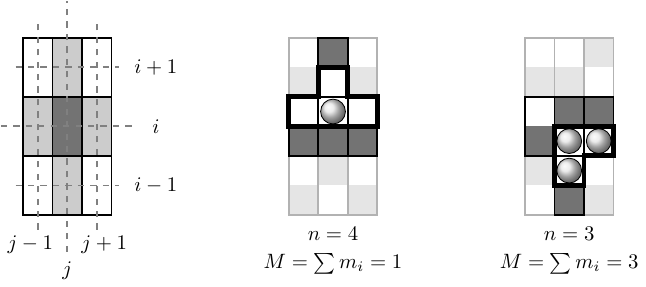}
  \caption{The left panel illustrates the neighborhood of actuators aligned in 
the square-lattice. Bright gray 
  actuators control the subsequent state of the actuated dark gray actuator. 
Middle and right panel: Two situations which are thought
  to illustrate the locally concerned volume $V=n\Delta V$ and the involved 
number of mucus droplets $M$. Bleached actuators are not involved to 
  the current interaction.}
  \label{fig:n}
\end{figure}

The interactions are quantified using a simple scheme intended to model in a 
crude fashion 
the transient entropic elasticity and relaxation of entangled mucin chains.  We 
view the actuator 
as a piston acting against the pressure $p$ exerted by mucus droplets contained 
in a certain volume $V$.  
For simplicity we assume that the interactions are mediated only through the 
set 
of $n$ direct neighbors 
of the targeted site. As a result the volume to be compressed by the action of 
the piston is 
$V = n \Delta V$ (see Fig.\ref{fig:n}).
By moving the actuator, this volume changes by $-\Delta V$ to $V' = (n-1) 
\Delta 
V$, what requires 
the work $\Delta w = p \Delta V$. 
(Strictly speaking we should consider the pressure difference $\delta p = p_f - 
p_b$ between the front and back of 
the actuator. We apologize for forgetting  the back pressure $p_b$ in the 
present set of simulation data, what 
fortunately does not change the main conclusions.) 

To determine the pressure $p$ we employ the standard thermodynamic relation:
$p / T = \partial S(V) / \partial V$,
where $S(V) = k_B\text{ln}(W_V)$ is the Boltzmann entropy of the 
mucus droplets enclosed within $V$. Replacing the differentials by discrete 
differences 
$\partial S \rightarrow \Delta S = k_B\text{ln}(W_{V'}/W_V)$ and $\partial V 
\rightarrow -\Delta V$ yields
\begin{equation}
 \Delta w = p\Delta V = k_B T \text{ln}\big(\frac{W_n}{W_{n-1}}\big)~.
 \label{eq:deltaw}
\end{equation}
Subsequently, we set $k_BT = 1$, i.e., energy is measured in  units of $k_BT$.
A plausible expression for the multiplicity ("thermodynamic probability") $W_n$ 
can be deduced as follows:
Random deposition of $M$ mucus droplets on $n$ available sites is equivalent to 
rolling an $n$-sided dice. Thus, the  
result of the deposition of $M$ droplets is a sample from the multinomial 
distribution, with equal probabilities for hitting a field unoccupied by an 
actuator. 
Thus, the multinomial 
coefficient 
$C_m = M!/(m_1!m_2! \cdot\cdot\cdot m_n!)$ is a good 
candidate for the multiplicity 
in Boltzmann entropy. 
($m_i$ denotes the number of droplets deposited on a site $i$.)
However, after the deposition, prior to an attempted move of an 
actuator, we allow 
the distribution of droplets to relax to a "thermodynamic equilibrium", i.e., 
into a state of 
maximum multiplicity, where the droplets are most uniformly distributed on the 
available sites, so that 
$|m_i - m_j| \leq 1$, for $i \neq j$. 
(In other words, we invoke the Maximum 
Entropy Principle \cite{Hanel2014}.)
Therefore, 
we define the multiplicity involved in Eq. \ref{eq:deltaw} as 
$W_n=\text{min}\big(m_1!m_2!\cdot\cdot\cdot m_n!\big)$ and 
$W_{n-1} = \text{min}\big(m_1'!m_2'!\cdot\cdot\cdot m_n'!\big)$, 
where $m_i$ and $m_i'$ 
are subject to the constraints $\sum_{i=1}^{n} m_i = \sum_{i=1}^{n-1} m_i'= M$.

To complete the specification of the mucociliary interactions 
we must specify the actuators energy $u$. 
At this point we introduce an additional stochastic element, 
assuming a certain distribution $f_U(u)$ of the actuator energy. 
Prior each attempt an actuator's energy $u$ is obtained by reverse sampling, 
according to: $u = F_{U}^{-1}(r)$, where $r$ represents a 
uniformly distributed 
random number and $F_U$ the cumulative energy distribution.
Since a cumulative distribution is a monotonous function of its 
argument, the sampling can be concisely included into the boolean update 
equation 
for $c_{ij}^t$ as:
\begin{equation}
\psi_{ij}^{t+1} = \text{XOR}\big\{ \psi_{ij}^t, 
\text{NOR}(l_0,l_1) \cdot
\Theta \big[ r - F_U(\Delta w) \big]  \big\}.
\end{equation}  

\begin{figure}[!phbt]
  \includegraphics[trim= 0cm 0cm 0cm 
0cm,clip,width=0.85\linewidth]{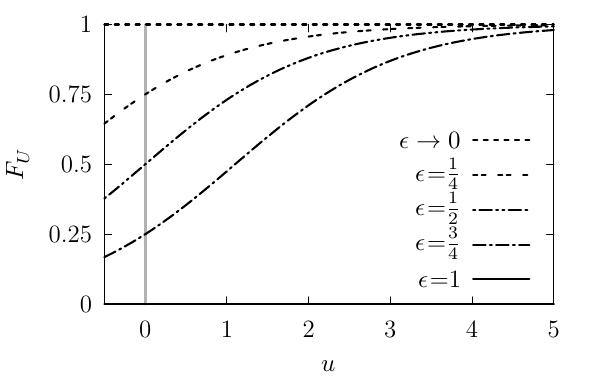}
  \caption{The curves show the actuators' cumulative energy distribution
  $F_U(u)$ for different values of the energy parameter: 
  $\epsilon \in \{\frac{1}{4}, \frac{1}{2}, \frac{3}{4}, 1 \}$ and for 
$\epsilon 
\to 0$.}
  \label{fig:epsilon}
\end{figure}

In the present simulations we used 
(for certain ``historical reasons'') the simple expression 
\begin{equation}
	F_U(u) = \frac{1}{\frac{\epsilon}{1-\epsilon}\exp(-u)+1}~, 
 	\label{eq:actenergy}
\end{equation}
where $u \geq 0$ and $\epsilon$ parameterizes the actuators' energy supply.
This function is illustrated in Fig.\ref{fig:epsilon}. 

\begin{figure}[!phbt]
  \centering 
  \includegraphics[scale=1.1]{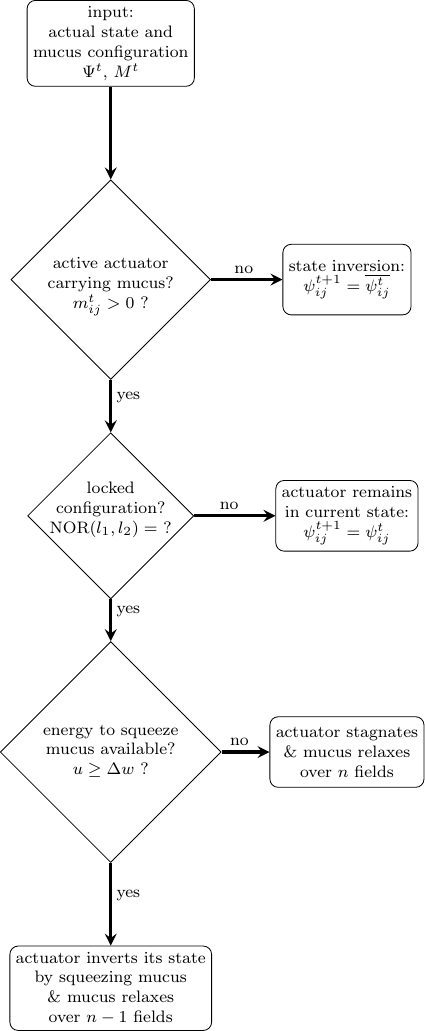} 
  \caption{The flow chart illustrates the update of an actuator's state as well 
as its local mucus distribution.}   	
  \label{fig:flow}
\end{figure}

Finally, we summarize the update of an actuator's state 
and its local mucus configuration
in the form of a flow chart in Fig.\ref{fig:flow}.
\section{Simulations and Results}
\label{sec:results}
\subsection{Initial State}
Prior to simulation the state of the virtual epithelium has to be initialized. 
First, an array of $N$ actuators
with random states is generated, i.e. $\Psi^0$, where $\psi_{ij} \in \{0,1\}$ 
are uniformly distributed.
Second, as the fraction $f$ represents the proportion
of unciliated cells, $f\cdot 
N$ randomly chosen sites are set to $NAN$.
A certain amount of mucus droplets is then seeded on a randomly chosen site 
(uniformly distributed).
For the sake of comparability between open and toric boundary conditions 
the amount of mucus droplets was held constant during the simulation.  
Therefore, when applying open boundaries the mucus droplets
leaving the modeling area were refed on a randomly chosen site. 
This way, we principally assumed the mucus excretion to be proportional to the 
mucus transport, which would of course need some kind of internal regulation 
mechanism in a real trachea.   
\subsection{Parameter Study} 
In order to assess under which circumstances our model self-organizes towards
a self-cleaning virtual epithelium and how it reaches its properly functioning 
states dynamically, 
the influence of the model parameters, introduced in the previous chapter, on 
the network dynamics has been studied.
The main parameter study 
encompasses the variation of the six model parameters in the following ranges. 
All cell alignments (USL, UHL, BHL, BHL+L), all update schemes 
(DAU, RAU1, RAU2, SRAU1, SRAU2) and all boundary 
conditions (OP, HC, VC, TO) have been used. The 
amount of mucus lumps has been varied in the range 
of 0.5\%, 1\%, 2\%, 4\%, ..., 256\% 
of the total number of actuators in the grid 
(an amount of 4\% corresponds to $0.04 \cdot N$ mucus droplets). 
The amount of unciliated cells has been set to 
0\%, 5\%, 10\%, 15\%, 20\%, 25\% and finally, the energy parameter has 
been set to 0, 0.25, 0.5, 0.75 and 1.   
Consequently, our main parameter study encompasses 24'000 simulation runs, which
have been iterated for $10^5$ time steps using a grid size of 50$\times$50 
cells. 
These simulations can be seen as a starting point of further simulations we 
conducted.  
Each simulation run is characterized by its corresponding parameter setting. A 
specific 
parameter set is denoted as a combination of the form: (grid size, cell 
alignment,
update scheme, boundary condition, mucus amount, energy parameter, amount of 
unciliated cells).  
\subsection{Observables}
\begin{figure*}[!phbt]
	\centering 
	\hspace{-0.4cm}	
    \begin{minipage}[t]{0.28\linewidth}
		a) 
        \includegraphics[trim= 1cm 1cm 0.8cm 
0.8cm,clip,width=\linewidth]{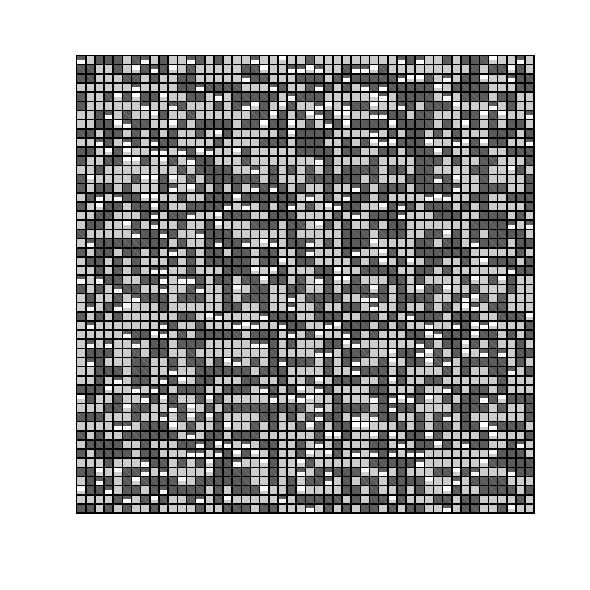}
        \includegraphics[scale=0.6]{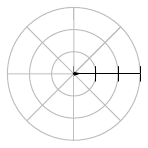}
    \end{minipage}
    \hspace{0.2cm}
    \begin{minipage}[t]{0.28\linewidth}
		b) 
        \includegraphics[trim= 1cm 1cm 0.8cm 
0.8cm,clip,width=\linewidth]{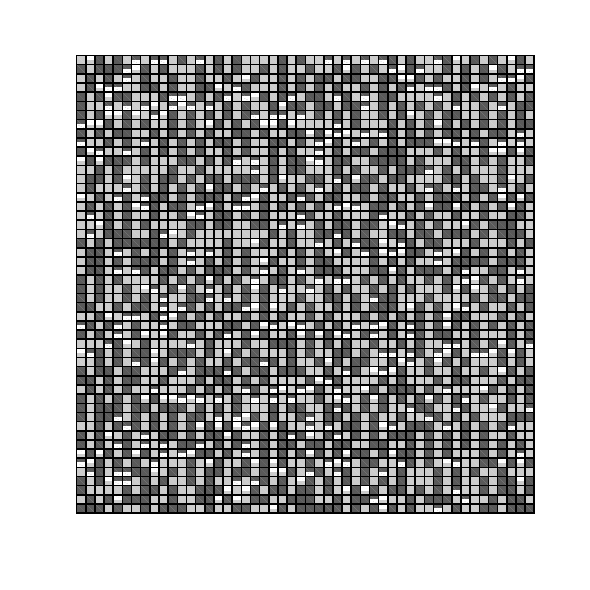}
        \includegraphics[scale=0.6]{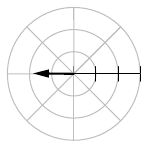}
    \end{minipage}
    \hspace{0.2cm}
    \begin{minipage}[t]{0.28\linewidth}
		c)
        \includegraphics[trim= 1cm 1cm 0.8cm 
0.8cm,clip,width=\linewidth]{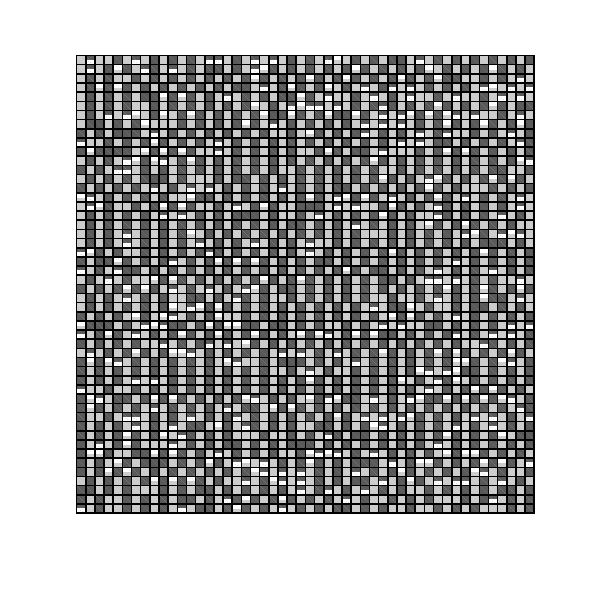}
        \includegraphics[scale=0.6]{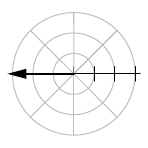}
    \end{minipage}
    \caption{The panels show the network state of an exemplary 
simulation run at three different stages of the 
self-organization process. The crosshairs in the 
second row visualize the mucus transport velocity $\vec{v_g}(t')$ 
for each stage (the radial tick interval 
corresponds to 0.2 (cells/it)). a) initial network 
state: randomly distributed states. b) network 
state after 100 iterations: due to the local interactions 
the emergence of order becomes visible. c) after 1500 iterations: 
the self-organized cooperative behavior of 
actuators forms spatio-temporal patterns and exhibits efficient 
self-organized transport.}
    \label{fig:3stages}
\end{figure*}

To provide a qualitative impression of the self-organizing 
character of the network model and to illustrate the 
meaning of the chosen observables, 
we present  first a typical simulation 
run. The parameters have been set to (50$\times$50, USL, DAU, TO, 16\%, 0.5, 
0\%). 
Fig.\ref{fig:3stages} shows the state of the network at three different
stages of the self-organization process. Fig.\ref{fig:3stages} 
represents a substitute for the temporal evolution
of the network's state, which is best visualized 
by \href{run:./anc/movie_S1.mp4}{movie\_S1} 
(see supplemental material \cite{supm}).
Actuators are colored in dark gray if $\psi_{ij}^t=1$ 
or in bright gray if $\psi_{ij}^t=0$. Mucus lumps are shown in white. 
Fig.\ref{fig:3stages}a shows the initial network state 
displaying the randomly generated initial configuration of states and mucus 
lumps. 
Fig.\ref{fig:3stages}b visualizes the state of the network after 100 
iterations, 
representing
an intermediate stage of the self-organization process, as the emergence of 
global order becomes clearly visible.
Finally, Fig.\ref{fig:3stages}c shows the
network state after 1500 iterations.
At this stage the fascinating self-organization process has almost completed,
and the actuators finally behave strongly coordinated at a global scale.
By examining the successive network
states in \href{run:./anc/movie_S1.mp4}{movie\_S1},
one can observe that the particles get initially moved around disorderly.
Quickly, the actuators start to cooperate by adjusting their oscillations to the
oscillations of the surrounding actuators until the particles get efficiently 
transported
into a well defined direction - what we call self-organized transport.
Consequently, in our models the self-organization process is actually twofold.
As in accordance with the emergence of spatio-temporal patterns, the
virtual epithelium exhibits self-organized transport.
This co-evolution of the network state and its associated mucus transport has 
been quantified 
in terms of several observables, which are introduced in the following. 
\subsubsection{Mucus Transport Velocity}
For each actuator at the position $ij$ at time $t$ 
we assign a local mucus velocity in terms of the local displacement of the 
center of mass (CM), which we 
denote as $\vec{v}_{ij}(t)$ and has been 
calculated (if $m_{ij}^t > 0$) according to:\\
\begin{equation}
	\vec{v}_{ij}(t) \doteq \frac{\sum_{p,q \in n_{ij}} (\vec{r}_{ij} - 
\vec{r}_{pq}) (m_{pq}^{t+1} - 
	m_{pq}^t)}{\sum_{p,q \in n_{ij}} m_{pq}^t}
	\label{eq:localv} 
\end{equation}
$\vec{v}_{ij}(t)$ measures the redistribution of mucus droplets in 
the neighborhood $n_{ij}$ of 
the activated actuator at $ij$ in terms of the local 
displacement of the CM. 
$\vec{r}_{ij} - \vec{r}_{pq}$ denotes the distance vector pointing from the 
activated actuator at ${ij}$  
to the neighboring actuators at $pq$. 
This relative distance vector gets weighted by 
the redistributed mucus droplets $(m_{pq}^{t+1}-m_{pq}^{t})$.\par
In order to quantify the global transport velocity 
$\vec{v}_g(t')$ (remind that $t' \doteq t/N$) we calculate 
the mucus-weighted 
average over the whole array of actuators of the formerly defined local 
velocity 
of the CM 
in Eq.(\ref{eq:localv}) according to: \\
\begin{equation}
	\vec{v}_g(t') \doteq \frac{\sum_{ij} \vec{v}_{ij}(t') \sum_{p,q \in 
n_{ij}} m_{pq}^{t'}}
	{\sum_{ij}\sum_{p,q \in n_{ij}} m_{pq}^{t'}}~.
	\label{eq:globalv}
\end{equation}
$\vec{v}_g(t')$ represents an approximation to 
the actual average mucus transport 
velocity, as droplets moving more than 
one field in one single timestep are neglected. 
As these movements only happen due to fluctuations 
originating from mucus relaxation, $\vec{v}_g(t')$ represents 
a good measure for 
the area averaged mucus transport velocity.\par
\begin{figure}[!phbt]
	 \centering
     \includegraphics[trim= 0cm 0cm 0cm 
0cm,clip,width=0.95\linewidth]{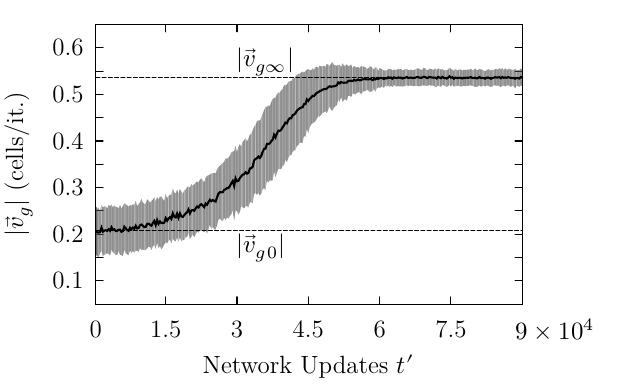}
     \caption{Temporal evolution of the average mucus transport speed of an 
ensemble consisting of 100 simulation 
runs. The typical limited growth behavior was omnipresent in all of our 
simulation runs displaying a self-organizing behavior.}
    \label{fig:transientrau1}
\end{figure}

  
Fig.\ref{fig:transientrau1} shows the temporal evolution of the 
global mucus transport speed $|\vec{v}_g(t')|$ for an ensemble 
consisting of 100 simulation runs differing only by their initial state.  
Each ensemble member corresponds to a simulation run 
for which the parameter setting (50$\times$50, BHL+L, RAU1, OP, 10\%, 1, 0\%) 
has been used.  
The curve shows the temporal evolution of the ensemble mean (solid line) and 
its 
standard 
deviation (shaded area) and displays the typical saturation-like behavior of 
the average mucus transport speed, which has 
been observed for each parameter setting showing a self-organizing behavior.   
Accordingly, we can define the initial mucus transport velocity 
$\vec{v}_{g0} \doteq \lim\limits_{t'\to 0}\vec{v}_g(t')$ and the final mucus 
transport velocity   
$\vec{v}_{g\infty} \doteq  \lim\limits_{t'\to\infty}\vec{v}_g(t')$. 
Consequently, the global average transport velocity can be expressed as  
\begin{equation}
	\vec{v}_g(t') = \vec{v}_{g0} + (\vec{v}_g(t') - \vec{v}_{g0}) \doteq 
\vec{v}_{g0} + \Delta\vec{v}_g(t')~,
	\label{eq:deltav}	
\end{equation} 
where $\Delta\vec{v}_g(t')$ can be seen as the effectively 
self-organized mucus transport velocity.
\subsubsection{Mucus Transport Direction} 
We observed that some parameter sets drive the model towards
well organized attracting network states 
exhibiting non-negligible area averaged transport speeds. 
However, some of these well organized states 
exhibit unrealistic velocity fields.\par
\begin{figure}[!phbt]
	\centering
	\hspace*{-0.8cm} 
	\includegraphics[trim= 0.5cm 1.5cm 0cm 
1.5cm,clip,width=0.68\linewidth]{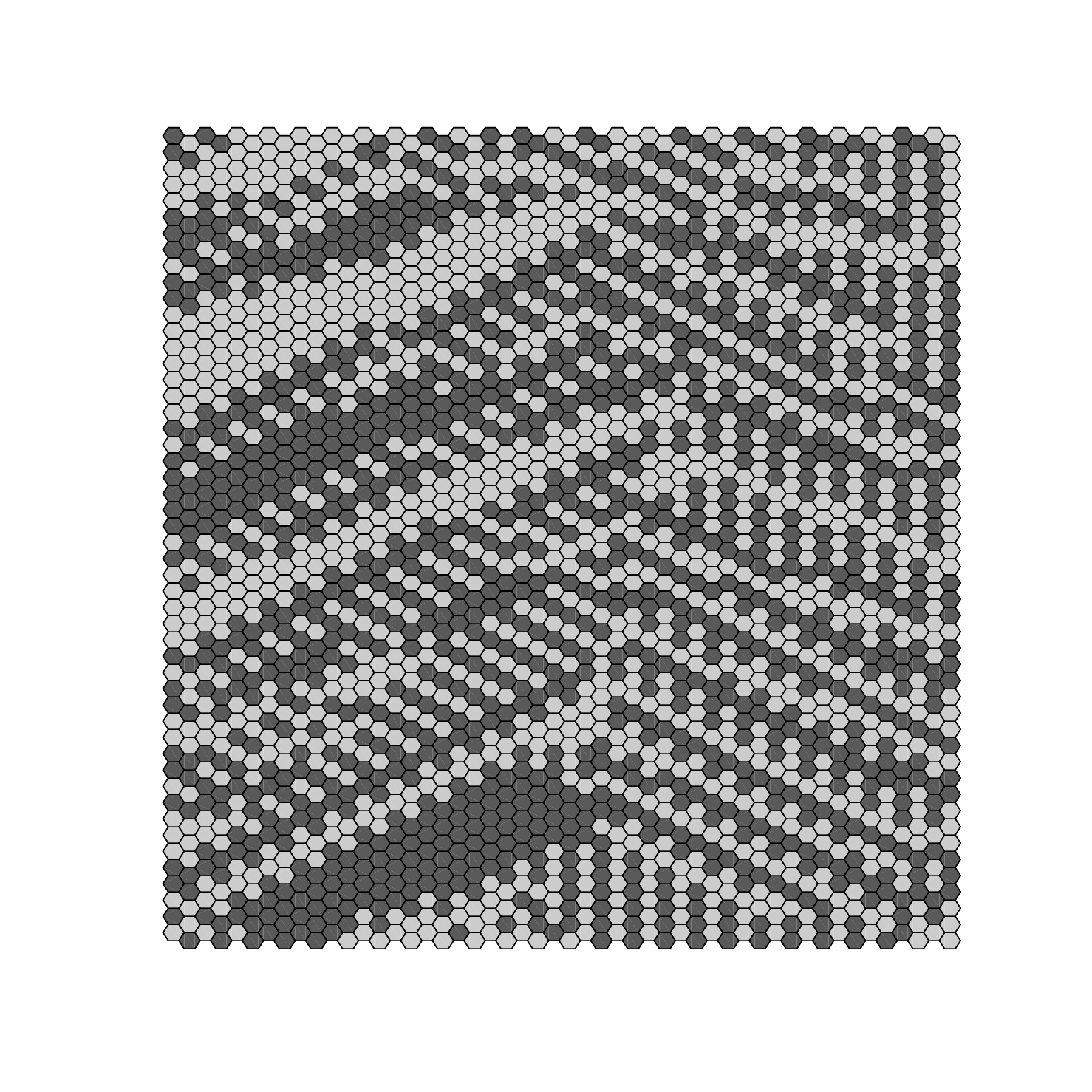}\\
	\includegraphics[trim= 0.8cm 0cm 0.8cm 
0cm,clip,width=0.85\linewidth]{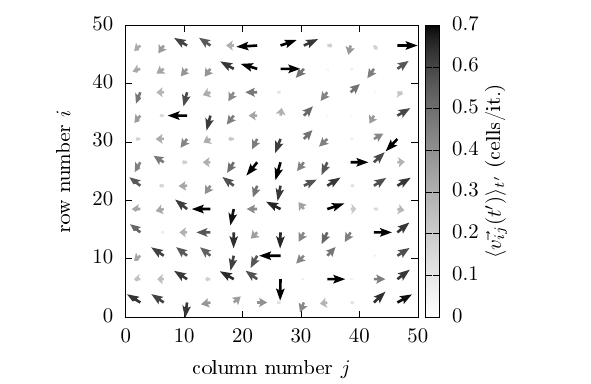}\\	
   	\caption{Example of a highly ordered attracting state (top)  
	         displaying a disordered (temporally averaged) velocity field 
(bottom). The gray level indicates 
                 the droplets' speed (cell/it.).}
    \label{fig:undir}
\end{figure}

 
The network state shown in Fig.\ref{fig:undir} 
(upper panel) and its corresponding (temporally averaged) 
velocity field (lower panel) has been generated by applying the 
parameter set (50$\times$50, UHL, DAU, OP, 64\%, 1, 0\%).  
It can be seen that even if the network state is well organized
the corresponding velocity field appears to be rather disordered. 
A closer look reveals that the network state as well as 
its corresponding velocity field is divided into two parts. 
The right half transports the mucus with a tendency to the right, 
while the transport on the left half tends to the left.
The global average velocity amounts to  
$\vec{v}_{g\infty} = (-0.14, -0.03)$ cells/it.\par
In order to classify parameter sets generating such odd 
velocity fields as malfuctioning, we measured 
the spread of the 
(temporally averaged) local velocity fields
in terms of 
$\langle \cos\theta\rangle \doteq \langle \cos\theta_{ij}\rangle_{ij}$, where 
$\langle... \rangle_{ij}$ indicates spatial averaging. 
We defined $\cos\theta_{ij}$ as: 
\begin{equation}
\cos\theta_{ij} \doteq \lim\limits_{t' \to 
\infty}\frac{ \langle\vec{v}_{ij}(t') \rangle_{t^{\prime}}\cdot 
\langle\vec{v}_g(t')\rangle_{t^{\prime}} }
{|\langle\vec{v}_{ij}(t')\rangle_{t^{\prime}}| \cdot 
|\langle\vec{v}_{g}(t')\rangle_{t^{\prime}} |}~,
\end{equation}
where $\langle...\rangle_{t^{\prime}}$ indicates temporal averaging 
(over the last $10^3$ iterations of each simulation).  
For the velocity field shown in 
Fig.\ref{fig:undir} $\langle\cos\theta\rangle$ amounts to 0.18
indicating not properly directed transport.    
\subsubsection{\label{sec:transienttime}Transient Time}
Since $\Psi^{t'}$ represents the state of the Boolean network
at time $t'$ containing $N$ Boolean variables, the set of all 2$^N$ possible 
network states forms the state space of the Boolean network.  
The successive network states $\Psi^0,\Psi^1, \Psi^2 ...$ form a trajectory in 
the state space. 
In the case of 
deterministic Boolean networks, a network sooner or later reaches 
a state, which has been reached before 
(due to the finite state space) and consequently, enters a cycle 
consisting of a subset of 
states of the state space, which is called an 
attractor. A more general definition for dynamical systems says 
that an attractor is a set of states to which the system 
evolves after a long enough time \cite{Zou2011}.   
The transient time $\tau$ is the number of states 
a network undergoes (starting from an initial state), before it reaches an 
attractor 
\cite{Wuensche1998, Gershenson2004, Greil2012}.\par
\begin{figure}[!phbt]
	\centering
     \includegraphics[trim= 0cm 0cm 0cm 
0cm,clip,width=0.95\linewidth]{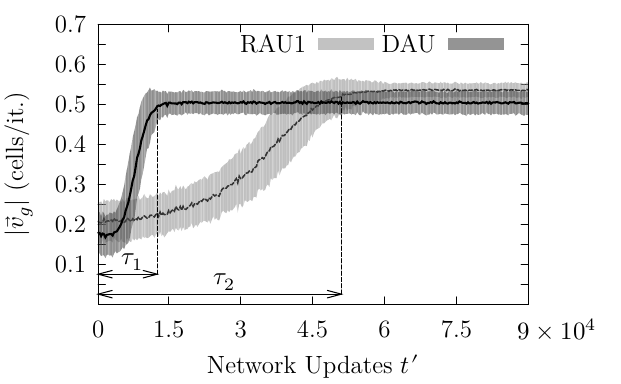}
     \caption{The curves show the temporal evolution of the average mucus speed 
				for two ensemble simulations differing only by 
their update scheme. The darker 
				and brighter curve correspond 
				to simulations for which DAU and RAU1 has been 
applied, respectively. 
                                Simulation runs initialized with the update 
scheme DAU reach roughly four times 
                                faster their attractors than those initialized 
with RAU1, which
 				is indicated by the corresponding transient 
times $\tau_1$ and $\tau_2$.} 
    \label{fig:transienttime}
\end{figure}

 
Fig.\ref{fig:transienttime} depicts the temporal evolution of two ensembles 
consisting of 
100 ensemble members differing only by their initial condition. 
The brighter band corresponds to the curve presented in 
Fig.\ref{fig:transientrau1} showing 
the evolution of the average mucus transport speed being generated with the 
parameter setting (50$\times$50, BHL+L, RAU1, OP, 10\%, 1, 0\%). 
The darker curve has been generated with exactly the same parameter settings 
apart from 
the update scheme. Instead of RAU1 the update DAU was used. 
Both ensembles show 
the typical saturation-like temporal evolution  
of the average transport speeds, which reflects the capturing of the system 
dynamics by attractors.  
We calculated the transient time based on the transient behavior of the average 
transport 
speed, which roughly follows an exponential behavior: 
$|\vec{v}_g(t')| = |\vec{v}_{g0}| + \exp(-t'/\tau') \cdot 
(|\vec{v_{g\infty}}| - |\vec{v_{g0}}|)$~. 
The transient time $\tau$ has been defined as $\tau \doteq 
3\cdot\tau'$. 
As one can notice in Fig.\ref{fig:transienttime} the transient time for the 
ensemble simulation using 
the deterministic update scheme DAU is roughly four times shorter than the one 
for which the update scheme 
RAU1 was used. This behavior may be more fundamental and shall be discussed 
later. 
\subsubsection{Autocorrelation}
\begin{figure}[!phbt]
	\centering
     \includegraphics[trim= 2cm 1cm 0cm 
1cm,clip,width=1.0\linewidth]{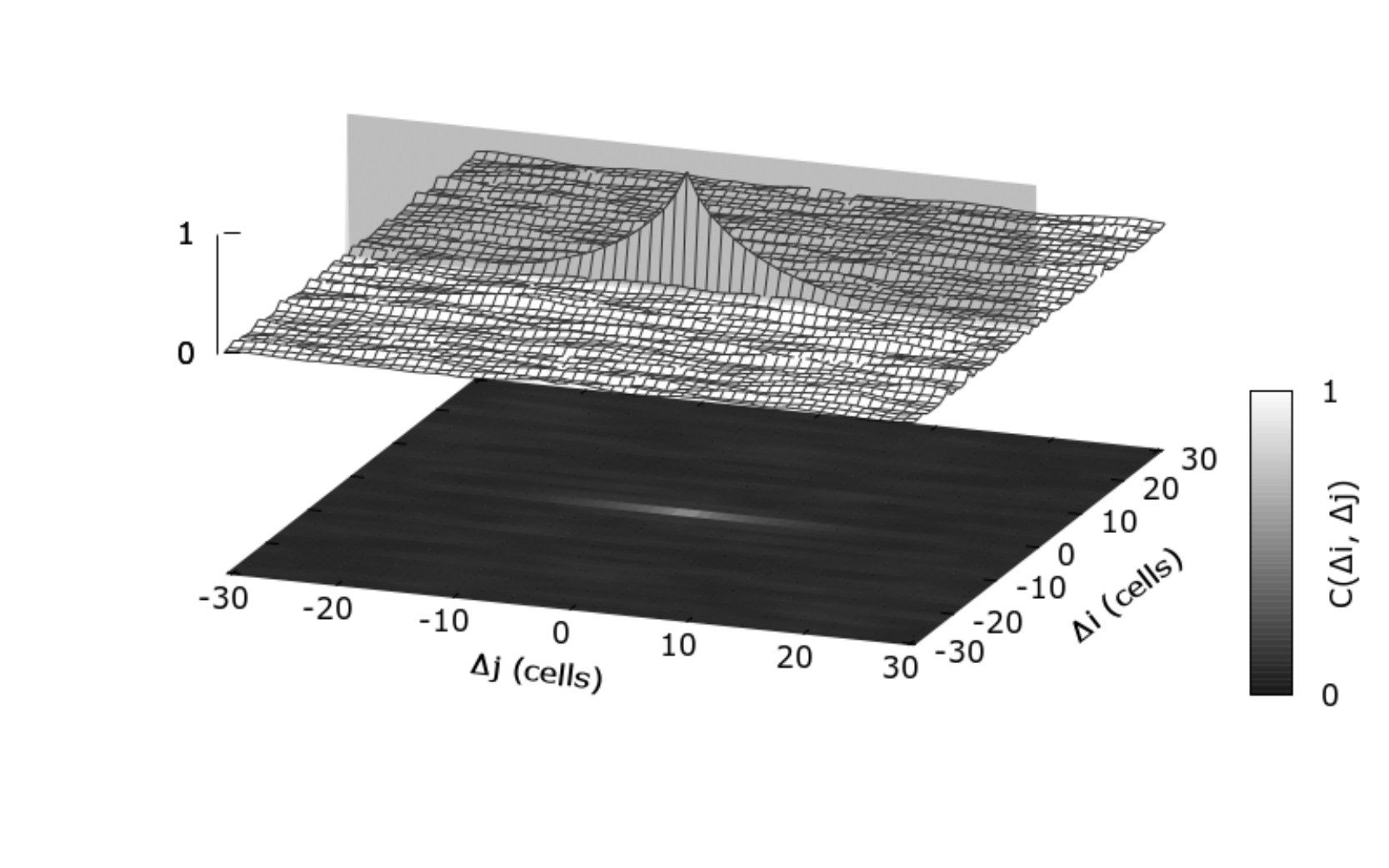}
     \caption{The autocorrelograms of attracting states typically have 
				an elongated shape along a certain direction 
(indicated by the bright gray vertical plane), 
along which we determined the autocorrelation-length $\rho_c$.}
    \label{fig:autocorr}
\end{figure}

In order to characterize the coordination amongst the 
actuators in a particular network state, we use the 
spatial autocorrelation 
function C$(\Delta i, \Delta j, t')=\sum\limits_{i,j} \Psi (i,j,t') \Psi 
(i+\Delta 
i,j+\Delta j,t')$ , where 
$\Delta i$ and $\Delta j$ denote the shifts into the $i$- and $j$-direction, 
respectively.
As we would like to compare different simulation runs with respect to the 
emergent spatial order,
we used the autocorrelogram to determine the spatial autocorrelation-length 
$\rho_c$, representing 
a measure for the degree of order in a network state. As 
most correlograms appear to be strongly elongated as illustrated in 
Fig.\ref{fig:autocorr},
we determined the autocorrelation-length along the direction of maximum 
correlation, which is indicated by the 
gray plane in Fig.\ref{fig:autocorr}. Note that in roughly 95\% of all properly 
self-cleaning attracting states 
the direction of maximum correlation coincides with the direction of mucus 
transport. 
As most correlograms show an exponential-like decrease along the
direction of maximum correlation, the autocorrelation-length $\rho_c$ 
has been determined according to:
\begin{widetext} 
\begin{equation}
\rho_c(t') = \text{max}\left\{ \frac{1}{2} \sum\limits_{v=v_1...v_2} 
\text{C}(v, 
\gamma(v), t') ~~ | ~~ 
\gamma(v) = a\cdot v + b ~;~ a,b,v \in \mathbb{R} ~;~ 0 \leq v_1,v_2 \leq I ; 0 
\leq \gamma(v) \leq J \right\} 
	\label{eq:autocorr}
\end{equation}
\end{widetext}
Eq.\ref{eq:autocorr} delivers the autocorrelation-length $\rho_c$ along the 
line 
of maximum correlation, which 
is represented by the parameterization $\gamma(v)$. 
\subsection{\label{sec:coevolution}Co-evolution of Spatio-Temporal Patterns and 
Transport}
\begin{figure}[!phbt]
    \centering
    \includegraphics[trim= 0cm 0cm 0cm 
0cm,clip,width=0.95\linewidth]{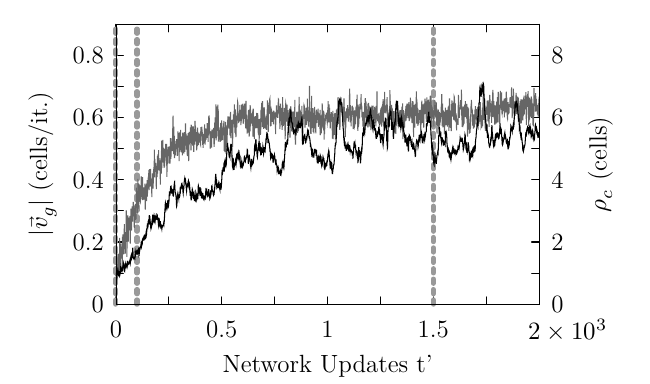}
    \caption{The graph illustrates the co-evolution of the auto-correlation 
length of the network
state along the direction of maximum correlation $\rho_c(t')$ (dark 
curve) and
the area averaged mucus transport speed $|\vec{v_g}(t')|$ (bright 
curve). The vertical dashed lines indicate the initial ($t'$=0), an intermediate
($t'$=100) and a final stage ($t'$=1'500) of the self-organization process.}
    \label{fig:vevol}
\end{figure}

The co-evolution of self-organized transport and spatio-temporal patterns
is illustrated in Fig.\ref{fig:vevol}. 
The brighter curve shows the area-averaged transport speed $|\vec{v_g}(t')|$
(according to Eq.\ref{eq:globalv}) illustrating the build up of self-organized 
transport.
The darker curve represents
the auto-correlation length of the network's state at time $t'$. 
As at the beginning of the simulation the mucus droplets get displaced almost 
erratically,
the corresponding area-averaged transport speed almost vanishes.
As time passes the average particle speed grows until the coordination of the 
actuators reaches a maximum.
The three vertical dashed lines indicate the moments at which the three network 
states
shown in Fig.\ref{fig:3stages} have been recorded.
Interestingly, all model settings leading to self-organized transport 
have shown a similar growth behavior with respect to the global average 
transport speed as the one depicted in Fig.\ref{fig:vevol}.  
This typical transient dynamical behavior reflects the restricted growth of 
the cooperation among the actuators. 
This sigmoid logistic-like dynamical behavior can be observed
in other studies investigating self-organizing processes as well, and appears
to be omnipresent for self-organizing processes. An example is provided
by Fig.2 in \cite{Niedermayer2008}, in which the self-organized synchronization 
and wave
formation in one-dimensional cilia arrays has been studied.\par 
Within the meaning of the state space concept, the curves in Fig.\ref{fig:vevol}
can be seen as the network's transient behavior. 
The saturation-like behavior reflects the capture
of the network dynamics by an attractor.
Accordingly, the strongly ordered network state shown in Fig.\ref{fig:3stages}
represents an attracting state.

Our model typically exhibits a co-evolution of the
network's state and its associated mucus transport. As indicated 
earlier, we suggest to see 
this two-fold self-organization in the context of adaptive boolean networks. 
Since mucus droplets functionally connect an actuator to its surrounding 
and disconnect an actuator when squeezed away, the network topology is exposed 
to 
dynamics and interrelated to the dynamics of the network's state. 
Consequently, we observe not only dynamics \emph{on the network}, represented 
by 
state
transitions, but also \emph{dynamics of the network}, caused by the 
transportation of the mucus droplets.
This means that changes in the network's state are affected by the network's 
topology and vice versa, forming 
a feedback loop between the topology and the state of the network. 
The resulting co-evolution of the network state and the network topology 
represents
the characteristic property of coevolutionary or adaptive 
(boolean) networks \cite{Gross2008, Rohlf2008}.\\
\subsection{Characterisitcs of Attractor States}
\subsubsection{Transport of Attracting States} 
\begin{figure*}[!phbt]
    \centering

    \begin{minipage}[t]{0.24\linewidth}
    USL \\
    \vspace*{0.65cm}\includegraphics[scale=0.9]{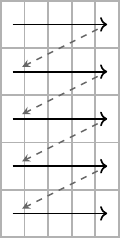}
    \end{minipage}
    \hfill
    \begin{minipage}[t]{0.24\linewidth}
    UHL \\
    \vspace*{0.5cm}\includegraphics[scale=0.9]{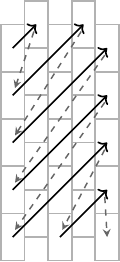}
    \end{minipage}
    \hfill
    \begin{minipage}[t]{0.24\linewidth}
    BHL \\
    \vspace*{0cm}\includegraphics[scale=0.9,angle=-45]{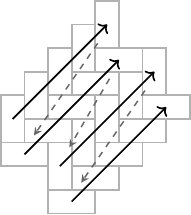}
    \end{minipage}
    \hfill
    \begin{minipage}[t]{0.24\linewidth}
    BHL+L \\
    \vspace*{0cm}\includegraphics[scale=0.9,angle=-45]{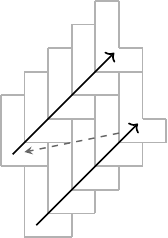}
    \end{minipage}

  	\begin{minipage}[t]{0.24\linewidth}
  	\vspace{0.15cm}
    $m\cdot\vec{v}_{g\infty}$
    \includegraphics[width=0.9\linewidth]{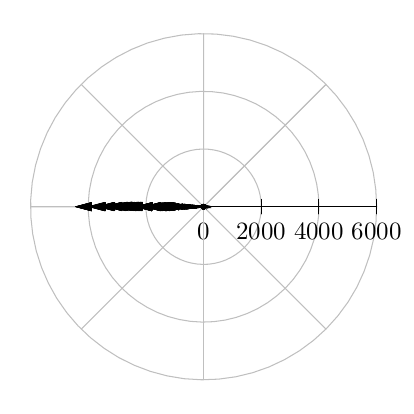}
    \end{minipage}
    \hfill
    \begin{minipage}[t]{0.24\linewidth}
    \vspace{0.15cm}
    $m\cdot\vec{v}_{g\infty}$
    \includegraphics[width=0.9\linewidth]{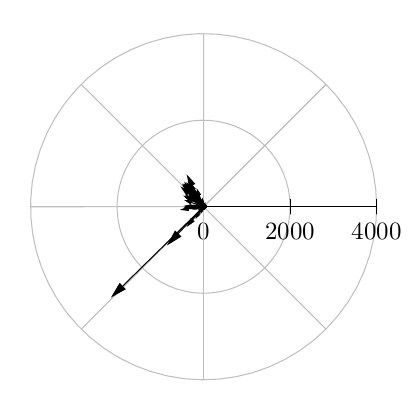}
    \end{minipage}
    \hfill
    \begin{minipage}[t]{0.24\linewidth}
    \vspace{0.15cm}
    $m\cdot\vec{v}_{g\infty}$
    \includegraphics[width=0.9\linewidth]{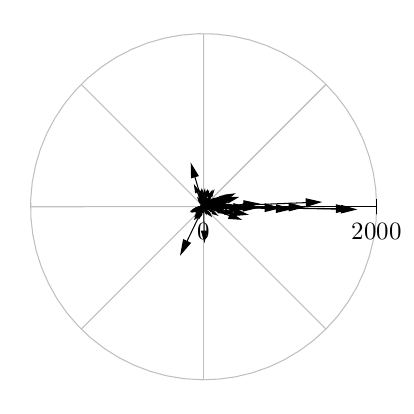}
    \end{minipage}
    \hfill
    \begin{minipage}[t]{0.24\linewidth}
    \vspace{0.15cm}
    $m\cdot\vec{v}_{g\infty}$
    \includegraphics[width=0.9\textwidth]{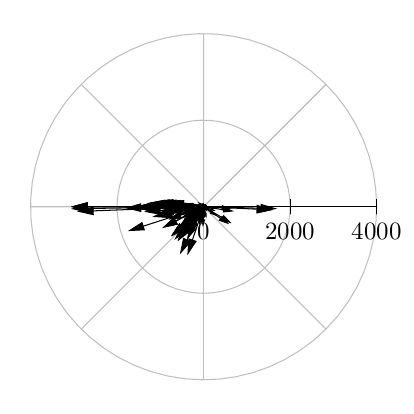}
    \end{minipage}

   	\begin{minipage}[t]{0.24\linewidth}
    $\vec{v}_{g\infty}$
    \includegraphics[width=0.9\linewidth]{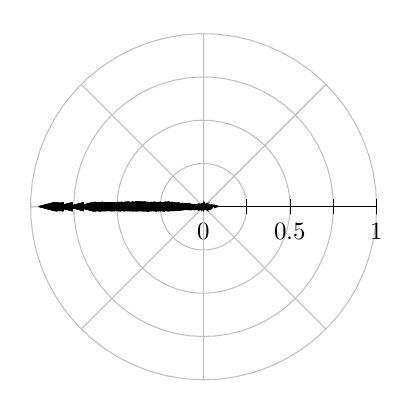}
    \end{minipage}
    \hfill
    \begin{minipage}[t]{0.24\linewidth}
    $\vec{v}_{g\infty}$
    \includegraphics[width=0.9\linewidth]{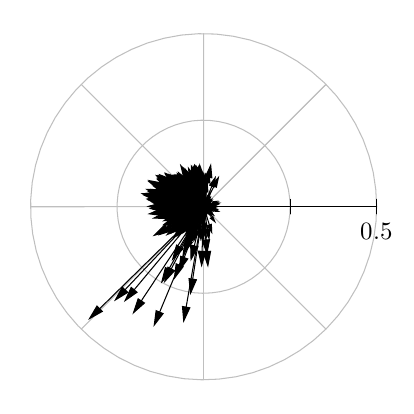}
    \end{minipage}
    \hfill
    \begin{minipage}[t]{0.24\linewidth}
    $\vec{v}_{g\infty}$
    \includegraphics[width=0.9\linewidth]{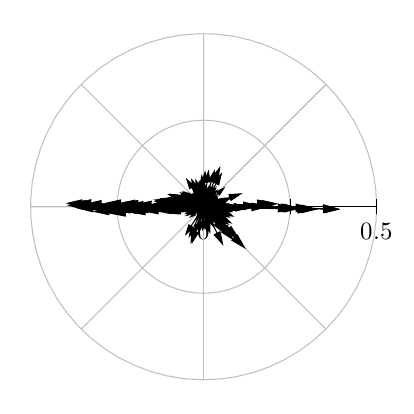}
    \end{minipage}
    \hfill
    \begin{minipage}[t]{0.24\linewidth}
    $\vec{v}_{g\infty}$
    \includegraphics[width=0.9\textwidth]{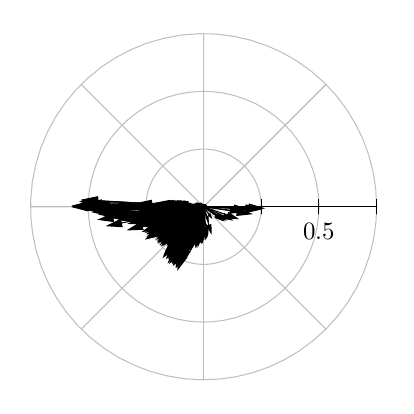}
    \end{minipage}

    \begin{minipage}[t]{0.24\linewidth}
    $\vec{v}_{g0}$
    \includegraphics[width=0.9\linewidth]{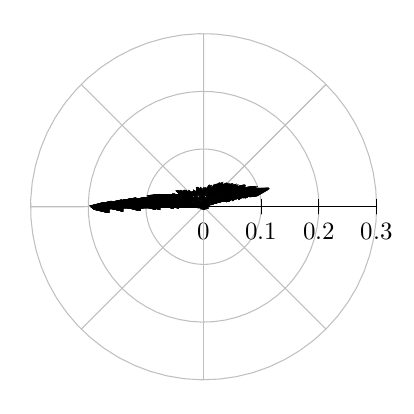}
    \end{minipage}
    \hfill
    \begin{minipage}[t]{0.24\linewidth}
    $\vec{v}_{g0}$
    \includegraphics[width=0.9\linewidth]{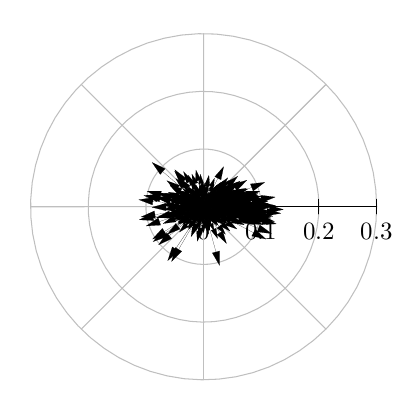}
    \end{minipage}
    \hfill
    \begin{minipage}[t]{0.24\linewidth}
    $\vec{v}_{g0}$
    \includegraphics[width=0.9\linewidth]{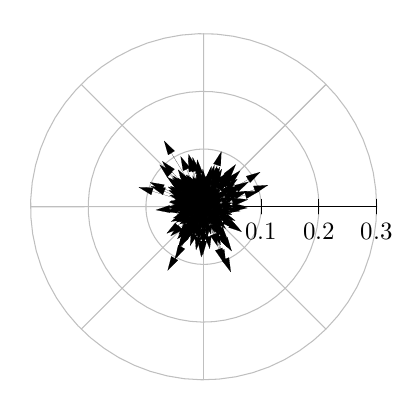}
    \end{minipage}
    \hfill
    \begin{minipage}[t]{0.24\linewidth}
    $\vec{v}_{g0}$
    \includegraphics[width=0.9\textwidth]{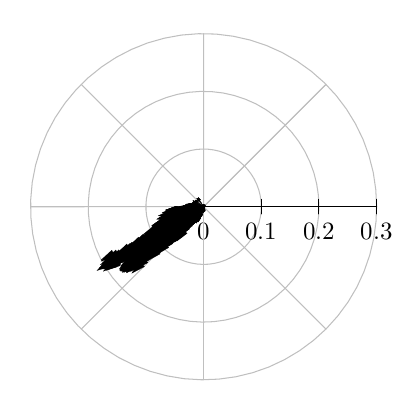}
    \end{minipage}
\caption{The rows present the distribution of the 24'000 mucus transport vectors
$m\cdot\vec{v}_{g\infty}$ (\#m$\cdot$cells/it.), average terminal velocities
$\vec{v}_{g \infty}$ (cells/it.) and the initial
velocities $\vec{v}_{g0}$ (cells/it.) for each alignment (column-wise).}
 \label{fig:polar}
\end{figure*}

Fig.\ref{fig:polar} presents an overview of the terminal mucus transport
($m\cdot\vec{v}_{g\infty}$), the terminal mucus transport velocity
($\vec{v}_{g\infty}$) and the initial mucus transport velocity
($\vec{v}_{g0}$) for each cell alignment (column-wise).
The first row indicates to which cell alignment each column corresponds to
and illustrates the sequences
after which the actuators have been activated when the
deterministic update scheme DAU has been applied.
Apparently, the globally averaged terminal transport velocities
have a clear preference considering their direction for
each cell alignment. Generally, the preferred transport direction
seems to be oppositely-oriented
to the direction of the deterministic update signal.
However, for a small set of the settings using BHL or BHL+L the
transport shares the direction of the update signal.
\subsubsection{Classification Attempt} 
\begin{figure}[!phbt]

    \includegraphics[trim=0cm 0.5cm 0cm 0.5cm,scale=0.9]{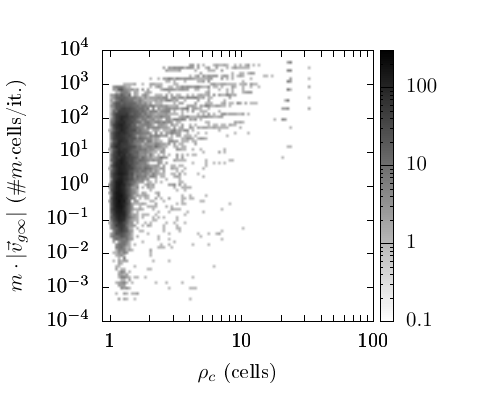}\\
    \includegraphics[trim=0cm 0.5cm 0cm 0.5cm,scale=0.9]{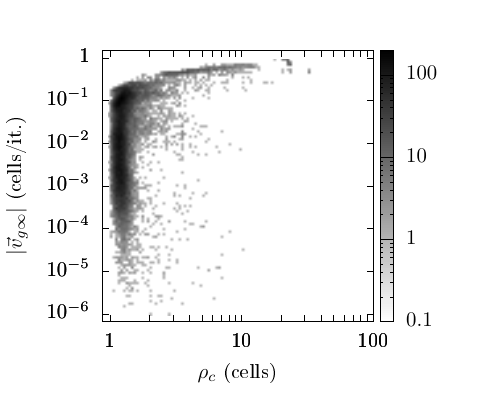}

    \caption{Distribution of the frequency densities of attracting
    states considering their transport rate (top) and transport
    speed (bottom) with respect to their corresponding correlation length.}
    \label{fig:transportcorr}
\end{figure}

Fig.\ref{fig:transportcorr} illustrates how the transport
capability (upper colormap) and the transport
speed (lower colormap) of attractor states is related to the
correlation length. The gray scale indicates the frequency density.\par
Considering the attractors' function and structure a rough
classification
into the following four classes (C1-C4) seems appropriate:
C1: nontransporting disorganized states,
C2: transporting disorganized states, 
C3: nontransporting structured states
and C4: transporting structured states. 
We measured the transport capability in terms of 
the transport speed
$|\vec{v_{g\infty}}|$ and the degree of the organization of 
the expression patterns in terms of $\rho_c$. 
States/settings exhibiting a higher $|\vec{v}_{g\infty}|$ than 
0.1 (cells/it.) are considered as transporting, while states/settings 
exhibting a $|\vec{v}_{g\infty}|$ of less than 0.01 (cells/it.) have 
been considered as nontransporting. 
States satisfying $\rho_c > 2$ (cells) have been classified as 
organized and states with a $\rho_c$ of less than 
1.5 (cells) as disorganized ($\rho_c$ of the randomly generated 
initial states were shorter than 1.6 cells).\par
An illustration of how the observables and the 
associated parameter  
values are distributed in each class, as well as 
a listing of the most striking conspicuities, 
can be found in the supplemental material \cite{supm}.
Here, we just note that attractor states of 
class 4 still contain many states exhibiting  
odd velocity fields, like bisected or disordered ones, 
which still reach considerable terminal transport speeds.
Consequently, only a subset of class 4 is considered 
as ``properly self-cleaning'' attractor states, whose 
velocity fields shall fulfill further conditions specified 
in the next section.   
\subsubsection{Effectively Self-Organized Properly Self-Cleaning 
Attractors} 
In order to find the parameter values allowing 
the network to self-organize  
towards properly self-cleaning attractor states, we 
finally classified the settings
as ``functioning'' or ''malfunctioning''. 
A certain parameter set is classified as functioning, if 
the following four conditions are simultaneously fulfilled after an 
integration time of $t'=10^5$. 
1) The globally averaged transport speed $|\vec{v}_{g\infty}|$ is faster than 
0.1 cells/it., which would 
roughly correspond to 20$\mu\text{m}/s$ in the real system (assuming a ciliary 
beat frequency of 10 Hz and 
the diameter of a ciliated cell being 10$\mu\text{m}$). 
2) We require the velocity field being sufficiently directed by: 
$\langle\cos\theta\rangle > 0.65$.
3) The velocity field must be self-organized and not simply being imposed by 
the choice of a parameter set leading to an initial 
tendency of the transport direction. 
Consequently, by requiring $|\Delta \vec{v}| \doteq 
|\vec{v_{g\infty}}-\vec{v_{g0}}| > 0.1$, we demand that 
the velocity field changes (at least its direction) in the course of a 
simulation.
4) The auto-correlation length $\rho_c$ has to be longer than two cells. 
The number of iterations may be seen as 
a fifth condition considering the efficiency of the self-organization process.  
As thus, the transient time $\tau$ has to be shorter than $10^5$ iterations. 
The classification of parameter sets as functioning and malfunctioning was 
applied 
to the parameter study encompassing 24'000 parameter sets.  
564 out of 24'000 settings ($\widehat{=}$ 2.4$\%$) have been classified as 
functioning. 
Table \ref{tab:tab1} shows how these functioning settings 
spread across the different cell alignments and update schemes (values are 
listed in $[\%]$). 
\begin{table}[hbt]
	\centering
	\caption{\label{tab:tab1}The table shows how the functioning states 
spread across the different topologies and 
		update schemes. The values represent the corresponding 
proportions of all functioning states in [\%].}
	\begin{ruledtabular}
		\begin{tabular}{d{1} | d{1} | d{1} | d{1} | d{1} | d{1} }

				&	\text{DAU}	&	\text{RAU1}	
	
&	\text{RAU2}		&	\text{SRAU1}	&	\text{SRAU2}	
	
\\	
\hline
\text{USL}		&	84.2		&	0.0			
	
&	0.0				&	0.0			&	
0.0				 	\\
\hline  
\text{UHL}		&	1.2			&	0.0		
	
	&	0.0				&	0.0			
	
&	0.0				  	\\
\hline
\text{BHL}		& 	0.5 		&	0.0			
	
&	0.0				&	0.0				
&	0.0					\\
\hline
\text{BHL+L}	& 	3.7 		&	1.6				
&	0.0				&	0.9				
&	7.8				 	\\

		\end{tabular}
	\end{ruledtabular}
\end{table} 
Table \ref{tab:tab1} summarizes the following observations. 
The cell alignment USL can cope with the largest set of parameter settings
and makes up $84.2\%$ of all functioning settings. 
$98.2\%$ of all functioning settings are made up by settings using either the 
cell alignment USL or BHL+L. 
The deterministic update scheme DAU can cope with each cell alignment and 
$89.6\%$ of all functioning 
settings were generated by this update scheme.  
None of the settings using the completely random update RAU2 evolves the 
network to a self-organized self-cleaning state. 
The cell alignment BHL+L (lowest row in table \ref{tab:tab1}) 
can cope with all update schemes except the fully 
random update scheme RAU2.
Apparently, a certain degree of local determinism  
is advantageous for an efficient self-organization towards properly transporting 
attractor states, since all functioning parameter settings 
either involve DAU or BHL+L.  
\subsubsection{Mean Wave Propagation Direction in Stereotypical 
Functioning States}
The set of functioning parameter values drives the model towards
families, or stereotypes, of attractor states
comprising very similar attractor states. 
In order to determine the mean direction of wave propagation in
these sterotypical attractors, we employed 
the concepts outlined in \cite{Ryser2007} and examined
sequences of sequential attracting states of  
one specific representative parameter set for each stereotype.
A more detailed discussion of these representative 
stereotypical attractor states
can be found in the supplemental material \cite{supm}. Here, we would 
like to report the summarizing comparison between the direction 
of the mean wave propagation and the direction of transport, what can be 
found in Table \ref{tab:tab2}. Note that mean $k$-vectors visualized by 
double arrows indicate that the mean direction of wave propagation 
could not be determined unambiguously, which comes from the 
boolean nature of the model (undersampling).    
\begin{table*}
	\caption{The table reports the paramater sets of stereotypical 
	representatives, the typical appearance
	 of attracting states, the transport speed and particularly,
	 the direction of transport as well as of the mean wave propagation 
	 is visualized.}
	\vspace*{0.3cm}
	\begin{tabular}{ l  l  l  l  c  c }
\hline
	& \quad Parameters & \quad Appearance & \quad $|\vec{v}_{g\infty}|$ & 
	\quad $\vec{v}_{g\infty}$ & \quad $\vec{k}$ 	
	
\\	
\hline

USL - stereo1 & \quad (DAU,TO,64\%,0.25,0\%) & \quad
``frozen'' & \quad 0.69 & \quad 
\raisebox{-0.45\totalheight}{\includegraphics[scale=0.6]{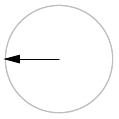}} 
& \quad \raisebox{-0.45\totalheight}{\includegraphics[scale=0.6]{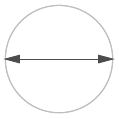}}
\\
\hline
USL - stereo2 & \quad (DAU,TO,64\%,0.25,0\%) & \quad ``re-organizing'' & 
\quad 0.5 & 
\quad \raisebox{-0.45\totalheight}{\includegraphics[scale=0.6]{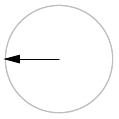}} & 
\quad
\raisebox{-0.45\totalheight}{\includegraphics[scale=0.6]{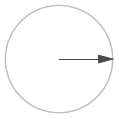}}
\\
\hline
UHL - stereo1 & \quad (DAU,VC,128\%,1,0\%) & \quad ``frozen'' &
\quad 0.38 & \quad
\raisebox{-0.45\totalheight}{\includegraphics[scale=0.6]{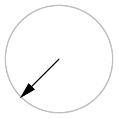}} & 
\quad
\raisebox{-0.45\totalheight}{\includegraphics[scale=0.6]{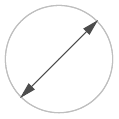}}
\\
\hline 
UHL - stereo2 & \quad (DAU,TO,256\%,1,0\%) & \quad ``gliders'' & 
\quad 0.13 & \quad
\raisebox{-0.45\totalheight}{\includegraphics[scale=0.6]{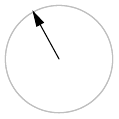}} &
\quad 
\raisebox{-0.45\totalheight}{\includegraphics[scale=0.6]{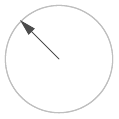}}
\\
\hline
BHL - stereo1 & \quad(DAU,TO,256\%,1,0\%) & \quad poorly organized & 
\quad 0.2 & \quad
\raisebox{-0.45\totalheight}{\includegraphics[scale=0.6]{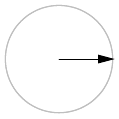}} & \quad
\raisebox{-0.45\totalheight}{\includegraphics[scale=0.6]{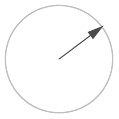}}
\\
\hline
BHL+L - stereo1 & \quad (RAU1,OP,256\%,1,0\%) & \quad frozen crystal &  
\quad 0.5 & \quad
\raisebox{-0.45\totalheight}{\includegraphics[scale=0.6]{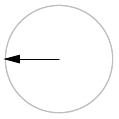}} & \quad
\raisebox{-0.45\totalheight}{\includegraphics[scale=0.6]{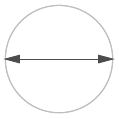}}
\\
\hline
BHL+L - stereo2 & \quad (DAU,OP,256\%,1,0\%) & \quad frozen & \quad 0.5 & \quad
\raisebox{-0.45\totalheight}{\includegraphics[scale=0.6]{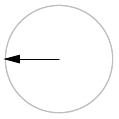}} & \quad 
\raisebox{-0.45\totalheight}{\includegraphics[scale=0.6]{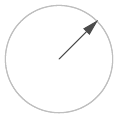}}
\\
\hline
BHL+L - stereo3 & \quad (DAU,TO,256\%,1,0\%) & \quad modular, re-organizing & 
\quad 0.3 & \quad
\raisebox{-0.45\totalheight}{\includegraphics[scale=0.6]{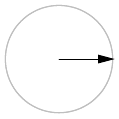}} & \quad
\raisebox{-0.45\totalheight}{\includegraphics[scale=0.6]{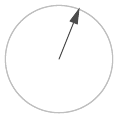}}
\\
\hline
	\label{tab:tab2}
	\end{tabular}
\end{table*} 

\subsection{Dynamical Characteristics}

\subsubsection{Imposed Asymmetry as a Starting Assistance 
For Self-Organized Self-Clearing}  
\begin{figure}[!phbt]
	\centering
	\includegraphics[trim= 0cm 0cm 0cm 
0cm,clip,width=0.85\linewidth]{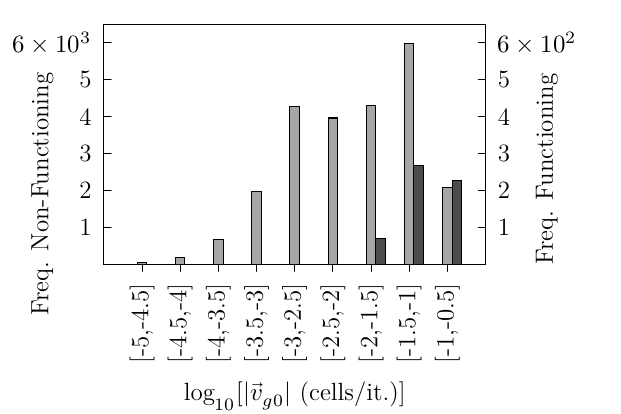}
	\caption{The histograms depict the frequency distributions 
of functioning (dark gray bars) and malfunctioning (bright gray bars) 
parameter sets considering their respective values of $|\vec{v}_{g0}|$.}
   	\label{fig:v0_1}
\end{figure}

In this section we discuss the potential role of imposed asymmetries as 
a control parameter. As shown in Table \ref{tab:tab1} functioning settings 
either include DAU or BHL+L, or both - all other settings don't result in 
self-organized self-clearing states. 
Consequently, these settings may reveal a common characteristic which 
ultimately 
leads to self-organized self-cleaning.
We realized that properly functioning parameter sets 
consistently impose an initial tendency of the transport direction.
The initial area averaged transport speed $|\vec{v}_{g0}|$ can be used in order 
to 
quantify the anisotropy of the transport at time $t'$=0. 
In order to estimate the expected initial transport 
speed at $t'$=0 of a certain parameter setting, we used a grid 
consisting of 1500$\times$1500 actuators,
which have been updated for a single timestep ($\Psi^0 \rightarrow \Psi^1$). 
This has been 
done for an ensemble of 100 simulation runs. 
The ensemble mean provides an estimate for the imposed initial transport speed 
$|\vec{v}_{g0}|$.
Fig.\ref{fig:v0_1} illustrates the frequency distribution 
of functioning (dark gray bars) as well as malfunctioning 
(bright gray bars) parameter sets 
consindering their corresponing $|\vec{v}_{g0}|$-classes
(10 logarithmically spaced $|\vec{v}_{g0}|$-classes between 
$10^{-5}$ and 1). 
Note that the scale of the two ordinates differ by one order 
of magnitude.    
It can be particularly seen that 
the relative frequency of functioning parameter sets increases 
with increasing $|\vec{v}_{g0}|$-values. 
Moreover, all functioning parameter sets    
correspond to $|\vec{v}_{g0}|$-values being faster than 0.01 cells/it.\par
\begin{figure*}[!phbt]

    \hspace*{1cm}  
    \begin{minipage}[t]{0.4\linewidth} 
    \includegraphics[trim=0cm 0.5cm 0cm 0cm,scale=0.77]{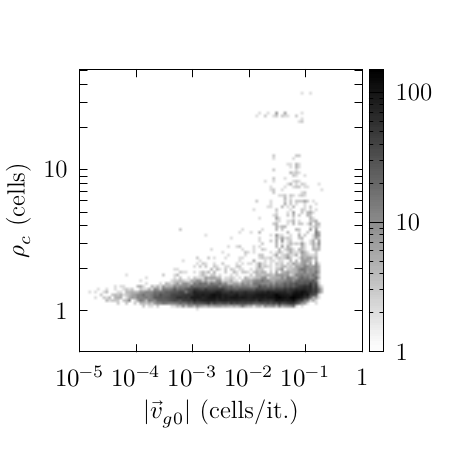}
    \end{minipage}
    \begin{minipage}[t]{0.4\linewidth} 
    \includegraphics[trim=0cm 0.5cm 0cm 0cm,scale=0.8]{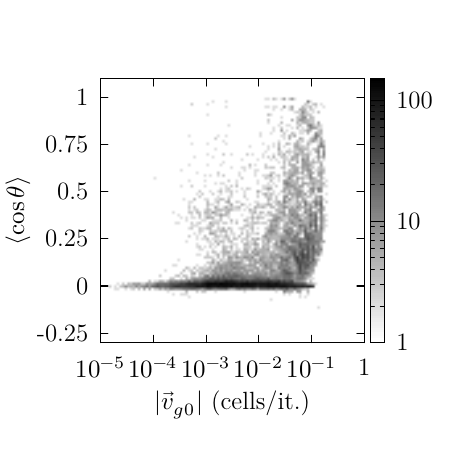}
    \end{minipage} 
    \hspace*{1cm}
    
    \hspace*{0.6cm}
    \begin{minipage}[t]{0.4\linewidth} 
    \includegraphics[trim=0cm 0.5cm 0cm 1.5cm,scale=0.8]{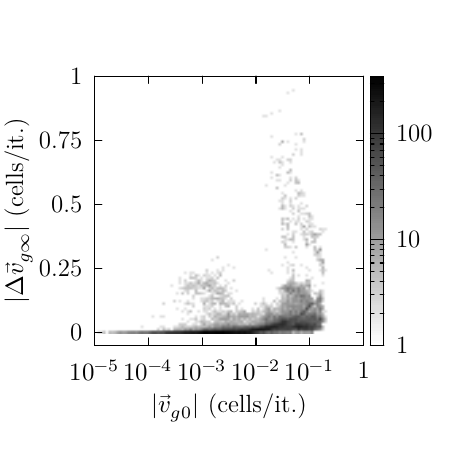}
    \end{minipage}
    \hspace*{0.1cm}
    \begin{minipage}[t]{0.4\linewidth} 
    \includegraphics[trim=0cm 0.5cm 0cm 1cm,scale=0.8]{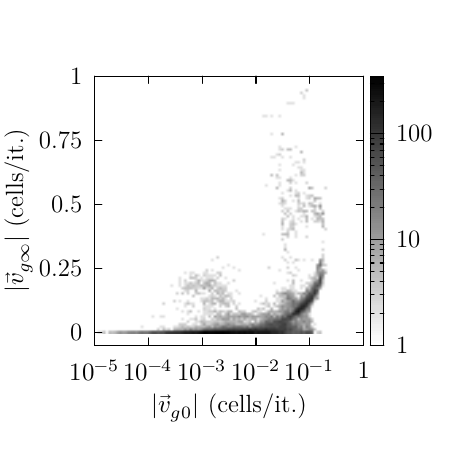}
    \end{minipage}  
    \hspace*{0.6cm}
    
	\caption{From top left to bottom right: Frequency densities (grayscales) 
	illustrating
	the tendency towards higher values for $\rho_c$, 
	$\langle\cos\theta\rangle$, $|\Delta\vec{v}_{g\infty}|$ and  
	$|\vec{v}_{g\infty}|$ for settings imposing larger 
	values of $|\vec{v}_{g0}|$.}
	\label{fig:triggv0}
\end{figure*}

Correspondingly, the measures $|\Delta\vec{v}|$, $\rho_c$, 
$\langle\cos\theta\rangle$ and $|\vec{v_{\infty}}|$
classifying the parameter sets tend to increase with increasing values of 
$|\vec{v}_{g0}|$, which is illustrated in Fig.\ref{fig:triggv0}.   
Consequently, Fig.\ref{fig:v0_1} and Fig.\ref{fig:triggv0} indicate 
that $|\vec{v}_{g0}|$ may trigger the model towards self-organized 
self-clearing.   
 
\begin{figure}[!phbt]
	\centering
	\begin{minipage}[t]{0.85\linewidth}
       \includegraphics[trim= 0cm 0cm 0cm 
0cm,clip,width=0.95\linewidth]{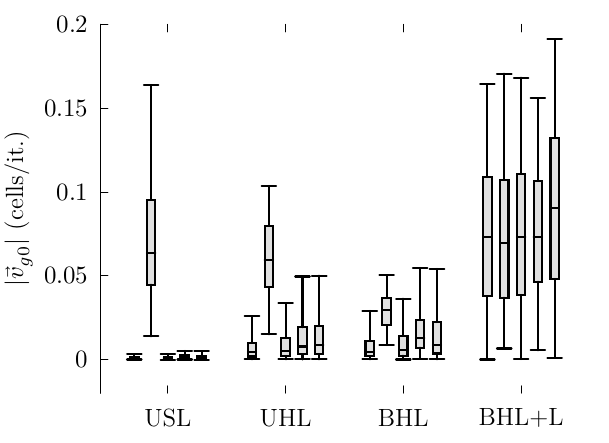}
	\end{minipage}
   \caption{Box plot diagrams to visualize the effect of the choice of the cell 
alignment and the update scheme on the 
			initial transport speed. The box plots of each cell 
alignment correspond to the respective update scheme, 
			according to the order: RAU1, DAU, RAU2, SRAU1, SRAU2.} 
   	\label{fig:boxv0}
\end{figure}

The boxplots shown in Fig.\ref{fig:boxv0} (Whiskers are set in order to 
indicate the range covered by 95\% of all values) display how 
the $|\vec{v}_{g0}|$-values 
distribute over the different update schemes and cell alignments. 
One can see that the highest values for $|\vec{v}_{g0}|$ are reached by 
settings using DAU or 
BHL+L. This clearly suggests that update schemes with local determinism impose 
an asymmetry among the 
local interactions, what triggers efficient self-organization towards 
self-clearing states.\\
\subsubsection{Effect of The Update Scheme on The Network Dynamics}
\begin{figure}[!phbt]
	\centering
    \includegraphics[trim= 0cm 0cm 0cm 
0cm,clip,width=0.95\linewidth]{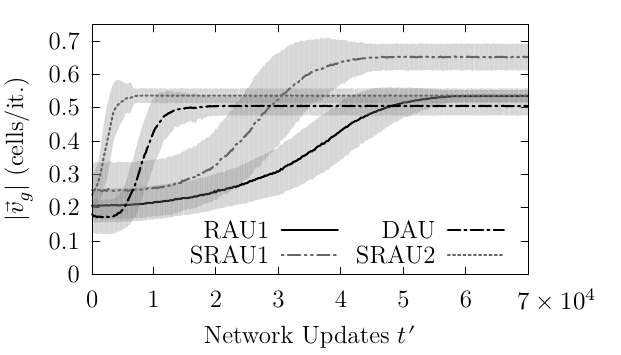}
    \caption{The curves depict the temporal evolution of the average transport 
speed 
		for four ensembles corresponding to different update schemes 
for 
which 
		the cell alignment BHL+L was chosen.}	
   	\label{fig:detvsrand}
\end{figure}

\begin{figure*}[!phbt]
	\centering
	\begin{minipage}[t]{0.3\linewidth}	
	a)
       \includegraphics[trim= 2.5cm 4.2cm 2.5cm 
2.5cm,clip,width=0.99\linewidth]{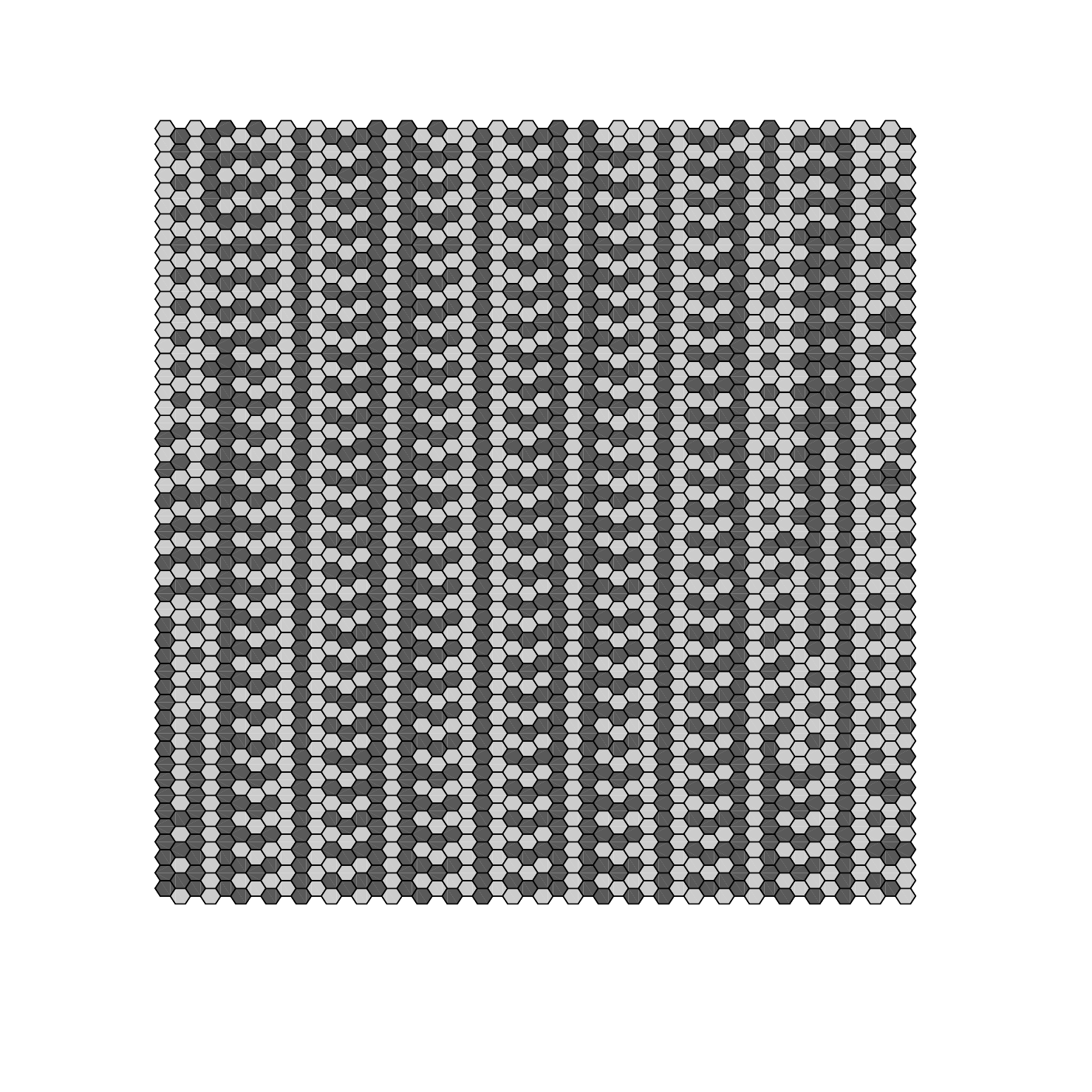}
       \includegraphics[scale=0.6]{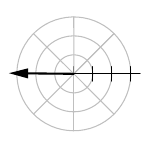}
	\end{minipage}
	\begin{minipage}[t]{0.3\linewidth}
	b)
       \includegraphics[trim=  2.5cm 4.2cm 2.5cm 
2.5cm,clip,width=0.99\linewidth]{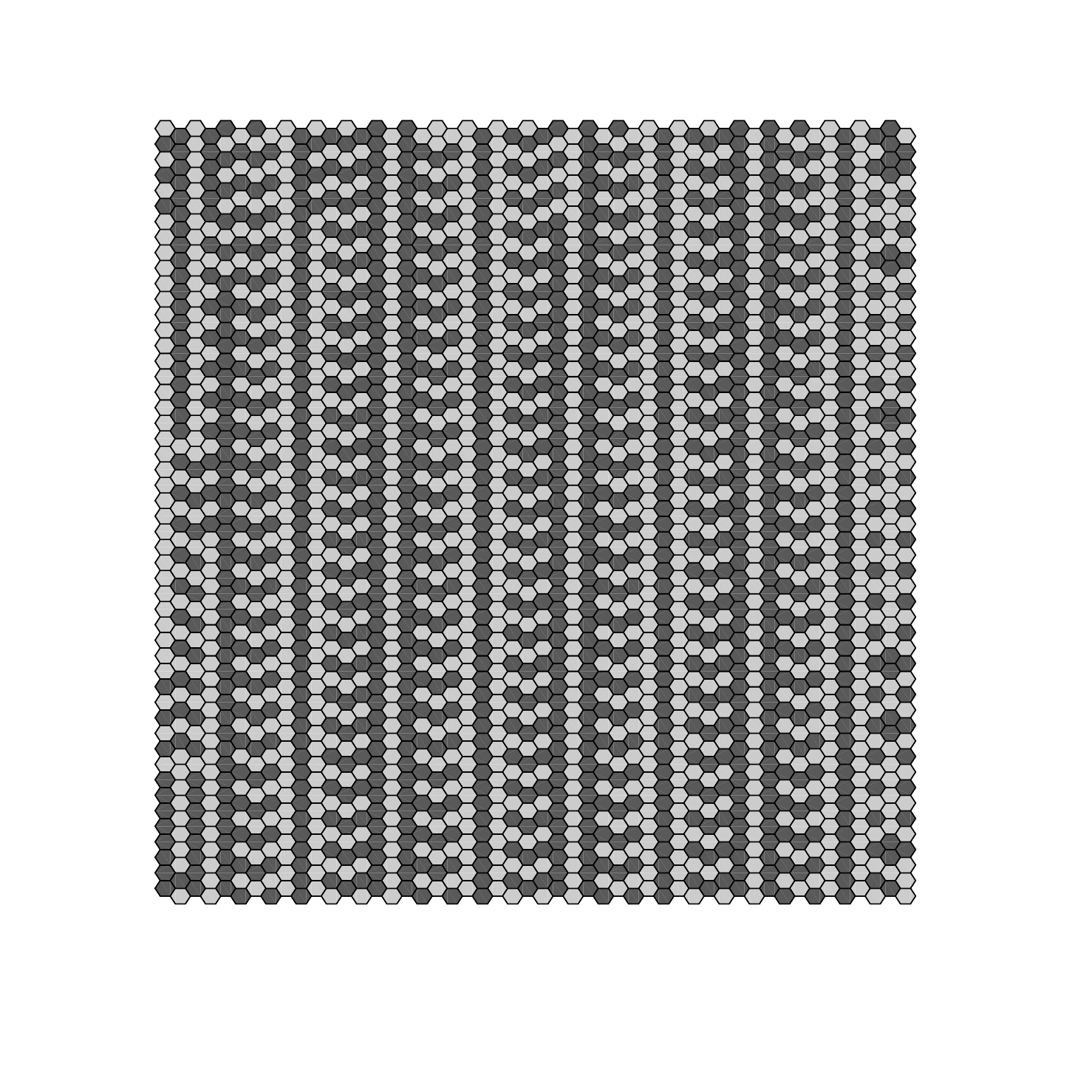}
       \includegraphics[scale=0.6]{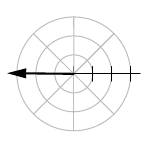}
	\end{minipage}
	\begin{minipage}[t]{0.3\linewidth}
	c)
       \includegraphics[trim=  2.5cm 4.2cm 2.5cm 2.5cm 
,clip,width=0.99\linewidth]{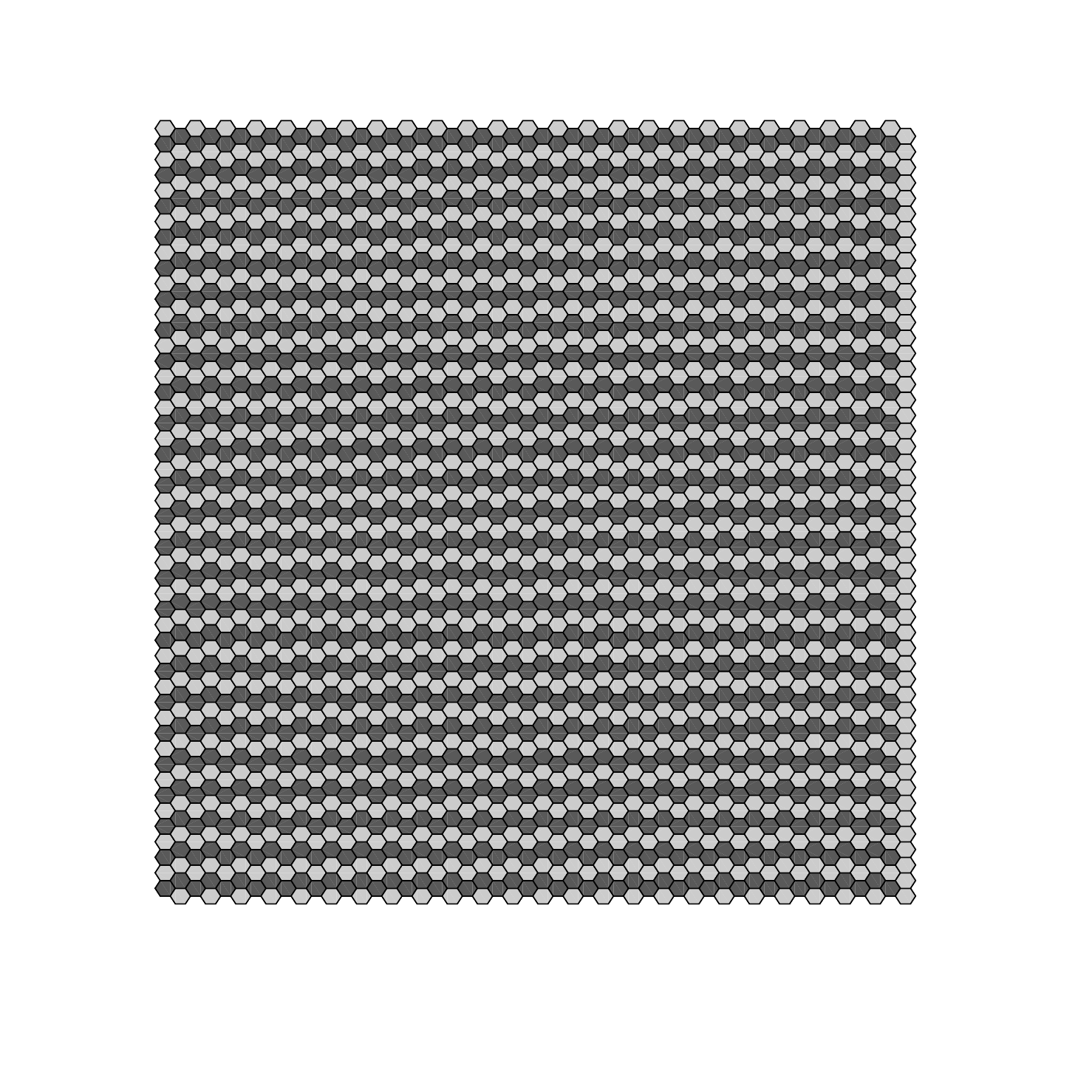}
       \includegraphics[scale=0.6]{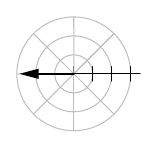}
	\end{minipage}
	\begin{minipage}[t]{0.3\linewidth}
	d)
       \includegraphics[trim=  2.5cm 4.2cm 2.5cm 2.5cm 
,clip,width=0.99\linewidth]{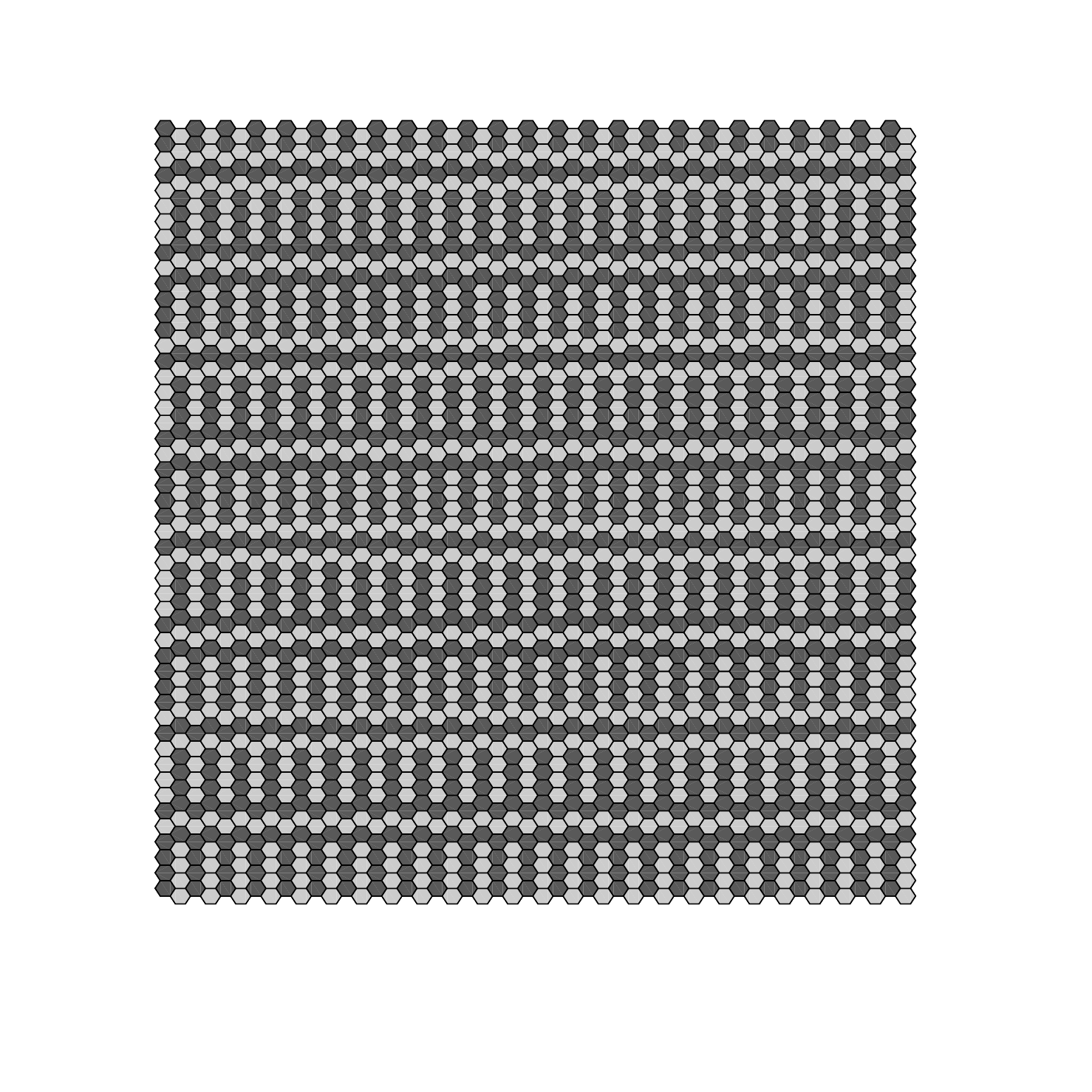}
       \includegraphics[scale=0.6]{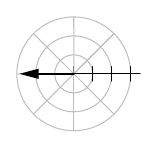}
	\end{minipage}
	\begin{minipage}[t]{0.3\linewidth}
	e)
       \includegraphics[trim=  2.5cm 4.2cm 2.5cm 2.5cm 
,clip,width=0.99\linewidth]{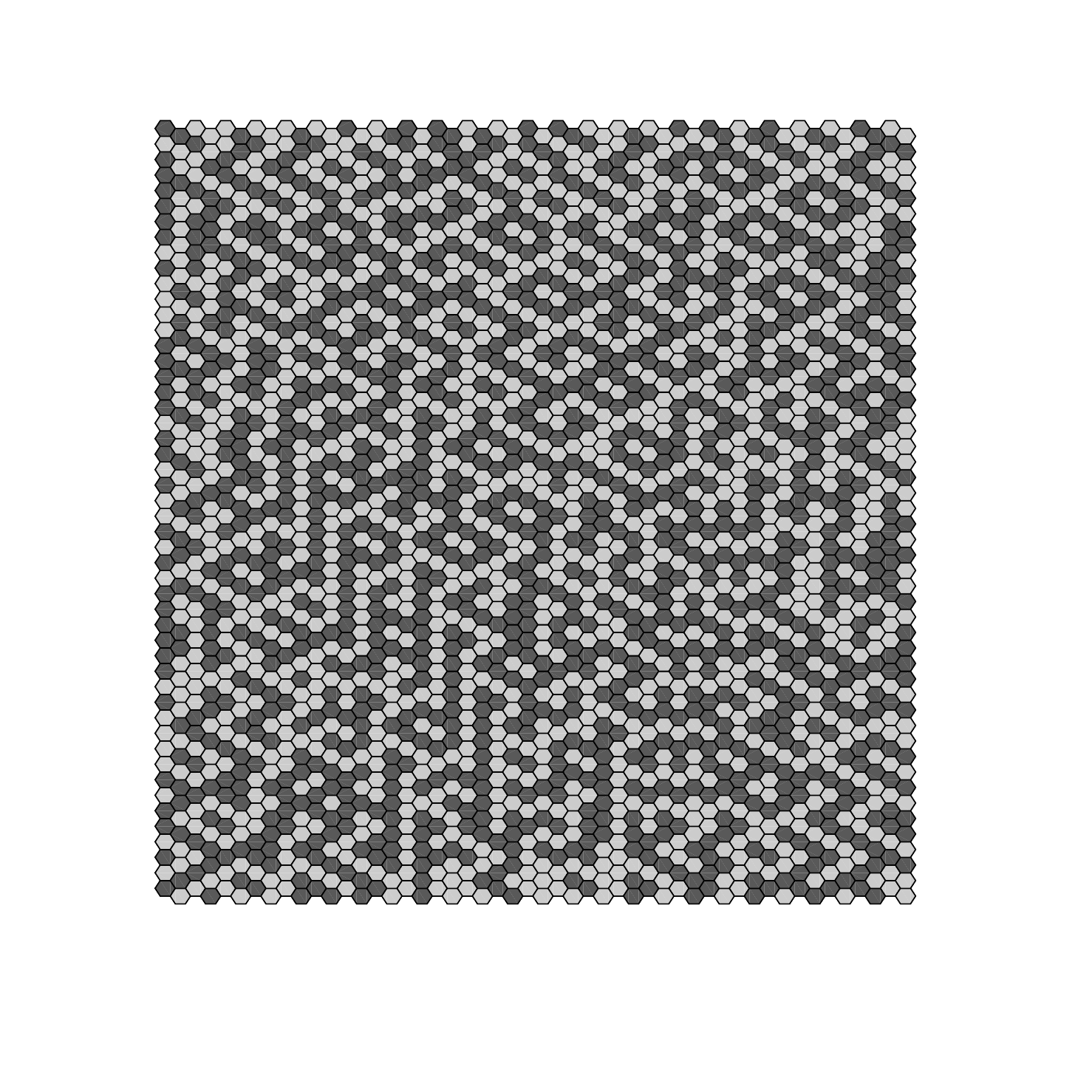}
       \includegraphics[scale=0.6]{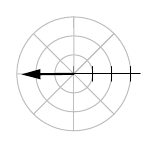}
	\end{minipage}
	\begin{minipage}[t]{0.3\linewidth}
	f)
       \includegraphics[trim=  2.5cm 4.2cm 2.5cm 2.5cm 
,clip,width=0.99\linewidth]{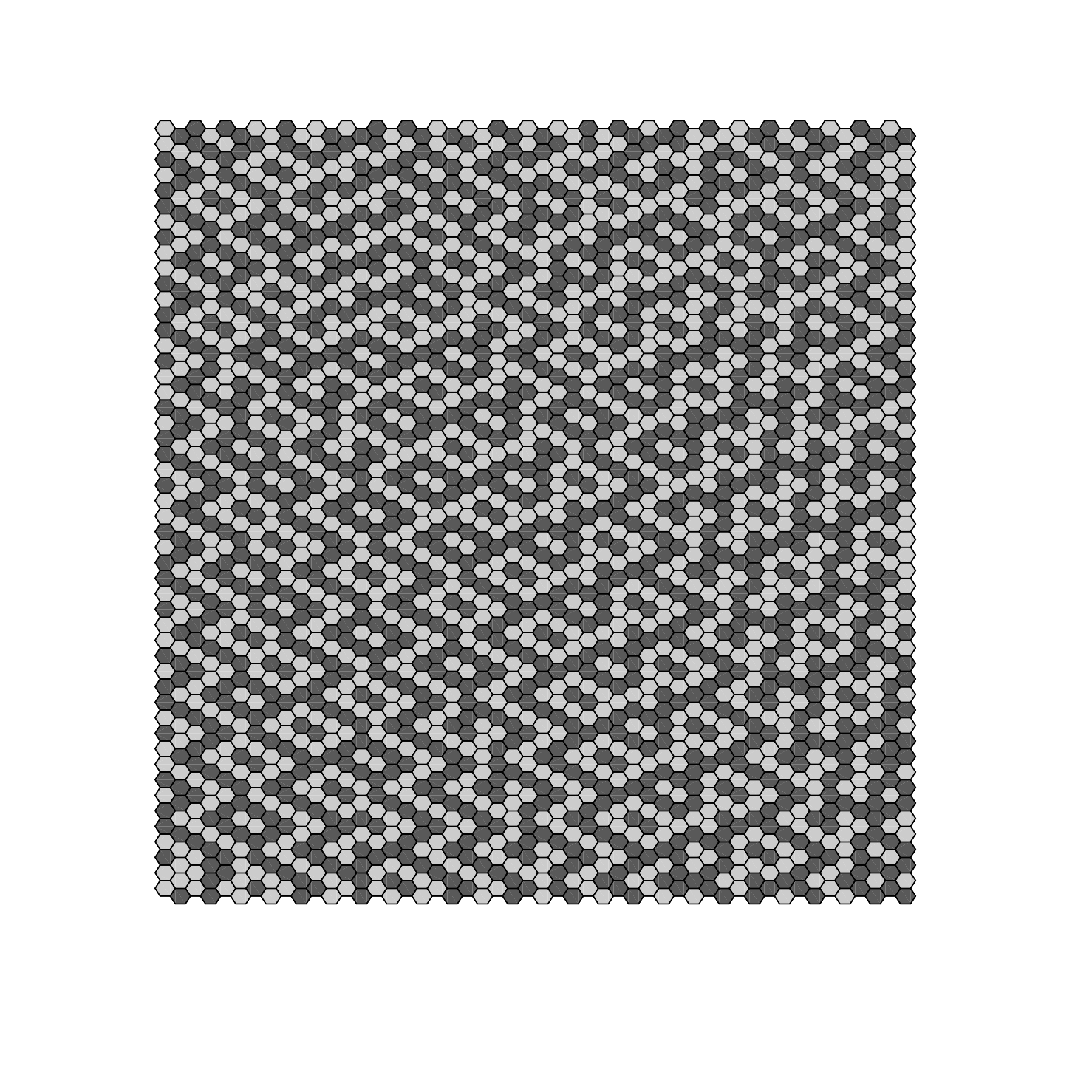}
       \includegraphics[scale=0.6]{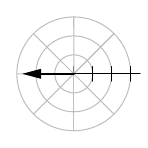}
	\end{minipage}
    \caption{Examples of attracting states reached by applying four different 
update schemes to the BHL+L alignment. 
	    From top left to bottom right: attracting state reached by a) 
SRAU1, 
b) SRAU1, c) RAU1, d) SRAU2, e) DAU, f) DAU.
	    For RAU1 and SRAU2 100 different initial states reached the same 
attracting network state shown in panel c) and d), respectively.
	    Whereas, applying SRAU1 and DAU lead to a diversity of attracting 
network structures, which in contrast to 
	    the perfect regular structures generated by RAU1 and SRAU2, show 
``defects''. 
	    The crosshairs visualize the mucus transport direction and its 
magnitude of the corresponding attractors (the radial tick interval is set to 
0.2 [cells/it.]), i.e. $\vec{v_{g,\infty}}$.}	
   	\label{fig:attr}
\end{figure*}

\begin{figure}[!phbt]
	\centering
    \includegraphics[trim= 0cm 0cm 0cm 
0cm,clip,width=0.9\linewidth]{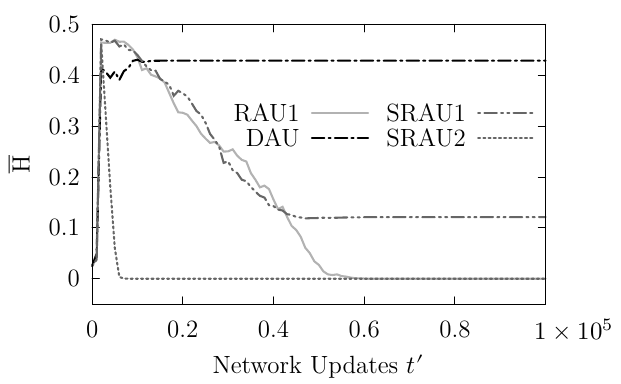}
    \caption{The curves show the average Hamming distance 
between the network state of a reference run 
and four simulation runs for which the initial state 
has been perturbed with respect to the reference run (1-4\% of 
the nodes have been inverted).} 
   	\label{fig:updyn}
\end{figure}

As settings using the locally prescribed four-phase sequence (BHL+L)
can cope with each update scheme, we compared the dynamic behavior produced by 
the different 
update schemes by simulating an ensemble of 100 simulation 
runs using the parameter settings: (50$\times$50,BHL+L, $\ast$, OP, 10\%, 1, 
0\%), where $\ast \in$ $\{$DAU, RAU1, SRAU1, SRAU2$\}$. 
Fig.\ref{fig:detvsrand} shows the transient behavior for the average mucus 
transport speed 
of the four ensembles corresponding to the different update schemes. 
Each setting produces a saturation-like behavior reflecting the capturing of 
the 
dynamics by attractors. 
As one can see, settings using SRAU2 and DAU reach their attractors 
considerably 
faster than settings using 
SRAU1 and RAU1.
In order to characterize the state space corresponding to each setting, we 
counted the 
number of attractors and the attractor periods. It turned out that each 
attractor 
has a period of four, corresponding to the prescribed cyclic four-phase 
sequence. 
Furthermore, SRAU2 and RAU1 drive each of the 100 different initial conditions 
towards the same 
attracting state, which is shown in Fig.\ref{fig:attr}. 
DAU and SRAU1 produce more realistic dynamics, as the 100 simulation runs were 
driven towards 
90 and 95 different attracting states, respectively. 
For DAU and SRAU1 two exemplary attracting states are depicted in 
Fig.\ref{fig:attr}.      
Even if SRAU1 drives different initial conditions towards different attracting 
states, 
the attracting states strongly resemble each other, as indicated in 
Fig.\ref{fig:attr}. 
DAU produces more different emerging structures. 
In order to verify this observation, we ran 5 simulation runs for each update 
scheme, 
which differed by their initial state. 
But this time we prescribed how much their initial state shall differ against a 
reference run. 
Namely, we switched the state for 1\%, 2\%, 3\% and 4\% of all actuators, 
respectively, and compared 
the networks' state at each timestep to the reference run in terms of the 
normalized Hamming distance \footnote{Oscillations 
of 0.5 in the normalized Hamming distance (caused by the XOR-function and the 
BHL+L setting) 
are suppressed by taking the minimum of four consecutive Hamming distances.} 
(see e.g. \cite{Zou2011}) see 
Fig.\ref{fig:updyn}). 
The average Hamming distance corresponding to SRAU2 and RAU1
decreases to zero, which means that all states reach the same attractor. On the 
other hand, 
SRAU1 and DAU produce different attractors, but the attractors 
produced by DAU show an almost maximum Hamming distance.
This behavior reflects the ability of the system to find new attractors 
in case of a perturbation and to conform to changes, which is an important 
characteristic for living systems.    
Consequently, if an attractor which has been reached by applying DAU gets 
perturbed, the 
system conforms to changes and runs into a new attractor.
\subsubsection{Open Boundaries: Marginal Nodes Guiding The Structure Emergence}
\begin{figure*}[!phbt]
\begin{minipage}[t]{0.3\linewidth}
a)
\includegraphics[trim= 2.5cm 2.5cm 2.5cm 
2.5cm,clip,width=0.99\linewidth]{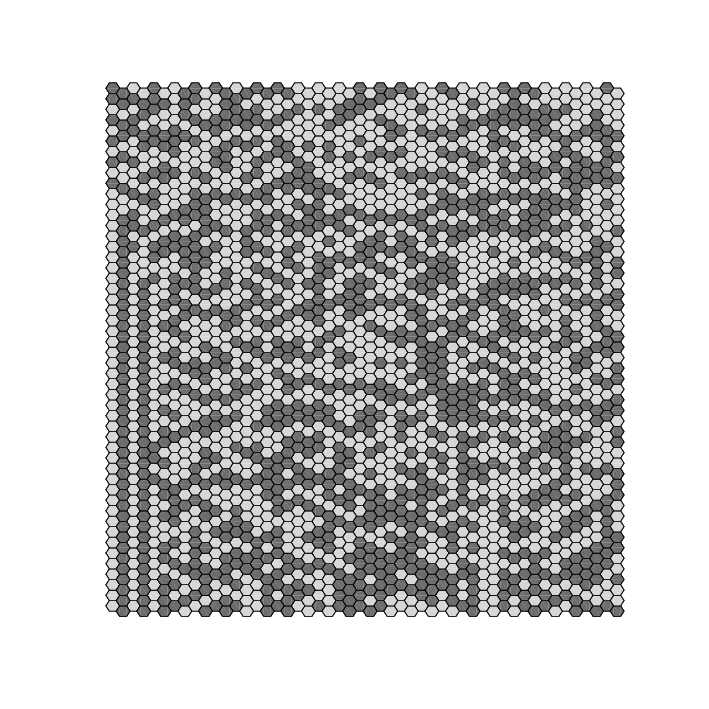}
\end{minipage}
\hfill
\begin{minipage}[t]{0.3\linewidth}
b)
\includegraphics[trim=  2.5cm 2.5cm 2.5cm 
2.5cm,clip,width=0.99\linewidth]{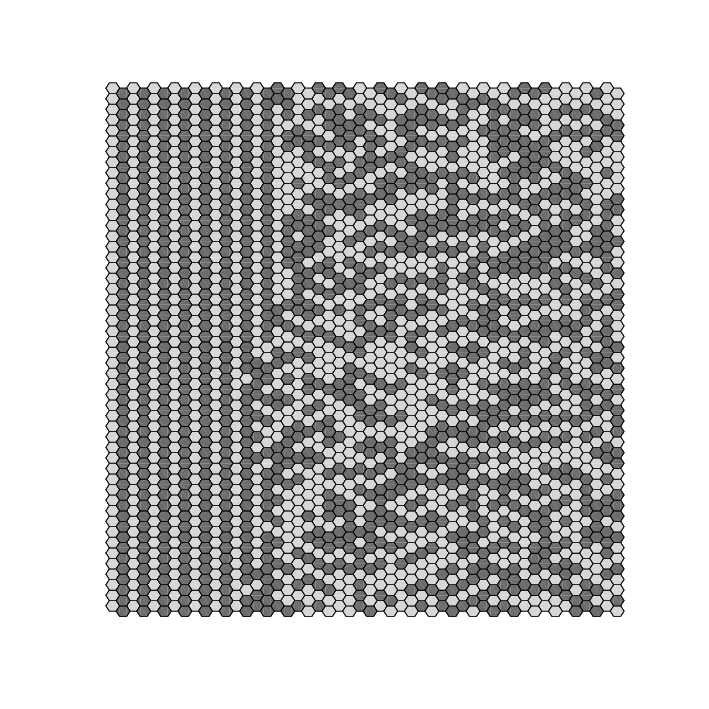}
\end{minipage}
\hfill
\begin{minipage}[t]{0.3\linewidth}
c)
\includegraphics[trim=  2.5cm 2.5cm 2.5cm 
2.5cm,clip,width=0.99\linewidth]{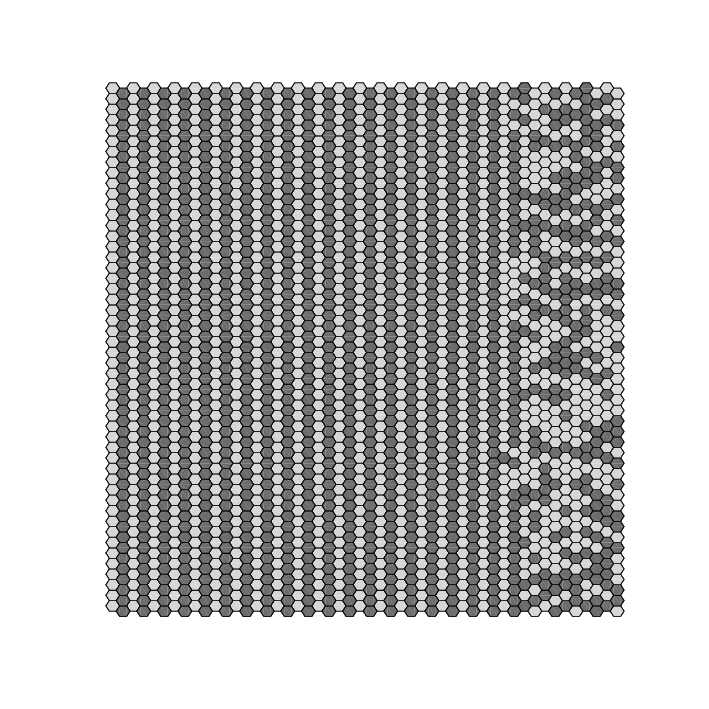}
\end{minipage}
\caption{Three snapshots of the network's state at three different stages of 
the 
self-organization process (time increases from panel a-c).
The network states were generated by applying RAU1 to the BHL+L alignment with 
open boundary conditions. The structure emergence
sets in at the left boundary and spreads to the right over the whole network.}
\label{fig:frombound}
\end{figure*}

First of all we consider a completely dense carpet of ciliated cells 
represented by a parquet of actuators.
In this ``densely ciliated carpet case'' we observe that settings with open 
boundaries 
yield a structure emergence always starting
at the same open boundary, from which it spreads over the whole network. 
This behavior is illustrated in Fig.\ref{fig:frombound}, where three snapshots
of the network state at three different stages of  
the self-organizing process are shown. 
The graphs have been generated by applying open boundaries, RAU1 and BHL+L.  
Further, we observed that settings leading to the ``crystallization-process'', 
shown 
in Fig.\ref{fig:frombound}, only show this self-organizing behavior as long as 
the boundary, 
from which the structure spreads, is open. This means that similar network 
dynamics have been 
observed for open and horizontal cylindrical boundaries. On the other hand, 
if we chose toric or vertical cylindrical boundaries, 
for which the ``structure-triggering'' border is glued to the opposing border, 
the model 
displays a completely different dynamical behavior with much lower 
self-organized transport speeds 
and much less well ordered network states. The only exception is given by the 
settings using the square-lattice 
alignment, which shows self-organized directed transport
under each boundary condition.

\href{run:./anc/movie_S2.mp4}{Movie\_S2},  
\href{run:./anc/movie_S3.mp4}{movie\_S3} and 
\href{run:./anc/movie_S4.mp4}{movie\_S4} (provided as supplemental material 
\cite{supm}) illustrate the structure emergence 
from an open boundary. They have been generated by applying 
(50$\times$50, USL, DAU, OP, 10\%, 0, 0\%),
(50$\times$50, BHL+L, DAU, OP, 15\%, 1, 0\%) and 
(50$\times$50, BHL+L, RAU1, OP, 15\%, 1, 0\%), respectively.  
\subsubsection{Leaders and Followers}
Since we observed for the ``densely ciliated carpet case'' when using open 
boundaries 
that the emergence of highly ordered structures 
always sets in from the same boundary, the actuators 
at the boundary seem to play an important role for the self-organizing process 
and consequently, 
it appears that there exists a kind of hierarchy among the actuators.\par
\begin{figure*}[!phbt]
	\centering
 	\includegraphics[trim= 0cm 0cm 0cm 
0cm,clip,width=0.8\linewidth]{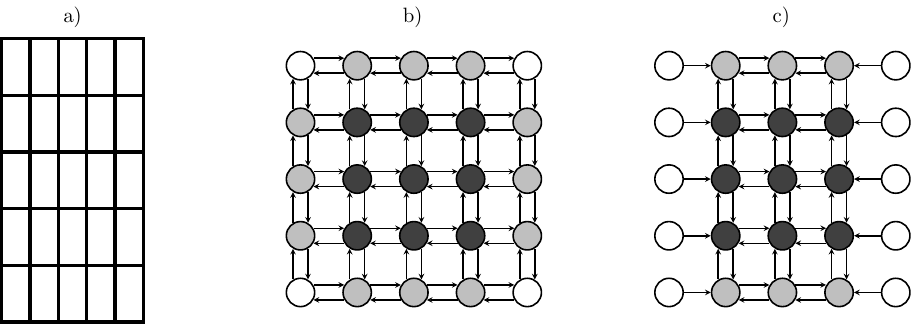}
	\caption{An array of 5$\times$5 actuators aligned in the square-lattice 
is shown in panel a).
	Actuators represent the nodes of the network and local interactions the 
links. Panel b) shows the network 
			topology considering the exchange of mucus particles 
among nodes. Panel c) illustrates 
			the network topology considering the state updates. 
Panel b) and c) correspond to open boundary conditions. 
			In panel c) it can be seen that open boundaries 
introduce ``leading nodes'' at the margin.}
	\label{fig:topo} 
\end{figure*}

 
This hierarchy among the nodes is caused by the 
underlying network topology and is primarily characterized by the distribution 
of in- and out-degrees. 
In Fig.\ref{fig:topo} we illustrate the underlying network topology for 
an array consisting of $5\times5$ cells arranged in a square lattice assuming 
open boundaries. 
The middle panel in Fig.\ref{fig:topo} illustrates the network topology 
considering the possible pathways of the mucus droplets. 
The right panel illustrates the network topology 
considering the possible locked configurations. 
Arrows entering a node indicate which
nodes may block its oscillation. Arrows leaving a node indicate 
for which nodes its state and mucus load is relevant to block the nodes the 
arrows are pointing at. 
The in-degree of a node is the sum 
of incoming arrows and correspondingly, the out-degree is the sum of outgoing 
arrows.
The grayscale in the graphs indicate the value of the in- and out-degrees and 
therewith
the prevalent hierarchy among the nodes.     
If we set open boundaries, 
marginal nodes in our networks are hardly influenced by their adjacent 
nodes. It is especially important that nodes at the left and right boundary 
have 
no in-degree
considering the locked-configurations-rule. Consequently, these nodes 
won't adapt their oscillatory 
motions according to their neighbors' motion and act as ``leading nodes''.\par 
Finally, we observed that some settings need a very regular topology 
with a prevalent hierarchy among the agents. If this ``leadership'' of 
a few nodes is abandoned, which happens when applying toric boundary conditions 
or importing unciliated cells, 
hardly any self-organization takes place. 
\subsubsection{Boundary-Driven Structure Emergence Hindered by Unciliated Cells}
Unciliated cells can only take up and release mucus droplets and are thus,
only passively involved in the state update of adjacent actuators.
This means that cells adjacent to unciliated cells   
obtain a lower in-degree and therefore ascend the hierarchy of actuators.
As we randomly distribute unciliated cells, ``leading nodes'' are no longer 
only 
found 
at the boundaries, but rather are distributed allover the network and 
consequently, 
the network topology becomes less regular.\par
\begin{figure*}[!phbt]
\begin{minipage}[t]{0.3\linewidth}
\centering
a)
\includegraphics[trim= 2.5cm 2.5cm 2.5cm 
1.5cm,clip,width=\linewidth]{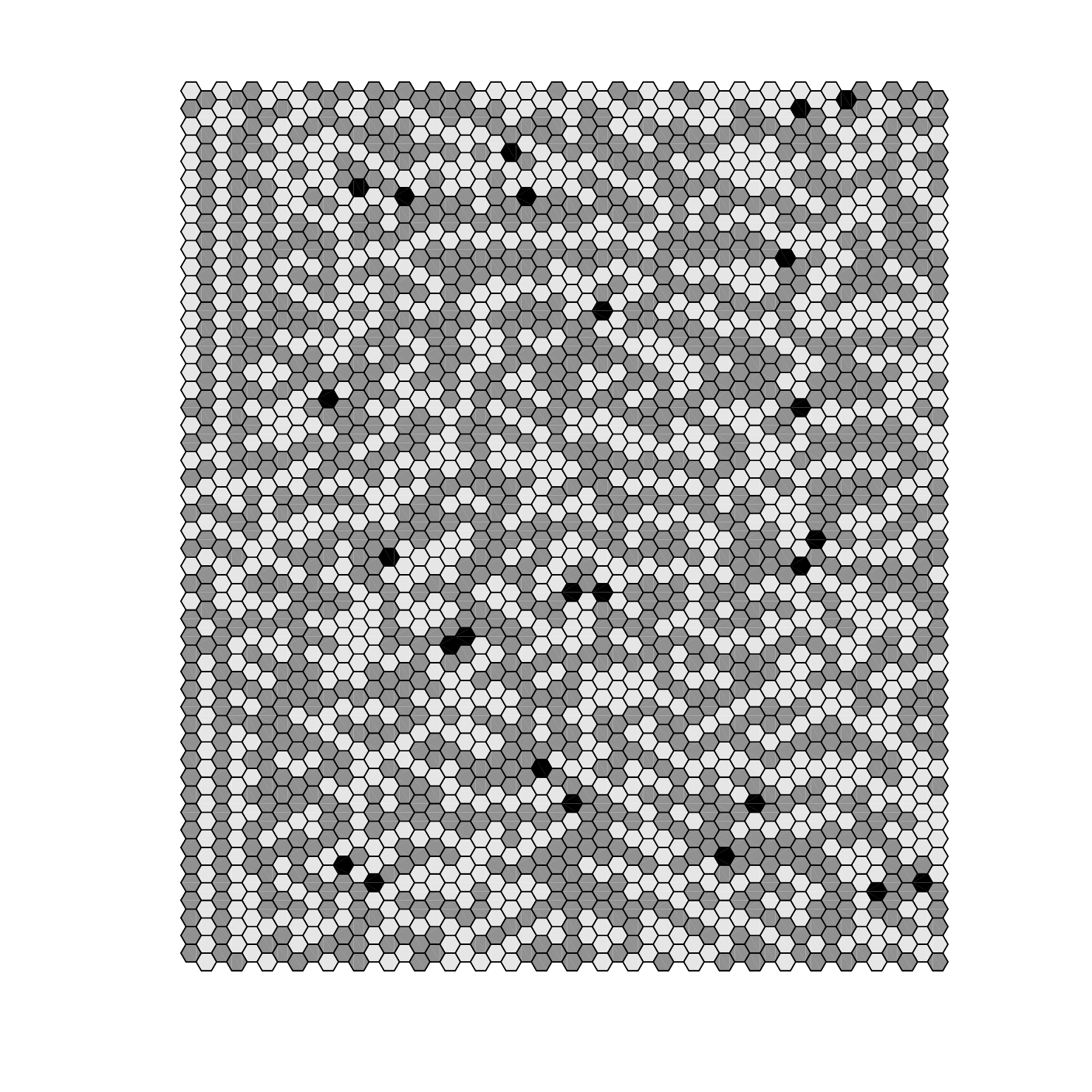}
\end{minipage}
\hfill
\begin{minipage}[t]{0.3\linewidth}
b)
\includegraphics[trim= 2.5cm 2.5cm 2.5cm 
1.5cm,clip,width=\linewidth]{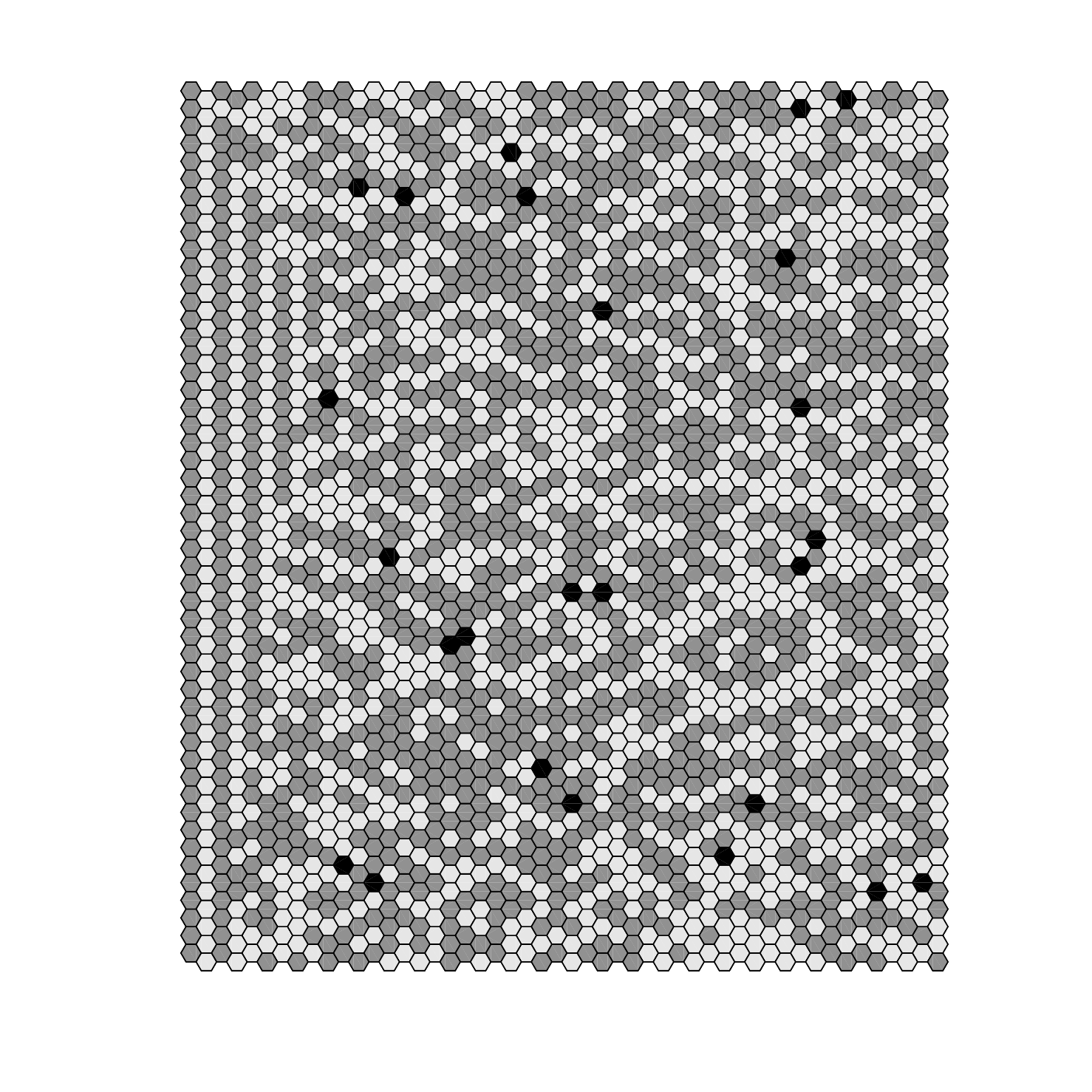}
\end{minipage}
\hfill
\begin{minipage}[t]{0.3\linewidth}
c)
\includegraphics[trim= 2.5cm 2.5cm 2.5cm 
1.5cm,clip,width=\linewidth]{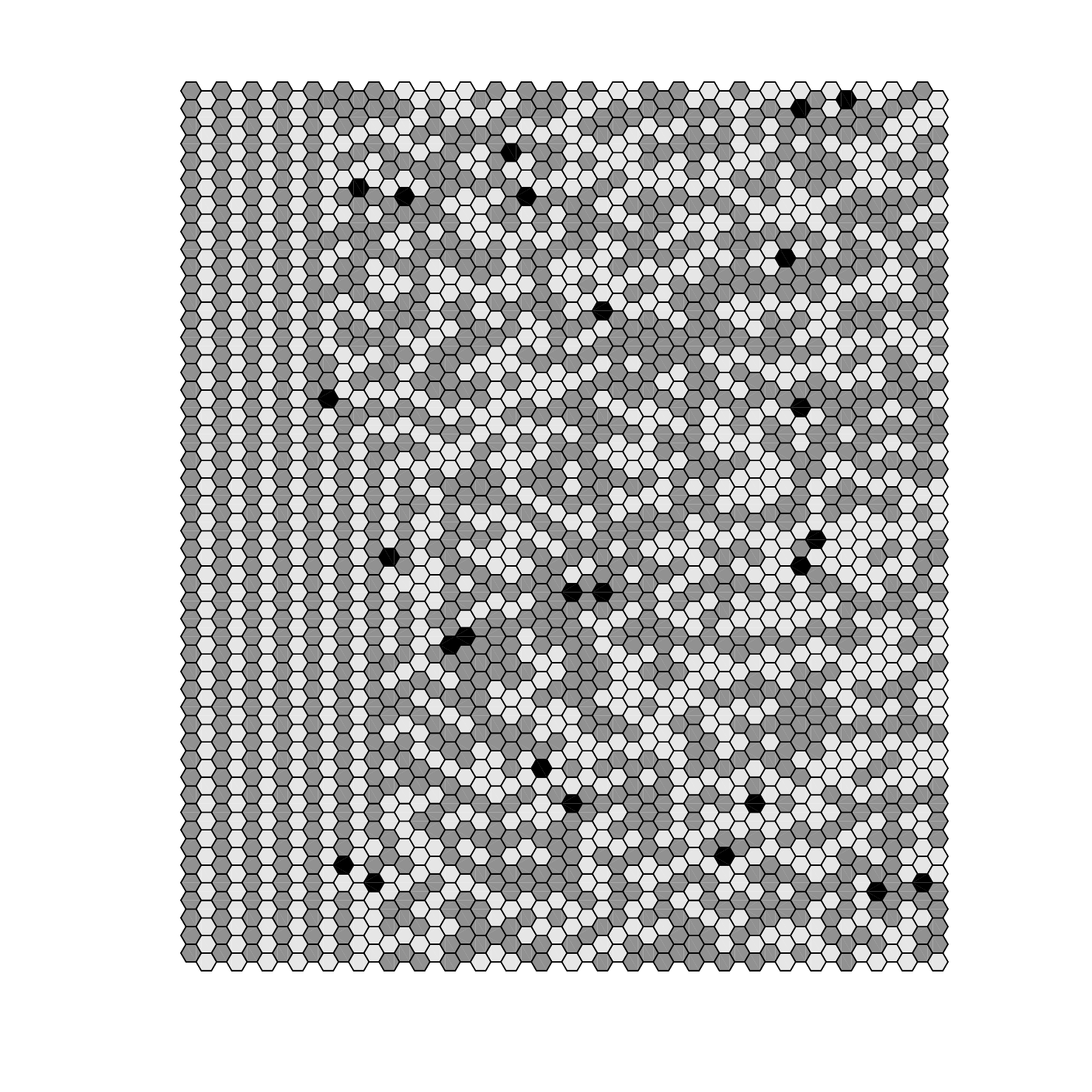}
\end{minipage}
\caption{Three snapshots of the temporal evolution of the model when using the 
BHL+L alignment (time increases from panel a-c), open boundaries, 
		RAU1 and 1$\%$ of randomly distributed unciliated cells. Note 
that the structure emergence is hindered  
		by unciliated cells (which are black colored).}
\label{fig:frombound2}
\end{figure*}

 
It has been observed that the introduction of unciliated cells 
hinder the spreading of the structure emergence for those settings 
for which the structure spreads strictly from an open boundary. 
Consequently, it seems that some settings 
require a very regular topology in order to be able to spread allover the 
network. 
The panels in Fig.\ref{fig:frombound2} show three stages of a simulation for 
which we introduced 1$\%$ of unciliated cells 
and otherwise applied the same 
settings as in Fig.\ref{fig:frombound}. 
Unciliated cells are shown in black. It's clearly visible that the 
structure emergence is hindered by the introduced unciliated cells, as the 
structure can't spread 
further than to the first unciliated cells, seen from left.
According to this simple observation one could imagine that the influence of 
the 
boundaries  
gets the more restricted the more unciliated cells are incorporated. Further 
simulations have confirmed this idea.  
\subsubsection{\label{sec:modself}Modular Self-Organization}
\begin{figure}[!phbt]
\begin{minipage}[t]{0.82\linewidth}
\includegraphics[trim= 0cm 0cm 0cm 0cm,
clip,width=\linewidth]{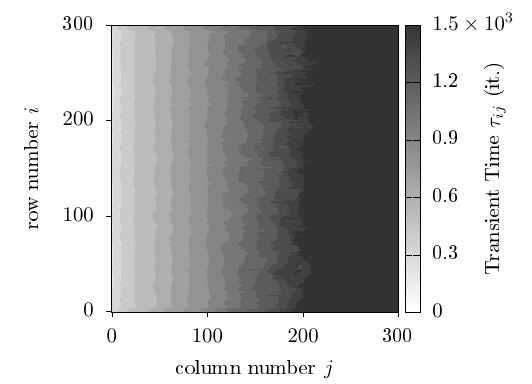}
\end{minipage}
\hfill
\begin{minipage}[t]{0.82\linewidth}
\includegraphics[trim= 0cm 0cm 0cm 0cm,
clip,width=\linewidth]{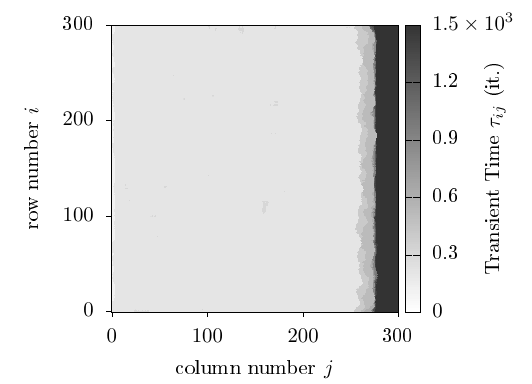}
\end{minipage}
\caption{The grayscale indicates the locally resolved transient time 
[network updates] (note that there are two different grayscales). 
		Both panels correspond to ensemble averages. 
		The upper panel corresponds to runs having 100\% ciliated 
cells, 
while the lower panel corresponds
		to runs with 90\% of ciliated cells. The homogeneity in the 
lower panel 
		indicates that boundary effects may be neglected in the 
network's interior, if unciliated cells are considered.}
\label{fig:grad}
\end{figure}

If we arrange the actuators in the square lattice (USL) and introduce a 
certain amount of unciliated cells, the array of actuators efficiently evolves 
to a self-cleaning epithelium.
In this case the network topology 
considering the state update gets strongly changed, what becomes obvious 
when thinking of the case of sparsely distributed ciliated cells.
One can imagine that a group of ciliated cells would be surrounded by 
unciliated 
cells forming thereby ``ciliated islands'' on an otherwise unciliated 
epithelium.
These islands would be hardly interconnected amongst each others when 
concerning 
the state updates.
How strongly these modules are interconnected depends on the density of 
ciliated 
cells.
In order to illustrate this modular character introduced by unciliated cells, 
the 
topology of the square lattice has been applied together with open boundary 
conditions and the update DAU for an 
ensemble with 100 members differing by their initial state.
The grid size has been set to $300\times300$ cells. 
Fig.\ref{fig:grad} illustrates the temporal evolution of the corresponding 
velocity fields. 
The grayscales illustrate the locally resolved transient time $\tau_{ij}$ 
(in effect 
the number of network updates it takes for an actuator 
to reach a local transport speed, which 
almost amounts to the final area averaged transport speed). 
The upper panel in Fig.\ref{fig:grad} corresponds to simulations ran with 
a dense mat of cells (100\% ciliated cells), while in the lower panel 10\% of 
randomly distributed 
unciliated cells were introduced. 
The panels represent the ensemble average of the spatially resolved transient 
times.  
The darker the color the longer it took 
until a certain actuator synchronized its movement.    
The discussed effect of the ``leading boundary'' is clearly visible in the 
upper 
panel of Fig.\ref{fig:grad}, 
as the gradient in brightness from the left to the right indicates the 
structure 
emergence, which sets in  
at the left boundary and spreads towards the right boundary. 
On the other hand, as soon as one introduces unciliated cells the 
self-organization process gets a modular character, as the 
structure emergence doesn't start at a certain point from which it spreads 
allover the network, 
but spreads at several locations simultaneously, which is illustrated by 
the relatively homogeneous distribution of grayscale in the lower panel in 
Fig.\ref{fig:grad}. 
The influence of the boundary is strongly 
restricted to the marginal nodes at both sides. 
The larger one chooses the grid size, or the more 
unciliated cells are introduced, 
the less important the ``dominance'' of the marginal actuators gets.\par
In conclusion, unciliated cells introduce a certain degree of modularity 
in the network topology as well as in the self-organizing process and 
therefore, the influence of the choice of boundary conditions can be neglected 
in the 
interior of the array, if the array size is chosen large enough and if a 
realistic amount of unciliated cells is considered.\par
\href{run:./anc/movie_S5.mpeg}{Movie\_S5}
(provided as suplemential material \cite{supm}) 
shows the evolution of the network's state of 
a simulation run generated 
with (100$\times$100, USL, DAU, OP, 20\%, 0.5, 10\%)   
exhibiting the typical modular self-organization process leading 
to modular expression patterns. 
  
\subsubsection{\label{sec:transsize}Transient Time vs. Network Size}
In this section we
point out that unciliated cells may strongly influence the 
dynamics of ciliated epithelia. As outlined in the previous section unciliated 
cells 
import topological modularity. Modularity  is an important and 
promising concept, which is inter alia studied in terms of modular random 
boolean networks. 
In \cite{Poblanno-Balp2011} it is claimed that topological modularity reduces 
the probability for 
damage spreading over the network, what promotes robustness.

In the following we show how unciliated cells affect the dynamics of 
our network model. As the square lattice is the only cell arrangement leading 
to 
a properly self-cleaning 
state, when introducing unciliated cells, all the results shown in this section 
refer to 
the square lattice arrangement.\par
\begin{figure}[!phbt]
    \centering
    \includegraphics[trim= 0cm 0cm 0cm 
0cm,clip,width=0.95\linewidth]{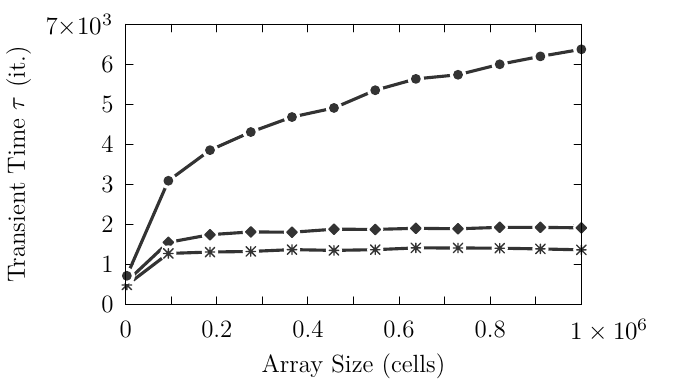}
    \caption{The graph shows the transient time [iterations] versus network 
size 
[cells]. The dots 
			correspond to a totally dense ciliated mat. 
Diamonds and stars 
			correspond to an array containing
			98\% and 90\% ciliated cells, respectively.}
    \label{fig:unctrans}
\end{figure}

The transient time has been determined 
for different array sizes in the range between $50\times50$ cells$^2$ and 
$1000\times1000$ cells$^2$ 
as well as for different fractions of unciliated cells (0\%, 2\% and 10\%). 
The collected data points are presented in Fig.\ref{fig:unctrans}. 
The lines are only guidelines for the eye. The dots correspond
to the dense mat case for which all cells are ciliated.  
Diamonds and stars correspond to 98\% and 90\% ciliated 
cells, respectively.
One can clearly see that the transient time is not only reduced when
introducing unciliated cells, but varies substantially different with 
increasing 
array size. While the transient time continuously grows
for an array representing a totally ciliated mat with increasing array size, 
the transient time runs into saturation with increasing array size if 
unciliated 
cells are present.
As discussed in the former section unciliated cells import not only topological 
modularity, but also cause the self-organization process to be modular, as the
emergence of structure and transport evolves simultaneously in different 
modules. 
This finding most probably explains  
the level off of the transient time with increasing array size 
if unciliated cells are considered. As the self-organization process 
takes place in a decentralized manner, it does not depend upon the size of 
the actuator-array. 
On the other hand, the transient time continues to increase
with increasing array size for an array purely consisting of ciliated cells, 
which is related to the nature of the structure emergence for these settings.  
According to the upper panel in Fig.\ref{fig:grad} the self-organization 
process 
mainly starts at the left boundary, from where it spreads to the right allover 
the network. 
Accordingly, one would expect that the growth of the transient time continues 
with 
increasing network size, as it is the case in Fig.\ref{fig:unctrans}.
(The transient time primarily depends on the length of the simulated 
actuator-array, as the ordered structures expand from the left to the right). 
As the transient time can be seen as a measure for robustness, our results 
suggest that the imported modularity promotes robustness.\par
\begin{figure}[!phbt]
    \centering
    \includegraphics[trim= 0cm 0cm 0cm 
0cm,clip,width=0.85\linewidth]{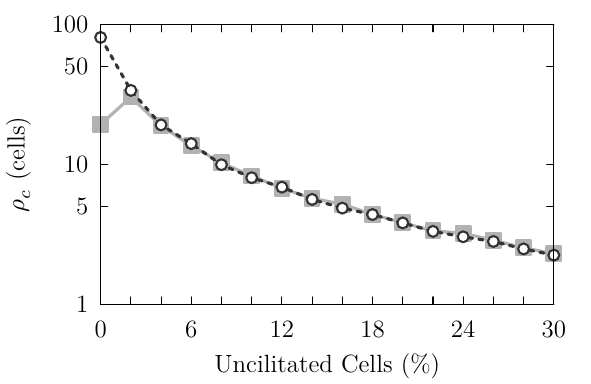}
    \caption{The graph shows the auto-correlation length [cells] vs. the 
relative amount of   
unciliated cells [\%] for different boundary conditions. 
The black circles and bright gray squares correspond to data points derived 
when 
using toric and open boundary conditions, respectively.}
    \label{fig:unccorr}
\end{figure}

Consequently, it seems that the modular topology  
imports modularity into the self-organization process, 
which means that the organization of a huge network gets decomposed into 
a simultaneous organization of submodules. 
These submodules can be recognized by examining the spatial structure of the 
network's state, as 
it has been observed that the modularity of the expression 
patterns clearly depend on the density of ciliated cells.
The auto-correlation length of attracting network states has been determined 
for settings differing by the amount of ciliated cells as well as the boundary 
conditions. 
As we are interested in the spatial structures of the expression patterns, 
the grid size has been chosen to 200$\times$200 cells. 
Fig.\ref{fig:unccorr} shows the auto-correlation length $\rho_c$ (cells) vs. the 
relative 
amount of   
unciliated cells (\%) for open and toric boundary conditions. 
It can be clearly seen that the auto-correlation length, decreases with an 
increasing portion of unciliated cells.
Furthermore, the auto-correlation length is roughly given by the mean distance 
of unciliated cells.  
Consequently, unciliated cells may play a further role in the self-organization 
process 
on the ciliated airway epithelium, as the appearance of the previously 
described 
patch-work character may be strongly influenced by the distribution of 
unciliated and ciliated cells.   
\section{Discussion}
\label{sec:discussion}
The aim of this study is twofold:
On the one hand, we want to make the self-organized spatio-temporally 
coordinated ciliary beat patterns 
as well as the self-organized fluid transport across multiciliated epithelia 
plausible. We suggest that the cooperation among ciliated cells emerges from 
locally interacting oscillating cilia bundles belonging to different ciliated 
cells. 
As our goal was to keep our model as simple as possible, we present a virtual 
self-cleaning epithelium model based 
on symmetrically interacting two-state actuators, 
which we formulate in terms of an adaptive boolean network. 
In the framework of adaptive boolean networks the oscillatory motion of 
ciliated cells can be represented by ``blinking nodes'' and 
discrete mucus droplets establish the local interactions and therefore the 
network's topology.
In Sec.\ref{sec:coevolution} we demonstrate the coevolution of the network's 
state and its topology, which 
is a characteristic property of adaptivity and represents the self-organized 
coevolution 
of ciliary beating patterns and associated fluid transport.\par 
On the other hand, we report our insights to our system's dynamics we gained 
by conducting parameter studies. 
In the following we discuss the observed effects of the update scheme, 
the boundary conditions and the amount of unciliated cells on the dynamics of 
our network model. 

As we formulate our epithelium model in terms of a discretized asynchronous 
multi-agent network, the 
question of how to update the network arises. The introduced update schemes are 
meant to represent different 
possible intercellular signaling mechanisms (membrane potentials and calcium 
waves). 
Only settings using either the deterministic asynchronous update (DAU) 
or the alignment BHL+L, or both, guide the network 
towards 
properly self-cleaning states (Table \ref{tab:tab1}). Settings using BHL+L or 
DAU introduce a local 
recurrent temporal coupling among adjacent actuators inducing an asymmetry of 
the local interactions.
Fig.\ref{fig:v0_1} suggests that the prevalent asymmetry, such as
the initial speed associated with a 
specific parameter setting, 
increases the probability 
for efficient self-organization towards self-cleaning states. 
In any case none of the roughly 11'000 simulations exhibiting 
an initial transport speed 
of less 
than $0.01$ cells/it self-organizes efficiently towards 
a properly self-cleaning state.
The fundamental role of asymmetric interactions on extended 
systems has been discussed previously. 
Asymmetry induced effects on the synchronization process of 
a pair of coupled fields have been reported in 
\cite{Boccaletti2005}, where it has been particularly argued that 
small changes in the asymmetry of the interactions could 
be used as an efficient way to synchronize or desynchronize 
the dynamics, as well as select the main statistical 
properties of the synchronized motion in 
ensembles of interacting units
and consequently, may have relevant 
consequences in natural systems. 
In \cite{Ghorbani2016} the synchronization process of a ciliary chain 
attached to a cylindrical
surface has been investigated. Each cilium is modeled in terms of a small 
sphere moving along an elliptic trajectory. It has been shown that 
an asymmetry in their orbits trigger the emergence of metachronal waves.   
Symmetrical settings have not shown any correlations in their beating patterns, 
what compares well to our results.  
\par 
Furthermore, the application of the deterministic update scheme (DAU) 
seems to generate less well ordered attracting network states. 
This behavior is exemplary illustrated in Fig.\ref{fig:attr} showing a 
perfectly 
ordered attracting state 
for the random asynchronous update (RAU1), slightly less well ordered network 
states for 
the semi-random update schemes (SRAU1 and SRAU2)  
and finally, the least ordered attracting states for the deterministic update 
scheme (DAU). 
All settings using the deterministic update scheme have consistently generated 
patterns showing 
``defects'' perpendicular to the direction of the update scheme and maximum 
correlation along the direction of 
the update. Fig.\ref{fig:autocorr} represents an autocorrelogram of an extreme 
case: maximum correlation is found 
into the direction of the primary update direction (from right to left), which 
is most probably caused by the
local temporal coupling among actuators, while there seems to be almost no 
local 
temporal coupling 
from the top to the bottom, what leads to typically elongated autocorrelograms 
if DAU is applied.\par
Finally, the less strict organization the network generates when using DAU may 
lead to more     
flexible dynamics as in the case of a perturbation the system does not simply 
recover its original 
attracting state, but conforms to the changes by running into a completely 
different attractor (see Fig.\ref{fig:updyn}).    

The effects of different update schemes on the dynamics of multi-agent systems 
are still being investigated.
In \cite{Cornforth2005} six different update schemes have been applied on 
one-dimensional cyclic cellular automata to compare the resulting dynamics. It 
has been concluded
that deterministic update schemes confer a degree of flexibility upon the 
system 
dynamics, what compares well 
to the observed conforming character of our model settings using a 
deterministic 
update scheme.  
Consequently, so far, the findings show evidence that in various 
asynchronous processes leading to self-organization, 
a deterministic update scheme leads to more realistic dynamics (flexibility and 
robustness) and may 
therefore be favored by evolution \cite{Gershenson2003}.  
Recall, however, that ``deterministic'' does not mean that actuators 
would displace the mucus in a pre-determined direction. Rather, the transport 
direction evolves through the interplay between the update asymmetry and 
the largely stochastic interactions.  

Finally, we would like to point out the possible effect of unciliated cells on 
the dynamics on ciliated epithelia. 
The topology of our network model is primarily given by the formulation of the 
local interaction rules, the choice 
of the boundary conditions and the amount of unciliated cells. 
As we have outlined in Sec.\ref{sec:modself} the amount of unciliated cells 
introduces a certain degree of topological modularity. 
The topological modularity in turn causes a modular self-organization, 
which means that the 
self-organization does not start at a specific point or boundary on the grid - 
as it has been observed for completely
dense mats of ciliated cells - but starts in each module simultaneously. 
This modular character of the self-organization leads to the size-independence 
of the transient time, reported in Sec.\ref{sec:transsize}.
Consequently, modularity may provide robustness even to networks as large as 
the 
human ciliated 
airway epithelium consisting of more than $10^9$ cells \cite{Mercer1994}, as 
perturbations quickly fade away.
Furthermore, the modular topology leads to modular expression patterns, the 
size 
of which are roughly given by the 
mean distance of unciliated cells (as shown in Fig.\ref{fig:unccorr}). Finally, 
the 
finding of a modular self-organization caused by the underlying modular 
topology 
providing a highly robust patch-work amongst actuators, provides a  
consistent explanation of the modular expression patterns previously reported 
in 
experimental studies aimed 
at the quantitative description of the modulation wave fields on the tracheal 
epithelium \cite{Ryser2007}.  

As very recently pointed out in \cite{Dey2017}, 
theoretical studies
investigating the collective dynamics of hydrodynamically 
interacting cilia have, so far,
usually considered homogeneous carpets of cilia. 
In \cite{Dey2017} the role of a spatial heterogeneous 
ciliary distribution on coherent ciliary beating using one dimensional arrays 
of cilia represented by rowers has been investigated.
It is particularly shown that the phase coherence of 
random clustered distributions of 
rowers are less sensitive to variations of the number density than
(homogeneously) random distributions of rowers.  
This finding might be seen as another specific dynamical 
phenomenon improving robustness by an underlying modular (network) topology.

We conclude that an intercellular signaling mechanism is probable on ciliated 
epithelia, as 
deterministic update schemes drive the model towards robust self-organized 
states, which 
can still conform to changes. We suggest that the patchy expression patterns 
of the modulation wave field observed on real ciliated epithelia may be the 
result of the 
underlying 
modular topology, which is primarily formed by the distribution of ciliated and 
unciliated cells. 
This patch-work character among ciliated cells may be highly robust due to a 
modular self-organization, 
which prevents perturbations to spread over the whole network. 
Furthermore, the boundary conditions may become irrelevant on epithelia being 
either large enough or having 
a low amount of ciliated cells. 

We close this study by hypothesizing that the modular organization of the 
dynamics on ciliated epithelia 
may be seen as a robust size-independent construction plan of nature, which 
leads to properly self-cleaning 
airways in organisms being as small as new born mice as well as in adult 
giraffes.  
\bibliography{article}

\begin{widetext}
\end{widetext}
\clearpage
\begin{center}
\textbf{\large{Supplemental Material}}
\end{center}
\setcounter{equation}{0}
\setcounter{figure}{0}
\setcounter{table}{0}
\makeatletter
\renewcommand{\theequation}{S\arabic{equation}}
\renewcommand{\thefigure}{S\arabic{figure}}

\section{Classification Attempt}
In the following we shall classify the states reached after $10^5$ iterations
(and their corresponding parameter settings) into four classes:
\begin{itemize}
    \item Class 1 (C1): nontransporting ($|\vec{v}_{g\infty}| < 0.01$)
    disorganized state ($\rho_c < 1.5$)
    \item Class 2 (C2): transporting ($|\vec{v}_{g\infty}| > 0.1$)
    disorganized state ($\rho_c < 1.5$)
    \item Class 3 (C3): nontransporting
    ($|\vec{v}_{g\infty}| < 0.01$) structured state ($\rho_c > 2$)
    \item Class 4 (C4): transporting
    ($|\vec{v}_{g\infty}| > 0.1$)
    structured state ($\rho_c > 2$)
\end{itemize}

\begin{figure}[!phbt]
  \centering
  \includegraphics[trim= 0cm 0cm 0cm 0cm,clip,
    scale=0.68]{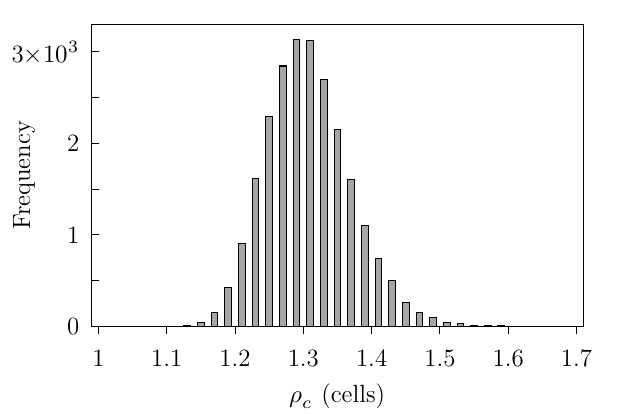}
  \caption{Histogram showing the distribution of the correlation
   lengths of randomly generated initial states.}
  \label{fig:initcorrls}
\end{figure}

Fig.\ref{fig:initcorrls}
shows how the correlation lengths of
the 24'000 randomly generated initial states are distributed
and can be
seen as a justification of the chosen classification into organized
and disorganized attracting states.
\par

\begin{figure*}[!phbt]

    \centering
    \begin{minipage}[t]{0.24\linewidth}
    \centering C1 : $\rho_c$
    \includegraphics[scale=0.9]{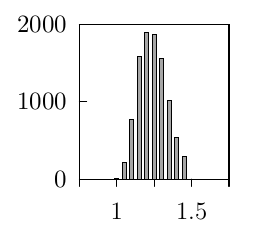}
    \end{minipage}
    \hfill
    \begin{minipage}[t]{0.24\linewidth}
    \centering C2 : $\rho_c$
    \includegraphics[scale=0.9]{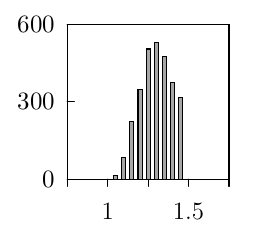}
    \end{minipage}
    \hfill
    \begin{minipage}[t]{0.24\linewidth}
    \centering C3 : $\rho_c$
    \includegraphics[scale=0.9]{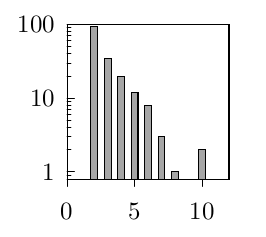}
    \end{minipage}
    \hfill
    \begin{minipage}[t]{0.24\linewidth}
    \centering C4 : $\rho_c$
    \includegraphics[scale=0.9]{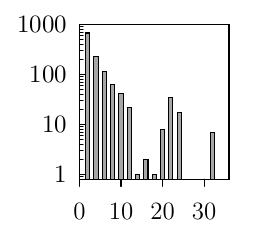}
    \end{minipage}

    \centering
    \begin{minipage}[t]{0.24\linewidth}
    C1 : $\langle\cos\theta\rangle$
    \includegraphics[scale=0.9]{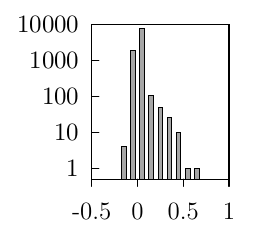}
    \end{minipage}
    \hfill
    \begin{minipage}[t]{0.24\linewidth}
    C2 : $\langle\cos\theta\rangle$
    \includegraphics[scale=0.9]{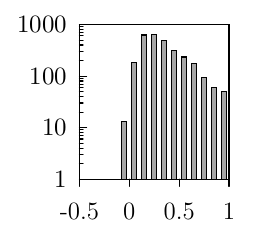}
    \end{minipage}
    \hfill
    \begin{minipage}[t]{0.24\linewidth}
    C3 : $\langle\cos\theta\rangle$
    \includegraphics[scale=0.9]{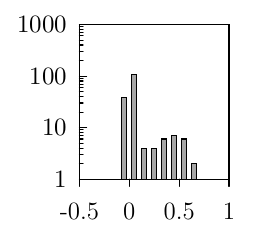}
    \end{minipage}
    \hfill
    \begin{minipage}[t]{0.24\linewidth}
    C4 : $\langle\cos\theta\rangle$
    \includegraphics[scale=0.9]{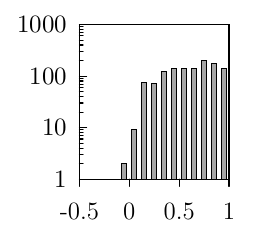}
    \end{minipage}

    \centering
    \begin{minipage}[t]{0.24\linewidth}
    C1 : $m\cdot|\vec{v}_{g\infty}|$
    \includegraphics[scale=0.9]{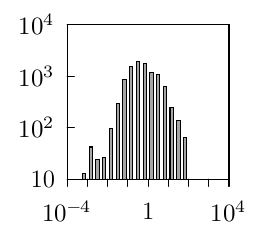}
    \end{minipage}
    \hfill
    \begin{minipage}[t]{0.24\linewidth}
     C2 : $m\cdot|\vec{v}_{g\infty}|$
    \includegraphics[scale=0.9]{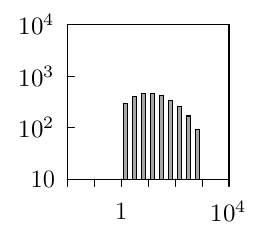}
    \end{minipage}
    \hfill
    \begin{minipage}[t]{0.24\linewidth}
     C3 : $m\cdot|\vec{v}_{g\infty}|$
    \includegraphics[scale=0.9]{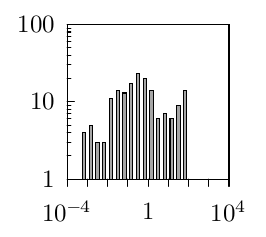}
    \end{minipage}
    \hfill
    \begin{minipage}[t]{0.24\linewidth}
     C4 : $m\cdot|\vec{v}_{g\infty}|$
    \includegraphics[scale=0.9]{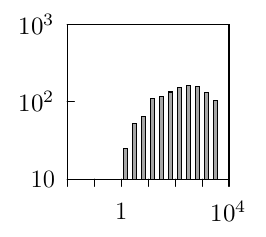}
    \end{minipage}

  	\centering
    \begin{minipage}[t]{0.24\linewidth}
    C1 : $|\vec{v}_{g\infty}|$
    \includegraphics[scale=0.9]{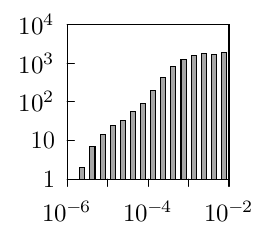}
    \end{minipage}
    \hfill
    \begin{minipage}[t]{0.24\linewidth}
    C2 : $|\vec{v}_{g\infty}|$
    \includegraphics[scale=0.9]{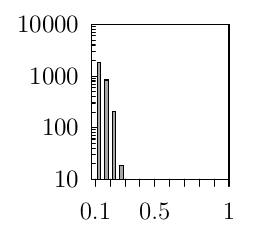}
    \end{minipage}
    \hfill
    \begin{minipage}[t]{0.24\linewidth}
    C3 : $|\vec{v}_{g\infty}|$
    \includegraphics[scale=0.9]{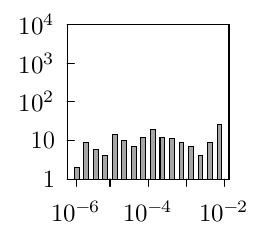}
    \end{minipage}
    \hfill
    \begin{minipage}[t]{0.24\linewidth}
    C4 : $|\vec{v}_{g\infty}|$
    \includegraphics[scale=0.9]{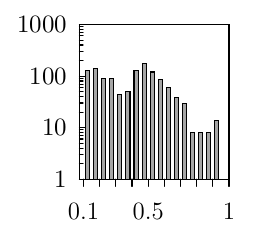}
    \end{minipage}

     \centering
    \begin{minipage}[t]{0.24\linewidth}
    C1 : $|\vec{v}_{g0}|$
    \includegraphics[scale=0.9]{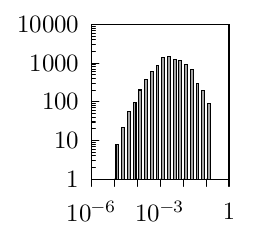}
    \end{minipage}
    \hfill
    \begin{minipage}[t]{0.24\linewidth}
    C2 : $|\vec{v}_{g0}|$
    \includegraphics[scale=0.9]{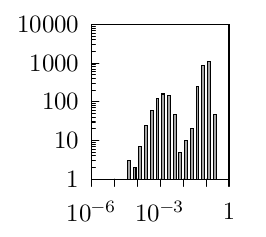}
    \end{minipage}
    \hfill
    \begin{minipage}[t]{0.24\linewidth}
    C3 : $|\vec{v}_{g0}|$
    \includegraphics[scale=0.9]{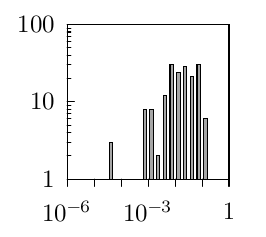}
    \end{minipage}
    \hfill
    \begin{minipage}[t]{0.24\linewidth}
    C4 : $|\vec{v}_{g0}|$
    \includegraphics[scale=0.9]{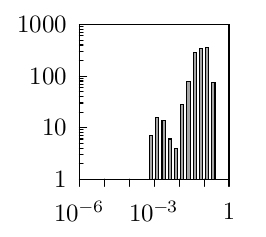}
    \end{minipage}

    \caption{Histograms illustrating the frequency
    distribution of the values for $\rho_c$,
    $\langle\cos\theta\rangle$, $m\cdot|\vec{v}_{g\infty}|$,
    $|\vec{v}_{g\infty}|$ and $|\vec{v}_{g0}|$ (row-wise) in
    in each class (column-wise).
    (Note the heterogeneity of x- and y-scales.)}
    \label{fig:histobs}
\end{figure*}

The histograms in Fig.\ref{fig:histobs} illustrate the distribution of
$\rho_c$-, $\langle\cos\theta\rangle$-, $m|\vec{v}_{g\infty}|$-,
$|\vec{v}_{g\infty}|$- and $|\vec{v}_{g0}|$-values (row-wise) in each
class (column-wise).

\begin{figure*}[!phbt]

    \centering
    C1-C4 : Cell Alignments

    \begin{minipage}[t]{0.24\linewidth}
    \includegraphics[scale=0.8]{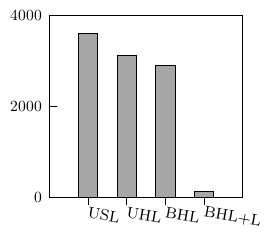}
    \end{minipage}
    \hfill
    \begin{minipage}[t]{0.24\linewidth}
    \includegraphics[scale=0.8]{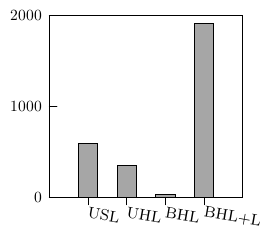}
    \end{minipage}
    \hfill
    \begin{minipage}[t]{0.24\linewidth}
    \includegraphics[scale=0.8]{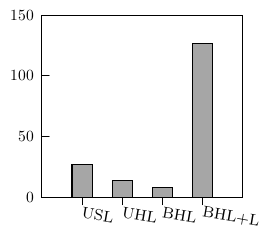}
    \end{minipage}
    \hfill
    \begin{minipage}[t]{0.24\linewidth}
    \includegraphics[scale=0.8]{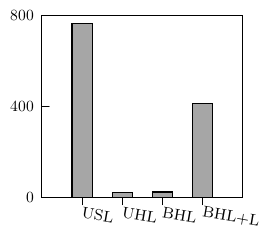}
    \end{minipage}
    \\
    \vspace{0.1cm}
    C1-C4 : Update Schemes

  \begin{minipage}[t]{0.24\linewidth}
    \includegraphics[scale=0.8]{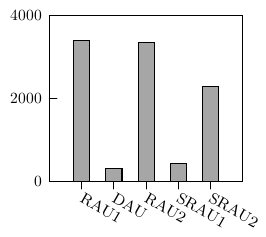}
    \end{minipage}
    \hfill
    \begin{minipage}[t]{0.24\linewidth}
    \includegraphics[scale=0.8]{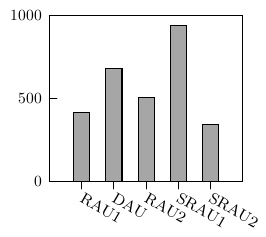}
    \end{minipage}
    \hfill
    \begin{minipage}[t]{0.24\linewidth}
    \includegraphics[scale=0.8]{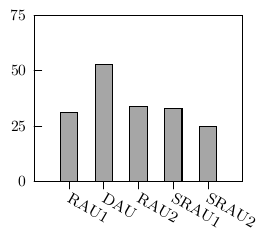}
    \end{minipage}
    \hfill
    \begin{minipage}[t]{0.24\linewidth}
    \includegraphics[scale=0.8]{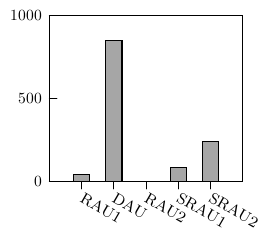}
    \end{minipage}
    \\

    \vspace{0.1cm}
    C1-C4 : Boundary Conditions

    \begin{minipage}[t]{0.24\linewidth}
    \includegraphics[scale=0.8]{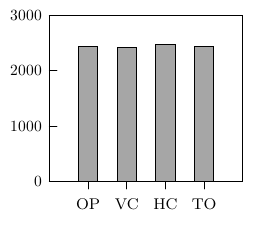}
    \end{minipage}
    \hfill
    \begin{minipage}[t]{0.24\linewidth}
    \includegraphics[scale=0.8]{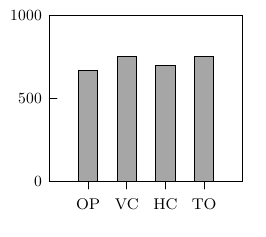}
    \end{minipage}
    \hfill
    \begin{minipage}[t]{0.24\linewidth}
    \includegraphics[scale=0.8]{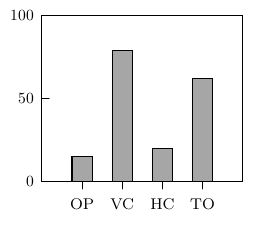}
    \end{minipage}
    \hfill
    \begin{minipage}[t]{0.24\linewidth}
    \includegraphics[scale=0.8]{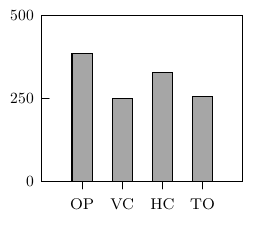}
    \end{minipage}

    C1-C4 : Mucus Amount (\%)

    \begin{minipage}[t]{0.24\linewidth}
    \includegraphics[scale=0.8]{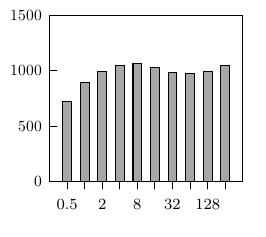}
    \end{minipage}
    \hfill
    \begin{minipage}[t]{0.24\linewidth}
    \includegraphics[scale=0.8]{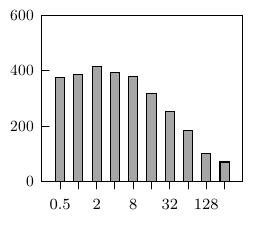}
    \end{minipage}
    \hfill
    \begin{minipage}[t]{0.24\linewidth}
    \includegraphics[scale=0.8]{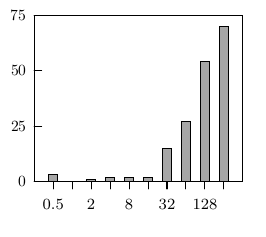}
    \end{minipage}
    \hfill
    \begin{minipage}[t]{0.24\linewidth}
    \includegraphics[scale=0.8]{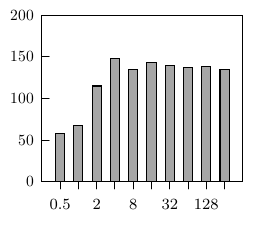}
    \end{minipage}

    C1-C4 : Energy Paramater $\epsilon$

    \begin{minipage}[t]{0.24\linewidth}
    \includegraphics[scale=0.8]{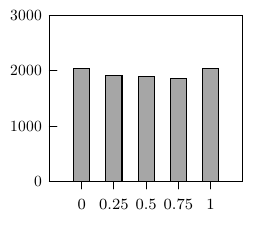}
    \end{minipage}
    \hfill
    \begin{minipage}[t]{0.24\linewidth}
    \includegraphics[scale=0.8]{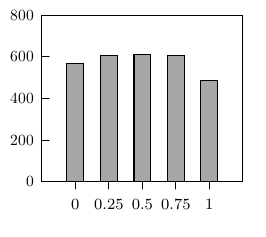}
    \end{minipage}
    \hfill
    \begin{minipage}[t]{0.24\linewidth}
    \includegraphics[scale=0.8]{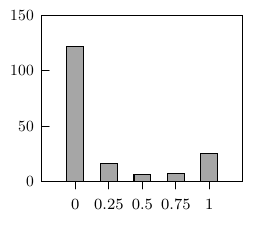}
    \end{minipage}
    \hfill
    \begin{minipage}[t]{0.24\linewidth}
    \includegraphics[scale=0.8]{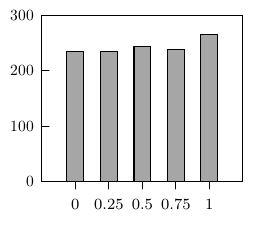}
    \end{minipage}

    C1-C4 : Unciliated Proportion (\%)

    \begin{minipage}[t]{0.24\linewidth}
    \includegraphics[scale=0.8]{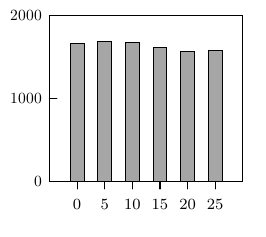}
    \end{minipage}
    \hfill
    \begin{minipage}[t]{0.24\linewidth}
    \includegraphics[scale=0.8]{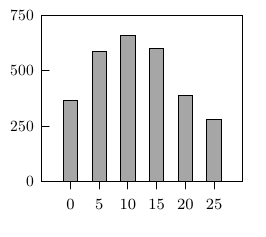}
    \end{minipage}
    \hfill
    \begin{minipage}[t]{0.24\linewidth}
    \includegraphics[scale=0.8]{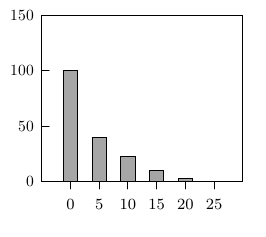}
    \end{minipage}
    \hfill
    \begin{minipage}[t]{0.24\linewidth}
    \includegraphics[scale=0.8]{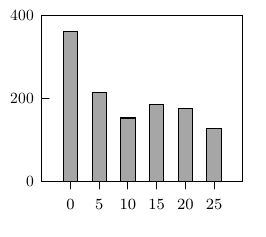}
    \end{minipage}

    \caption{The histograms are meant to illustrate the distribution
    of the respectively applied cell alignment, update scheme,
    boundary condition, mucus amount, $\epsilon$-value and
    the proportion of unciliated cells (row-wise) in each class
    (column-wise).}
    \label{fig:h2}
\end{figure*}

Accordingly, the histograms in Fig.\ref{fig:h2} illustrate
how the parameter values
are distributed in each class.

In the following we list the most striking conspicuities of each class
and explain some of the mechanisms accounting for the classification
of a certain parameter set.\\
\begin{center}
\textbf{Class1 : nontransporting disorganized states}
\end{center}
The very low $\langle\cos\theta\rangle$-values indicate
chaotic velocity fields. \par
Settings using BHL+L are scarcely found
in this class, which is due to their high initial transport speeds
generated by the pre-coordinated cyclic four phase motion.\par
Two thirds of the settings in this class are made up by settings using
either RAU1 or RAU2, which is most probably because these update schemes
do not impose any asymmetries among the local interactions.
\begin{center}
\textbf{Class2 : transporting disorganized states}
\end{center}
In some cases the states reach high $\langle\cos\theta\rangle$-values.
This means that organized velocity fields are possible even if the
states are disorganized.\par
Two thirds of the states in this class are generated by BHL+L, which
is due to their high initial
speeds ($|\vec{v}_{g0}| > 0.1$ cells/it.).\par
The terminal transport speeds in this class are higher than 0.1 cells/it.
and less than 0.3 cells/it. \par
This class is interesting because one might wonder if
it is possible that transport can develop without associated
structure emergence. A detailed look into the simulations
showing a considerable increase of transport speed,
has shown that spatial correlation always
slightly increases in tandem
during the course of a simulation.
\begin{center}
\textbf{Class3 : nontransporting organized states}
\end{center}
This class is characterized by very small terminal
transport speeds
as well as disorganized velocity fields
(low $\langle\cos\theta\rangle$-values), but nevertheless
highly organized states (two states even reach a correlation
length of $\approx$10 cells).\par
The initial transport speeds are remarkably high.\par
Three out of four states in this class are generated by settings
using BHL+L. \par
Further, high mucus amounts as well as low $\epsilon$-values
seem to constitute the majority of states/settings.

Visual examination of the simulations has shown that BHL+L settings
can generate ordered states before the actuators' motion gets
completely blocked by an overload of mucus.
This usually can happen with moderate mucus amounts and
$\epsilon < 1$, or less frequently, with
 high mucus amounts when $\epsilon$ is set to 1.

Settings using USL and $\epsilon = 1$ can generate structures.
However, during the course of a simulation the mucus particles
accumulate until reaching an unnatural state, which shifts
piles of mucus back and forth resulting in no net movement.

Settings using UHL reach fully blocked states as well as structured states
exhibiting velocity fields which are divided into two parts (similar to the
example shown in Fig.14 in our manuscript) and whose globally averaged
transport speeds vanish.
\begin{center}
\textbf{Class4 : transporting organized states}
\end{center}
In this class we observe the highest values for each observable.
The majority of parameter sets either involves USL or BHL+L
together with the deterministic update DAU. \par
Further analysis of this class has shown that
the direction of maximum correlation of roughly 90\% of the attractor
states coincides with the transport direction.\par
This class still contains many states exhibiting strange velocity
fields, which either appear disordered or bisected.
Especially bisected states can nevertheless reach
considerably high globally averaged transport speeds,
while their velocity fields are characterized by
low $\langle\cos\theta\rangle$-values.

\section{Superior States/Settings}

\begin{figure*}[!phbt]

    \hspace{-1.1cm}
    \begin{minipage}[t]{0.25\linewidth}
    \includegraphics[height=5cm]{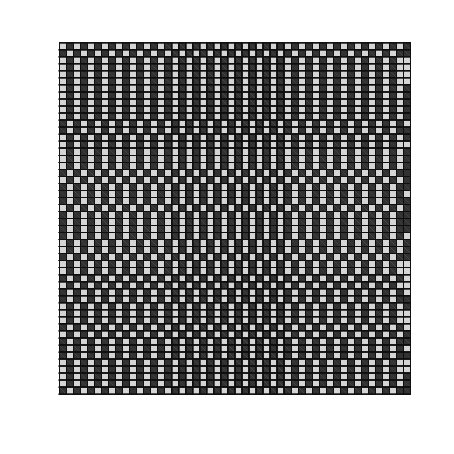}
    \end{minipage}
    \begin{minipage}[t]{0.28\linewidth}
    \includegraphics[height=5cm]{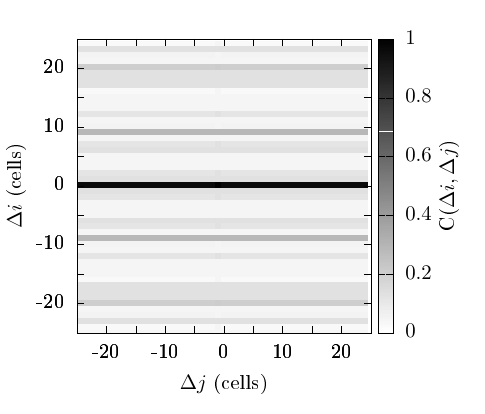}
    \end{minipage}
    \begin{minipage}[t]{0.34\linewidth}
    \includegraphics[height=5cm]{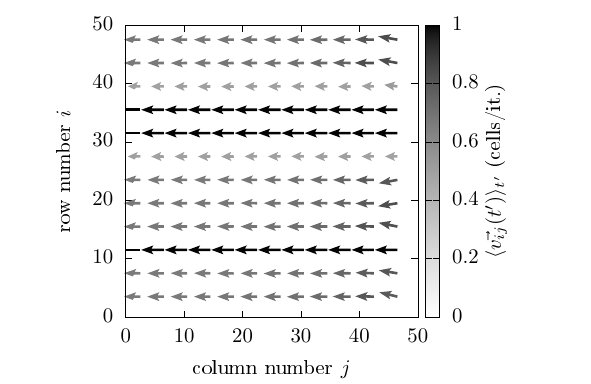}
    \end{minipage}

    \hspace{-1.1cm}
    \begin{minipage}[t]{0.25\linewidth}
    \includegraphics[height=5cm]{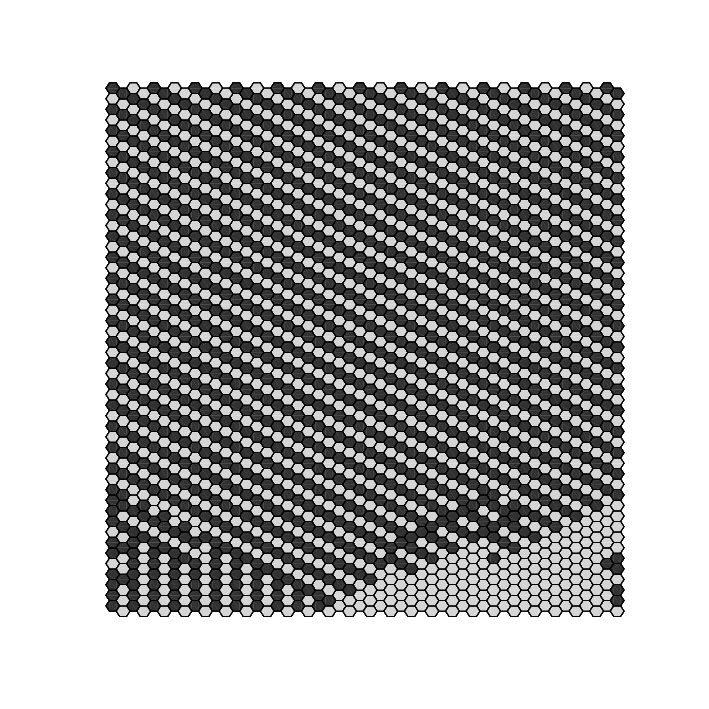}
    \end{minipage}
    \begin{minipage}[t]{0.28\linewidth}
    \includegraphics[height=5cm]{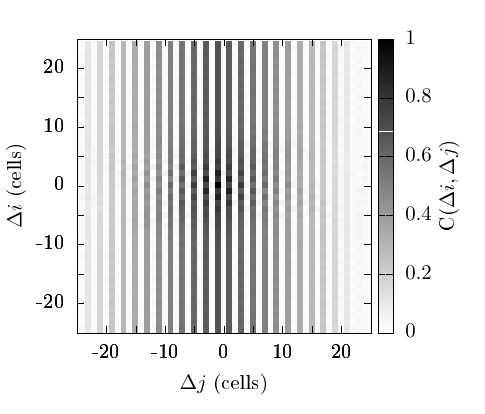}
    \end{minipage}
    \begin{minipage}[t]{0.34\linewidth}
    \includegraphics[height=5cm]{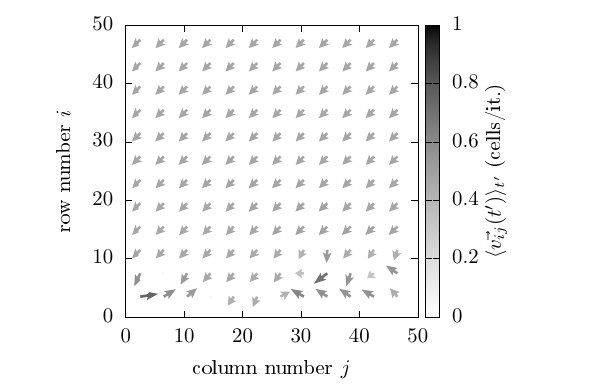}
    \end{minipage}

    \hspace{-1.1cm}
    \begin{minipage}[t]{0.25\linewidth}
    \includegraphics[height=5cm]{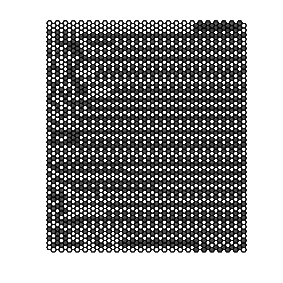}
    \end{minipage}
    \begin{minipage}[t]{0.28\linewidth}
    \includegraphics[height=5cm]{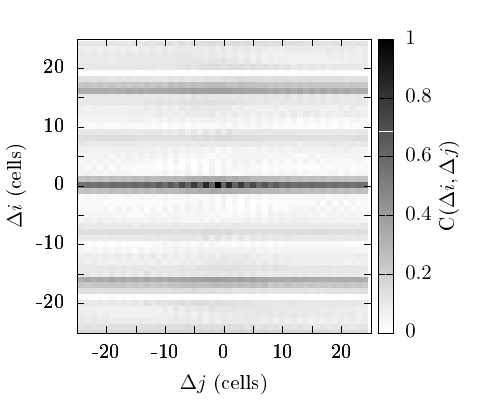}
    \end{minipage}
    \begin{minipage}[t]{0.34\linewidth}
    \includegraphics[height=5cm]{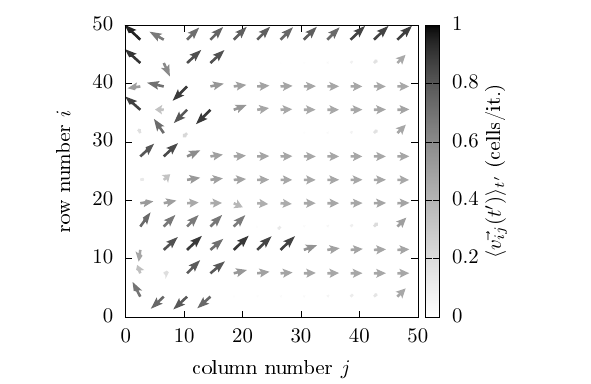}
    \end{minipage}

    \hspace{-1.1cm}
    \begin{minipage}[t]{0.25\linewidth}
    \includegraphics[height=5cm]{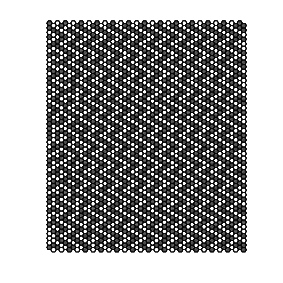}
    \end{minipage}
    \begin{minipage}[t]{0.28\linewidth}
    \includegraphics[height=5cm]{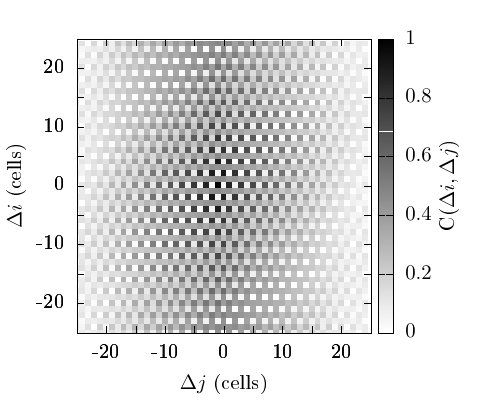}
    \end{minipage}
    \begin{minipage}[t]{0.34\linewidth}
    \includegraphics[height=5cm]{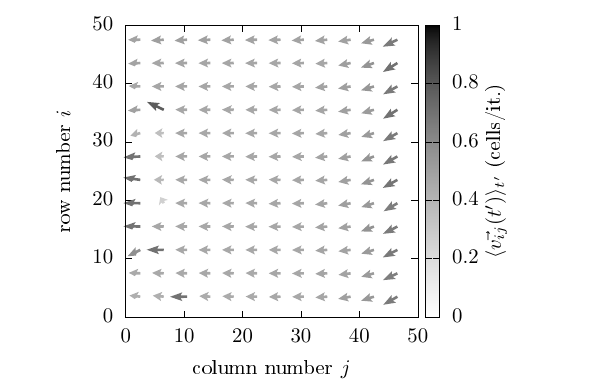}
    \end{minipage}

    \caption{Each row shows the best transporting
    attractor states with their
    corresponding
    autocorrelograms and velocity fields for
    USL, UHL, BHL and BHL+L (row-wise
    from top to bottom).}
    \label{fig:bestt}
\end{figure*}

\begin{figure*}[!phbt]

    \hspace{-1.1cm}
    \begin{minipage}[t]{0.25\linewidth}
    \includegraphics[height=5cm]{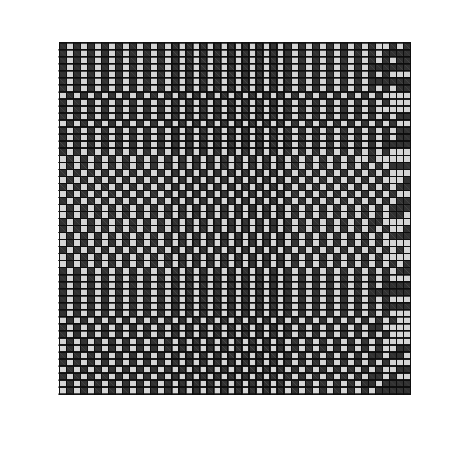}
    \end{minipage}
    \begin{minipage}[t]{0.28\linewidth}
    \includegraphics[height=5cm]{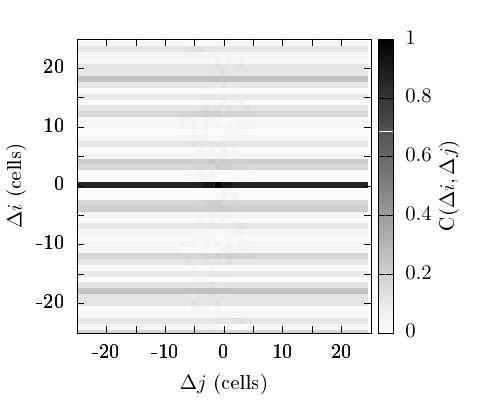}
    \end{minipage}
    \begin{minipage}[t]{0.34\linewidth}
    \includegraphics[height=5cm]{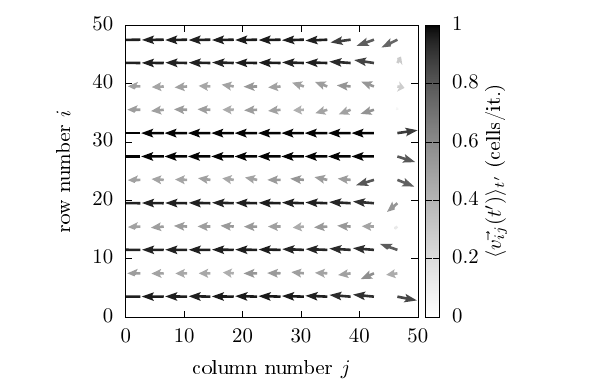}
    \end{minipage}

    \hspace{-1.1cm}
    \begin{minipage}[t]{0.25\linewidth}  
    \includegraphics[height=5cm]{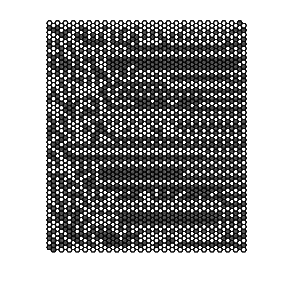}
    \end{minipage}
    \begin{minipage}[t]{0.28\linewidth} 
    \includegraphics[height=5cm]{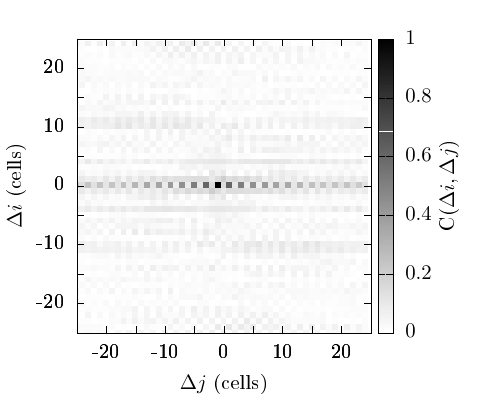}
    \end{minipage}
    \begin{minipage}[t]{0.34\linewidth} 
    \includegraphics[height=5cm]{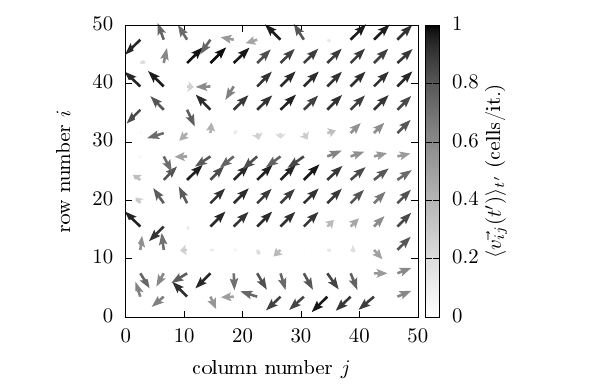}
    \end{minipage}
    
    \hspace{-1.1cm}    
    \begin{minipage}[t]{0.25\linewidth}  
    \includegraphics[height=5cm]{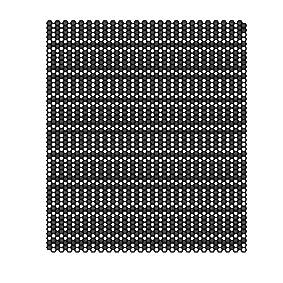}
    \end{minipage}
    \begin{minipage}[t]{0.28\linewidth} 
    \includegraphics[height=5cm]{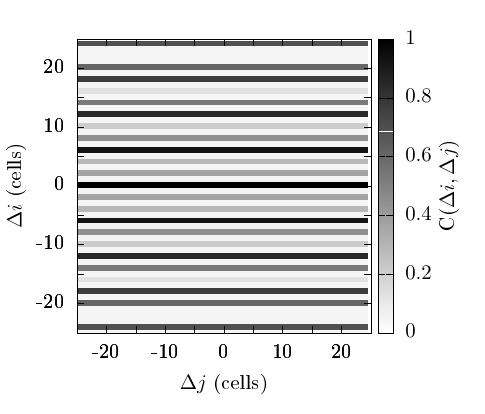}
    \end{minipage}
    \begin{minipage}[t]{0.34\linewidth} 
    \includegraphics[height=5cm]{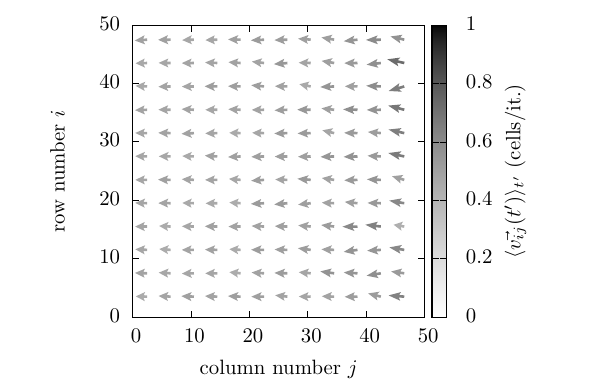}
    \end{minipage}

	\caption{Attractor states and their corresponding 
	auto-correlograms and velocity fields exhibiting 
	the fastest transport speeds 
	for settings using USL, BHL and BHL+L (row-wise).}
	\label{fig:maxvinf}
\end{figure*}

\begin{figure*}[!phbt]

    \hspace{-1.1cm}
    \begin{minipage}[t]{0.25\linewidth}  
    \includegraphics[height=5cm]{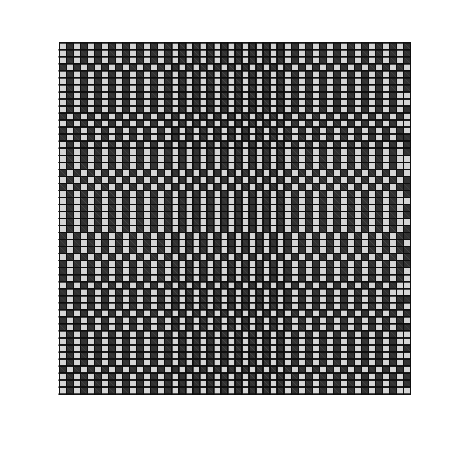}
    \end{minipage}
    \begin{minipage}[t]{0.28\linewidth} 
    \includegraphics[height=5cm]{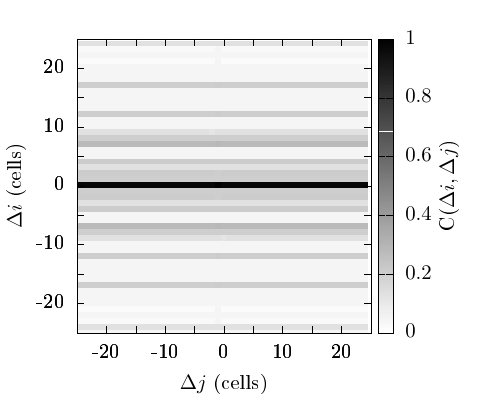}
    \end{minipage}
    \begin{minipage}[t]{0.34\linewidth} 
    \includegraphics[height=5cm]{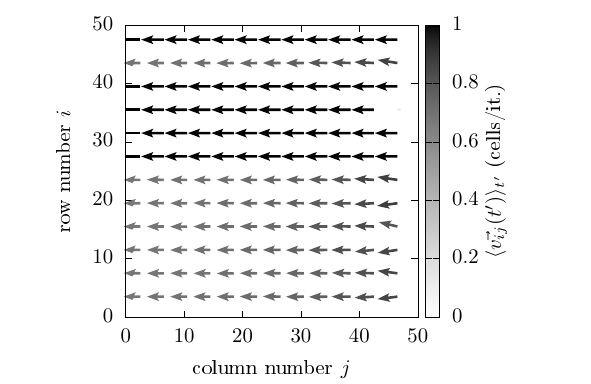}
    \end{minipage}

	\hspace{-1.1cm}
    \begin{minipage}[t]{0.25\linewidth}  
    \includegraphics[height=5cm]{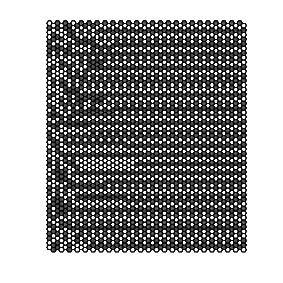}
    \end{minipage}
    \begin{minipage}[t]{0.28\linewidth} 
    \includegraphics[height=5cm]{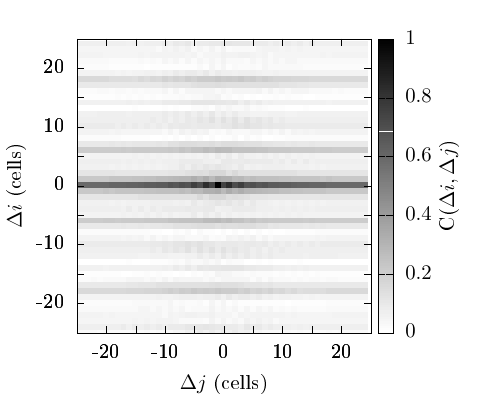}
    \end{minipage}
    \begin{minipage}[t]{0.34\linewidth} 
    \includegraphics[height=5cm]{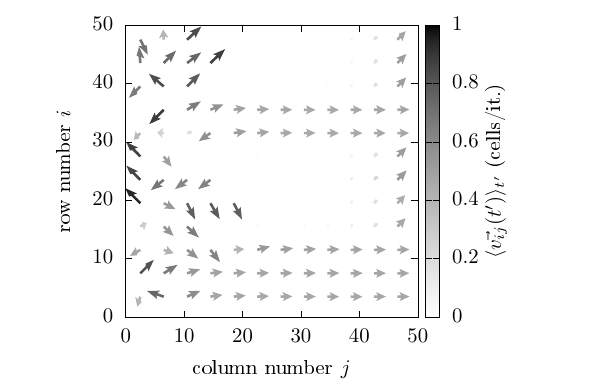}
    \end{minipage}

	\hspace{-1.1cm}
    \begin{minipage}[t]{0.25\linewidth}  
    \includegraphics[height=5cm]{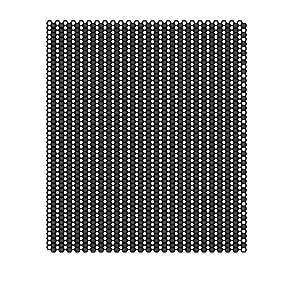}
    \end{minipage}
    \begin{minipage}[t]{0.28\linewidth} 
    \includegraphics[height=5cm]{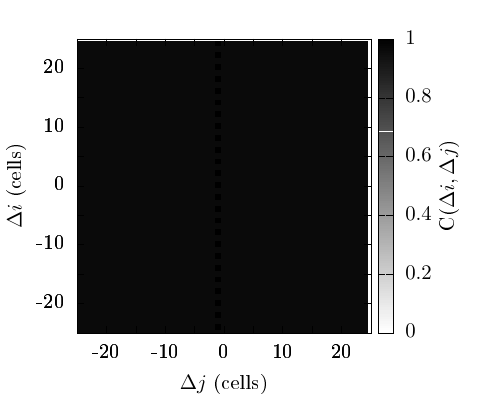}
    \end{minipage}
    \begin{minipage}[t]{0.34\linewidth} 
    \includegraphics[height=5cm]{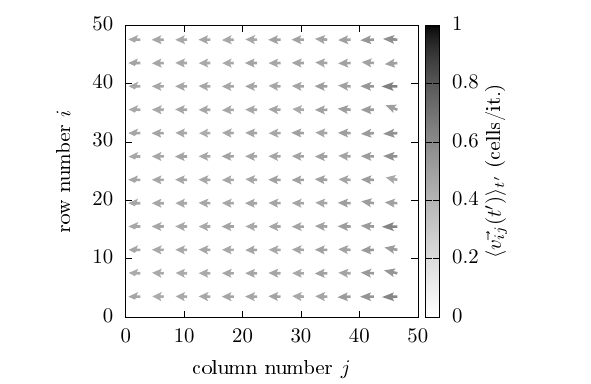}
    \end{minipage}

	\caption{From top to bottom: 
	snapshot of the most ordered attractor state 
	together with its corresponding 
	auto-correlogram and velocity field for settings reached 
	by USL, BHL, and BHL+L (row-wise).}
	\label{fig:maxcorr}
\end{figure*}

In Fig.\ref{fig:bestt} we present the state, its corresponding
autocorrelogram and its associated
(temporally and spatially averaged)
velocity field of the state exhibiting maximum transport
(maximum $m\cdot|\vec{v_{g\infty}}|$) for each cell
alignment (row-wise).
Correspondingly, Fig.\ref{fig:maxvinf} and Fig.\ref{fig:maxcorr}
show the state, its autocorrelogram and its velocity field
of the settings
exhibiting maximum transport speeds and displaying the highest
correlation lengths, respectively.
Table \ref{tab:t1} lists the applied parameter sets
and Table \ref{tab:t2} the associated observable values.
\begin{table*}[hbt]
\caption{\label{tab:t1} Parameter settings leading to
highest transport rates, highest transport speeds and most structured
states for each cell alignment.}
\begin{tabular}{l l l l l l}
\hline
 & \qquad Update\qquad\qquad & Boundary \qquad\qquad & Mucus (\%)
 \qquad\qquad & Energy $\epsilon$ \qquad\qquad  &
 Unciliated (\%) \qquad\qquad  \\

\hline

                   & \qquad DAU  & HC  & 256 & 0.25 & 0 \\
USL \qquad\qquad   & \qquad DAU  & HC   & 4   & 0.25 & 0 \\
                &  \qquad DAU   & HC    & 128 & 0.75 & 0 \\

    \hline

UHL  & \qquad DAU & VC & 256 & 1.0 & 0 \\

    \hline

    & \qquad DAU & OP & 256 & 1.0 & 0 \\
BHL & \qquad DAU & OP & 32   & 1.0 & 0 \\
    & \qquad DAU & HC & 256 & 1.0 & 0 \\

    \hline

        & \qquad DAU    &   OP  & 256   & 1.0   &   0 \\
BHL+L   & \qquad SRAU2  &   HC  & 1     & 1.0   &   0 \\
        & \qquad RAU1  &   HC   & 64    & 1.0   &   0 \\
    \hline

    \end{tabular}
\end{table*}
\begin{table}[hbt]
    \centering
    \caption{\label{tab:t2} Observable values of the best
    transporting, the fastest transporting and the best ordered
    state for each cell alignment.}
     \begin{tabular}{l l l l l l }

\hline
 &  $\rho_c$  & $\langle\cos\theta\rangle$  &
 $m\cdot|\vec{v}_{g\infty}|$  & $|\vec{v}_{g\infty}|$  &
 $|\vec{v}_{g0}|$  \\
    \hline

                    & 24.0 \qquad\qquad & 0.95 \qquad\qquad &
                    $4\cdot10^3$ \qquad\qquad & $0.69$\qquad\qquad & 0.02 \\
USL \qquad    & 21.7    & 0.85  & $10^2$ & 0.95 & 0.10 \\
                    & 24.2  & 0.95  & $2\cdot10^3$ & 0.72 & 0.05 \\

    \hline

UHL   \qquad &  10.6 & 0.86 & $3\cdot10^3$ & 0.46 & 0.10 \\

    \hline

     \qquad &  15.4 & 0.51 & $2\cdot10^3$   & 0.27 & 0.02 \\
BHL  \qquad & 5.9  & 0.14 & $3\cdot10^2$  & 0.39 & 0.15 \\
     \qquad &  16.3 & 0.57 & $2\cdot10^3$   & 0.27 & 0.01 \\

    \hline

        & 11.7 & 0.93 & $3\cdot10^3$ &  0.47    & 0.06 \\
BHL+L   &  24.9 & 0.82 & $10$       &   0.57    & 0.21 \\
        & 34.0 & 0.98 & $10^3$      &   0.48    & 0.10 \\
\hline
    \end{tabular}
\end{table}

\section{Discussion of Stereotypical Self-Organized Self-Cleaning Attractors}  
In the following we discuss the properties of selected representatives 
of ``families'' of attractor states classified as functioning in the 
accompanying manuscript.

\subsection{Wave Field Analysis}
In order to determine the wave propagation direction in attractor 
states we shall employ the concepts outlined in great detail in 
\cite{Ryser2007}.
The algorithms presented in \cite{Ryser2007} 
were developed for the analysis of experimentally
collected videos of the mucus modulation wave fields
generated by the coordinated movements of 
underlying tracheal cilia.\par
Here, we adopt the definition of the spatial power spectrum 
P$(k_j,k_i)$
as well as of the space-time correlation 
function C$(\Delta j, \Delta i, \Delta t)$.\par  
The spatial power spectrum P$(\vec{k})$ represents the distribution 
of the wave-vectors $\vec{k} = (k_j,k_i)$ and is defined as: \\
\begin{equation}
	P(k_j,k_i)	= \sum_{f=0}^{f=f_N} |F(k_j,k_i,f)|^2~.
	\label{eq:kdist} 
\end{equation}  
In Eq.\ref{eq:kdist} $F(k_j,k_i,f)$ denotes the distribution of 
complex Fourier amplitudes, which was determined by applying a
three-dimensional fast Fourier transform (FFT) to a sequence 
of attractor states ($\Psi(i,j,t')$, where t' $> \tau$).
\par
The space-time correlation function 
C$(\Delta j, \Delta i, \Delta t)$
is given as:\\
\begin{equation}
	\text{C}(\Delta j, \Delta i, \Delta t) = 
	\sum_{i,j,t'} \Psi(i,j,t') \cdot 
	\Psi(i-\Delta i, j-\Delta j, t'+\Delta t)~,
\end{equation}
where $t' > \tau$.
The velocity of the ``mean wave'' $\vec{v}_{wave}$ 
can then be measured by 
\begin{equation}
	\vec{v}_{wave} = \lim_{\tau \rightarrow 0} 
	\frac{\vec{\sigma}(\tau)}{\tau}~,
\end{equation}
where $\vec{\sigma} = (\sigma_j,\sigma_i)$ denotes 
the displacement of the correlation maximum 
with respect to the origin $\Delta j = 0,~ \Delta i = 0$.

\subsection{USL}
\subsubsection{Stereotype 1}  
The parameter setting 
(50$\times$50, USL, DAU, HC, 256\%, 0.25, 0\%) 
generates the attractor state with the highest transport rate,
which amounts to
$m\cdot|\vec{v}_{g\infty}| \approx 4\cdot10^3$ 
(\#$m\cdot$cells/it.).  
The terminal velocity reached amounts to:
$\vec{v}_{g\infty}=(-0.69,0)$.\par

\begin{figure}[!phbt]

  \includegraphics[trim= 0cm 0cm 0cm 0cm,clip,
	scale=0.85]{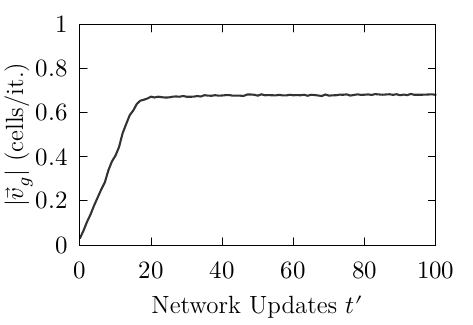}
  \caption{The curve illustrates the transient behaviour of
  the transport speed  
  $|\vec{v}_{g}(t')|$ generated by the parameter set 
  (50$\times$50, USL, DAU, HC, 256\%, 0.25, 0\%). 
  Very quickly, a stable attractor state gets reached, which 
  exhibits (apart from tiny fluctuations originating from 
  the local stochastic interaction component) 
  a constant transport speed.}  
  \label{fig:trans1}
\end{figure}

A snapshot of the considered attractor state 
is shown in Fig.\ref{fig:bestt}. 
Once reached this pattern remains stable. Strictly speaking, 
assuming the state shown in Fig.\ref{fig:bestt} is
represented by  
$\Psi(i,j)$, the reached attractor state alternates 
bettween $\Psi(i,j)$ and its (logical) negation 
$\overline{\Psi(i,j)}$.
Accordingly, once rached the terminal transport velocity
remains constant, as is illustrated in Fig.\ref{fig:trans1}.  

Consequently, a sequence of attractor states appears 
like a 'binary wave' propagating horizontally.
However, this binary wave may actually represent a standing wave.
This leads to a temporally constant spatio-temporal 
correlation function C($\Delta i, \Delta j, \Delta t$)
and thus consisits of a sequence of the same 
autocorrelogram, which is shown 
in Fig.\ref{fig:bestt}.\par 
Consequently, this attractor state may either represent a 
(binary) standing wave or it might be constituted of waves 
propagating either to the right or to the left, which 
can not be determined unambiguously. \par
Finally, all functioning attractor states reached by USL, for which 
either an open boundary or horizontal cylindric boundary 
conditions have been used, closely resemble each other
and particularly reach a stable state representing either 
a horizontally propagating or a standing wave. 
\subsubsection{Stereotype 2} 
\begin{figure*}[!phbt]
	\vspace*{-0cm}
	\hspace*{-0.6cm}
    \begin{minipage}[t]{0.4\linewidth}
    	a)\\   
  		\includegraphics[trim= 0cm 0cm 0cm 0cm,clip,
		scale=0.4]{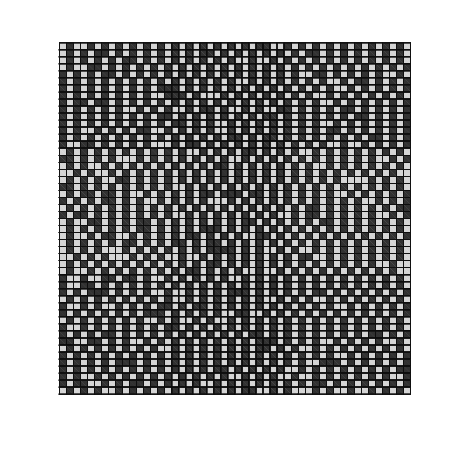}
	\end{minipage}
	\hspace*{1.4cm}
	\begin{minipage}[t]{0.4\linewidth} 
		b)\\  
  		\includegraphics[trim= 0cm 0cm 0cm 0cm,clip,
		scale=0.9]{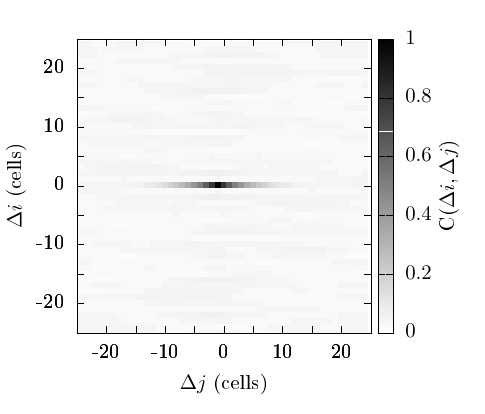}
	\end{minipage}
	
	\hspace{-1.1cm}
	\begin{minipage}[t]{0.4\linewidth}
		~~~~~ c)\\   
  		\includegraphics[trim= 0cm 0cm 0cm 0cm,clip,
		scale=0.9]{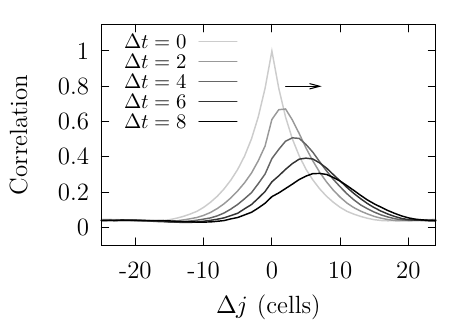}
	\end{minipage}
	\hspace{1.4cm}
	\begin{minipage}[t]{0.4\linewidth}  
		~~~~~~ d)\\ 
  		\includegraphics[trim= 0cm 0cm 0cm 0cm,clip,
		scale=0.9]{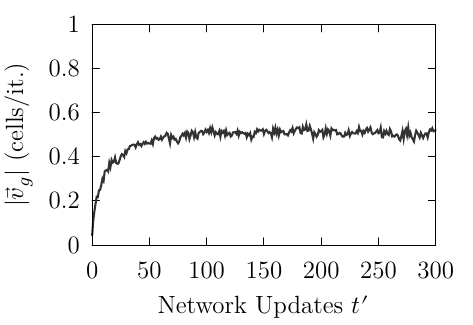}
	\end{minipage}

  	\caption{A snapshot of the second stereotypical attractor 
  	state is shown in panel a. Its modularly organized character
  	is reflected in the correspodning mean autocorrelogram 
  	shown in b. Panel c illustrates the displacement of 
  	the maximum of the correlation function with an increasing 
  	time lag. Panel d depicts the temporal course of the 
  	corresponding transport speed.}  
 	\label{fig:stereo2}
\end{figure*}

The parameter setting 
(50$\times$50, USL, DAU, TO, 64\%, 0.25, 0\%)
generates another representative of a
stereotypical attractor state exhibiting 
well-directed efficient 
transport.     
The pattern reached by this setting does not remain constant 
over time. Instead, it constantly reshapes while keeping 
its modular appearance. 
A snapshot of a particular pattern of this complex attractor 
is provided by Fig.\ref{fig:stereo2}a~.
Fig.\ref{fig:stereo2}b shows the corresponding 
mean auto-correlogram C$(\Delta i,\Delta j, 0)$. 
Similar to the first stereotypical pattern, the autocorrelogram 
is again elongated in parallel to the transport, which 
is directed from right to left: 
$\vec{v}_{g\infty} = (-0.5,0)$. 
However, this stereotype 
exhibits a spatially as well as temporally decreasing 
correlation function C$(\Delta i, \Delta j, \Delta t)$. 
The temporal decrease of the correlation function is 
illustrated in Fig.\ref{fig:stereo2}c, which shows the 
profile of the correlation function along the 
horizontal (given by $\Delta i = 0$) 
for increasing time lags. 
Before the correlation is lost
the maximum of the correlation function 
shifts its position along the horizontal to the right, 
as indicated by the arrow. 
We observed that a higher amount of ``interaction particles'' 
(or a lower value for $\epsilon$) causes the mean wave to 
move faster to the left as well as a faster
de-correlation.\par
Finally, Fig.\ref{fig:stereo2}d 
shows the transient behaviour of 
the average transport speed $|\vec{v}_g|(t)$. 
It can be seen that $|\vec{v}_g|(t)$ 
keeps varying around a typical average transport speed, which 
reflects the permanent re-organization of its associated state. 

\subsection{UHL}
\subsubsection{Stereotype 1}
\begin{figure*}[!phbt]

    \begin{minipage}[t]{0.4\linewidth}
    	\centering
    	a)\\   
  		\includegraphics[trim= 0cm 0cm 0cm 0cm,clip,
		scale=0.25]{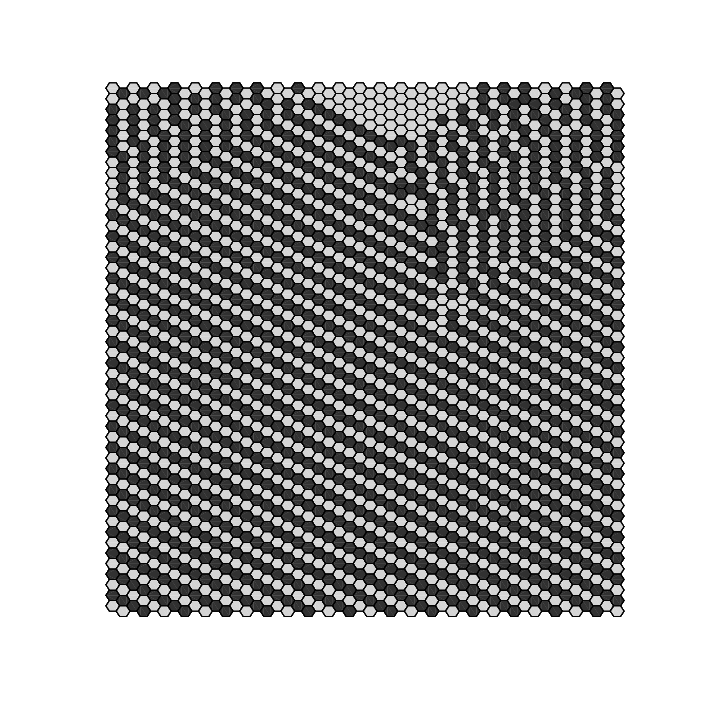}
	\end{minipage}
	\hspace{0.8cm}
	\begin{minipage}[t]{0.4\linewidth} 
		\centering
		~~~~~~~~~~ b)\\  
  		\includegraphics[trim= 0cm 0cm 0cm -0.4cm,clip,
  		scale=0.9]{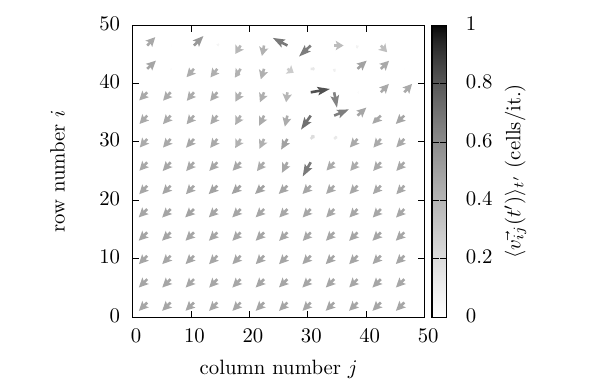}
	\end{minipage}
	\hspace*{1cm}
  	\caption{A snapshot of a stereotypical attractor state, 
  	which gets reached when using UHL, vertical cylindrical 
  	boundary condition and the deterministic update signal DAU, 
  	is shown at the left. The right panel illustrates the 
  	corresponding velocity-field indicating transport to bottom
  	left.}  
 	\label{fig:stereo1UHL}
\end{figure*}

A representative of the first stereotype when 
using the cell alignment UHL 
gets generated with the parameter set 
(50$\times$50, UHL, DAU, VC, 128\%, 1, 0\%).  
A snapshot of the corresponding attractor state 
is shown in Fig.\ref{fig:stereo1UHL}. 
Once reached this state remains stable (besides its 
alternation between the illustrated state and its logical 
negation) and its corresponding terminal transport 
velocity amounts to $\vec{v}_{g\infty} = (-0.26, -0.28)$.  
The binary nature of the considered attractor state 
does not allow us to determine the propagation direction 
of the mean wave. Consequently, the 
dominating wave-like binary structure either represents 
a wave propagating along or against the transport direction, 
or it might actually represent a standing wave.   

\subsubsection{Stereotype 2}
\begin{figure*}[!phbt]
	\hspace*{-1.7cm}
    \begin{minipage}[t]{0.38\linewidth}
    	a)\\   
  		\includegraphics[trim= 3.5cm 1.5cm 2.5cm 1.5cm,clip,
		scale=0.25]{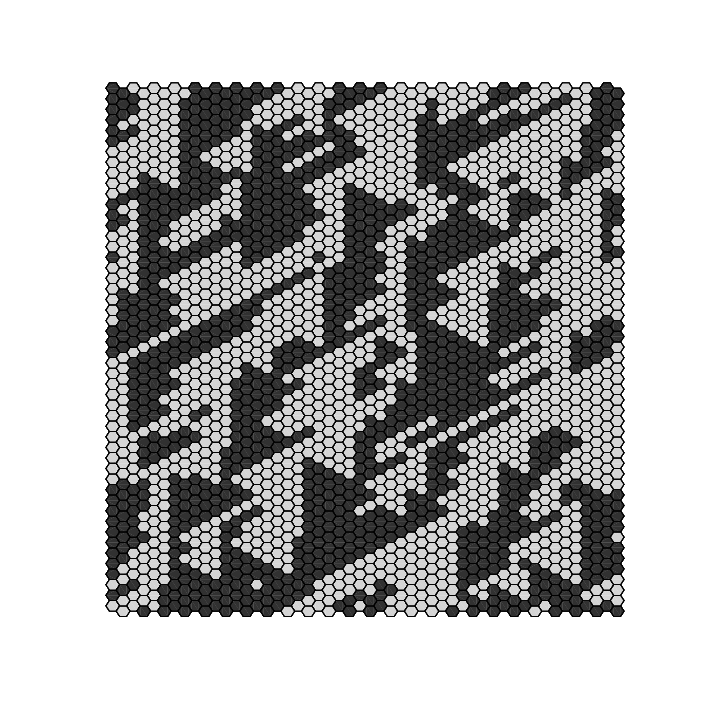}
	\end{minipage}
	\hspace*{0cm}
	\begin{minipage}[t]{0.38\linewidth} 
		~~~~~~~~~~~~~~~~~~~~~~ b)\\  
  		\includegraphics[trim= 0cm 0cm 0cm 0cm,clip,
  		scale=0.93]{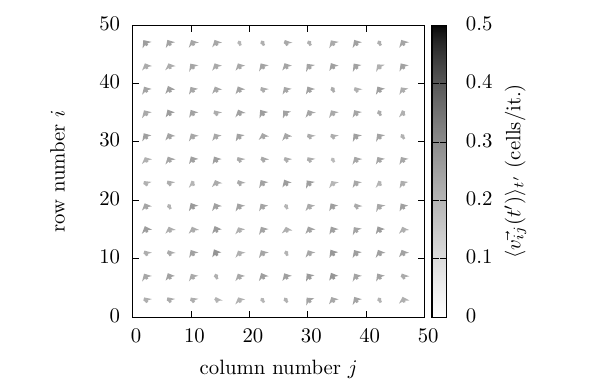}
	\end{minipage}
	
	\hspace{-1.0cm}
	\begin{minipage}[t]{0.38\linewidth} 
		~~ c)\\  
  		\includegraphics[trim= 0cm 0cm 0cm 0cm,clip,
  		scale=0.92]{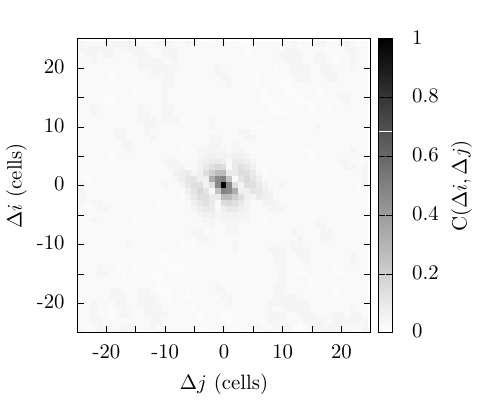}
	\end{minipage}
	\hspace{1.1cm}
	\begin{minipage}[t]{0.38\linewidth} 
		~~~~~~~ d)\\  
  		\includegraphics[trim= 0cm 0cm 0cm 0.15cm,clip,
  		scale=0.98]{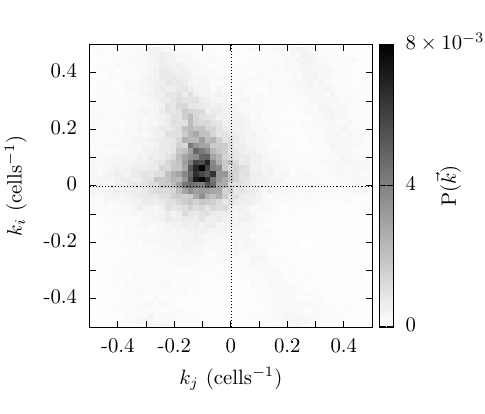}
	\end{minipage}

  	\caption{The application of toric boundaries and DAU 
  	to the unidirectional hexagonal lattice leads to triagnular 
  	appearing gliders as indicated by the snapshot of an attractor
  	shown in panel a). The corresponding velocity field is visualized in panel b). Panel c) and d) represent thecorresponding 
  	distribution of wave vectors and the mean autocorrelogram of the attracting states, respectively.}  
 	\label{fig:stereo2UHL}
\end{figure*}

The second stereotype gets generated with 
toric boundary conditions and otherwise 
similar parameter values, an example 
is provided by:  
(50$\times$50, UHL, DAU, TO, 256\%, 1, 0\%).
The attractor reached by this setting is 
characterized by mono-phase areas 
gliding over the network. These areas/gliders typically 
have a triangular-like shape, as indicated by the
snapshot provided in Fig.\ref{fig:stereo2UHL}a.\par
Even if the terminal transport is   
slow: $\vec{v}_{g\infty} = (-0.06, 0.11)$, 
the correspoding velocity field is well-directed
(see Fig.\ref{fig:stereo2UHL}b).\par
The strong pixelation of the correlation function
made it difficult to track the maximum of the
spatio-temporal  
correlation function. Therefore,  
the direction of the weighted mean wave vector
provides a more accurate estimate for the wave propagation direction. The weighted mean wave vector amounts to
$\langle\vec{k}\rangle = (-0.05,0.05)$.
The mean autocorrelation function and the 
relative power spectral density P$(\vec{k})$
is shown in Fig.\ref{fig:stereo2UHL}c and 
Fig.\ref{fig:stereo2UHL}d, respectively.  
\subsection{BHL}
\begin{figure*}[!phbt]
	\hspace{-1.2cm}
	\begin{minipage}[t]{0.38\linewidth}
  	\includegraphics[trim= 0cm 0cm 0cm 0cm,clip,
	scale=0.6]{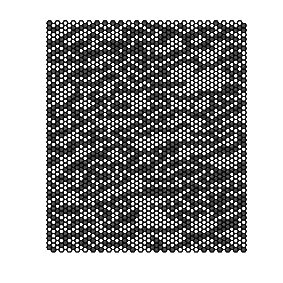}
	\end{minipage}
	\hspace*{0.5cm}
	\begin{minipage}[t]{0.38\linewidth}
  	\includegraphics[trim= 0cm 0cm 0cm 0cm,clip,
	scale=0.9]{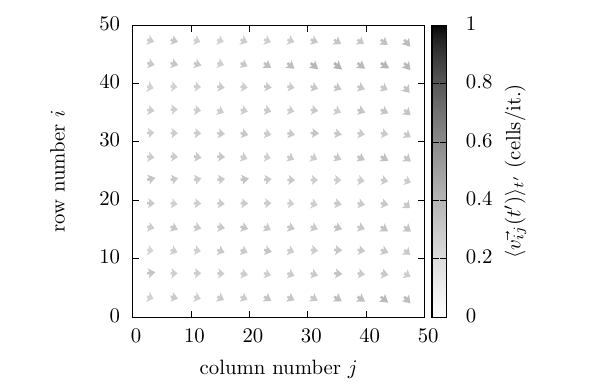}
	\end{minipage}
  	\caption{The left panel shows a snapshot of a stereotypical 
  	functioning attractor state reached by BHL. The associated 
  	transport velocity field is shown in the right panel.}  
 	\label{fig:stereoBHL}
\end{figure*}

Settings using the bidirectional hexagonal alignment
generate very similar functioning attractor states with vertical 
cylindric and toric boundaries. 
The best transporting state classified as functioning 
is generated by 
(50$\times$50, BHL, DAU, TO, 256\%, 1, 0\%).
A snapshot of the generated attractor is shown in 
Fig.\ref{fig:stereoBHL}. 
The state as well as sequences of states look rather chaotic, 
which is reflected in the broad 
distribution of $\vec{k}$-vectors. 
However, a slight tendency of waves propagating 
to the top right can be observed by the bare eye. 
Correspondingly, the expectation value 
amounts to $\langle\vec{k}\rangle = (0.04, 0.03)$.
Even if the wavefield appears rather disordered, the 
velocity field is well organized, since the 
(temporally averaged) local velocities tend to head to 
the right
(see Fig.\ref{fig:stereoBHL}b).
The terminal transport velocity is rather slow and 
amounts to: $\vec{v}_{g\infty} = (0.2,0.0)$.
\subsection{BHL+L} 
\subsubsection{Stereotype 1} 
\begin{figure}[!phbt]
  	\includegraphics[trim= 0cm 0cm 0cm 0cm,clip,
	scale=0.6]{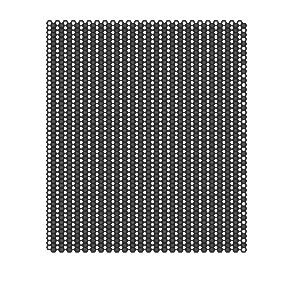}
	\vspace{-0.5cm}
  	\caption{Snapshot of the crystal-like appearing attractor 
  	reached by the application of horizontal cylindric or open 
  	boundaries with the update RAU1 to BHL+L. }  
 	\label{fig:stereo1BHLL}
\end{figure}

The usage of open or horizontal cylindric boundaries 
and the update scheme RAU1 leads to 
a crystal-like structure. A particular 
example generated by the parameter set 
(50$\times$50, BHL+L, RAU1, OP, 256\%,1,0\%)
is shown in Fig.\ref{fig:stereo1BHLL}.\par
Since settings including BHL+L impose a prescribed 
four-phase cycle on two actuators forming an ``L'', 
we do not further have to deal with binary images and 
can exploit the accessibility of four phases 
for the analysis of the wavefield.\par
The particular stereotypical attractor corresponding to 
the snapshot shown in Fig.\ref{fig:stereo1BHLL} 
exhibits a plane wave, whose propagation direction 
can not be determined unambiguously, since in this particular state 
the exhibited wave remains 
binary, despite the accessibility of four L-phases 
(L's along the horizontal remain sperated by 
phase shifts of $\pi$). 
The terminal transport velocity amounts to: 
$\vec{v}_{g\infty} \approx (-0.5,0)$ (cells/it.)~.  
\subsubsection{Stereotype 2}
\begin{figure}[!phbt]
	\vspace{-1cm}
	\centering
  	\includegraphics[trim= 0cm 0cm 0cm 0cm,clip,
  	scale=0.8]{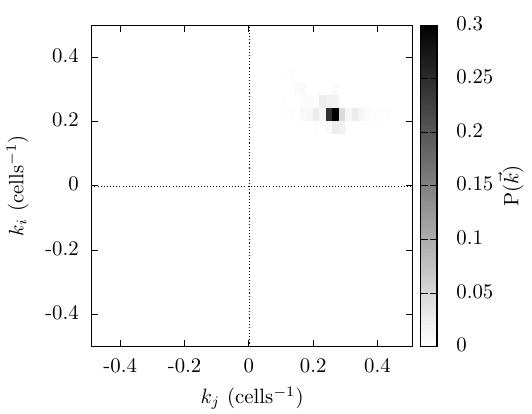}
  	\caption{Power spectral density of the attractor 
  	state reached by 
  	settings using BHL+L together with the deterministic update
  	update scheme DAU.}  
 	\label{fig:stereo1BHLLkxky}
\end{figure}

The parameter set 
(50$\times$50, BHL+L, DAU, OP, 256\%, 1, 0\%) 
exhibits the highest transport rate when
BHL+L is used  and represents another stereotypical 
attractor state. A snapshot of an attractor 
is shown in Fig.\ref{fig:bestt}. Such attractors
exhibit a transport velocity of $\approx (-0.5,0)$ .
The corresponding distribution of the
relative power spectral density P$(k_j,k_i)$ is shown in 
Fig.\ref{fig:stereo1BHLLkxky}. The weighted mean 
wave vector is directed to the right top and amounts to:
$\langle\vec{k}\rangle \approx (0.25,0.25)$.\\
\subsubsection{Stereotype 3,4}
When using BHL+L together with SRAU1 and SRAU2 the model
self-organizes towards two further stereotypical 
stable attractor states. 
In the accompanying paper one representative snapshot of each
type has been illustrated. \par 
Since SRAU1 as well as SRAU2 impose a wave-like activation 
of the actuators the wave velocity is purely imposed by the 
activation scheme, which is why we shall skip the discussion 
considering the characteristics of the exhibited wave field
for these two stereotypes. 
\subsubsection{Stereotype 5}
\begin{figure}[!phbt]

	\centering	
  	\hspace{-0.5cm}\includegraphics[trim= 0cm 0cm 0cm 0cm,clip,
  	scale=0.6]{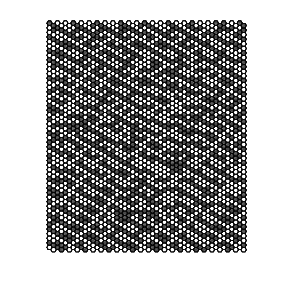}
  	\includegraphics[trim= 0cm 0.15cm 0cm 0cm,clip,
  	scale=0.83]{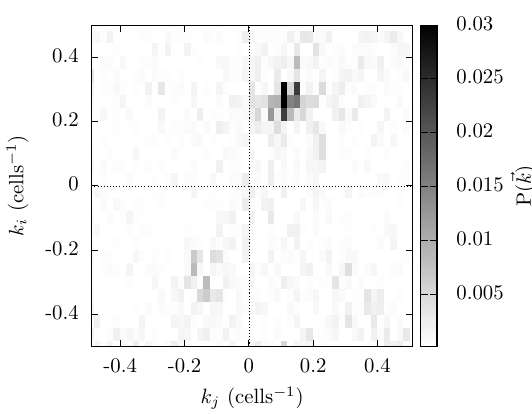}

  	\caption{Settings 
  	using the prescribed cyclic four-phase motion BHL+L reach 
  	a rather modular appearing  
  	attractor state (snapshot shown on the left) 
  	exhibiting defects, which keeps re-organizing, 
  	when applying toric boundary conditions. The right panel 
  	shows the corresponding spatial power spectrum.}  
 	\label{fig:stereo5}
\end{figure}

A small subset of settings generates functioning 
attractors with toric boundary conditions. 
A representative is given by the parameter set 
(50$\times$50, BHL+L, DAU, TO, 256\%, 1, 0\%). 
In contrast to the other stereotypes, reached 
by settings using the prescribed cyclic four-phase motion,
these states keep re-organizing and thus, the 
spatio-temporal correlation function 
decreases with increasing time lags.
The corresponding attractors display many defects
(as indicated by the snapshot provided in 
Fig.\ref{fig:stereo5})
and a sequence of states appears rather chaotic, while 
the propagation of organized wavelets 
heading towards the right top is clearly recognizable.
Accordingly, the relative power spectral density 
illustrated in Fig.\ref{fig:stereo5} displays a clear 
peak wave vector: $\vec{k}_{peak} = (0.1, 0.25)$. 
The exhibited terminal transport velocity amounts to  
$\vec{v}_{g\infty} = (0.3, 0)$ and therefore encloses an 
angle of $\approx \pi/3$  with the peak wave vector.

\end{document}